\documentclass[fleqn,usenatbib]{mnras}
\usepackage{newtxtext,newtxmath}
\usepackage{graphicx}	
\usepackage{etoolbox}
\usepackage{xspace}

\newcommand{\lya}{\ifmmode {\rm Ly}\alpha \else Ly$\alpha$\fi\xspace}

\newcommand{\Gizmo}{{\small GIZMO}\xspace}

\newcommand{\Music}{{\small MUSIC}\xspace}

\newcommand{\Msun}{\rm M_{\odot}}
\newcommand{\Mbh}{M_{\rm BH}}
\newcommand{\Mseed}{M_{\rm seed}}

\newcommand{\Mdot}{\dot{M}_{\rm BH}}
\newcommand{\Mstar}{M_\star}
\newcommand{\Lsun}{\rm{L_{\odot}}}

\newcommand{\kms}{\rm{km s$^{-1}$}}

\graphicspath{{./figures/}}

\title[The Seeds and Growth of the First Quasars]{The Formation  of the First Quasars. I. The Black Hole Seeds, Accretion and Feedback Models}
\author[Q. Zhu et al.]{Qirong Zhu$^{1}$\thanks{E-mail: qirongz@andrew.cmu.edu}, 
Yuexing Li$^{2, 3, 4}$, 
Yiting Li$^{5}$, 
Moupiya Maji$^{6}$, 
Hidenobu Yajima$^{7}$, \newauthor
Raffaella Schneider$^{4,8}$,  
and Lars Hernquist$^{9}$  
\vspace{0.15cm}\\
$^{1}$ Department of Physics, Carnegie Mellon University, 5000 Forbes Avenue, Pittsburgh, PA 15213, USA \\
$^{2}$ Department of Astronomy \& Astrophysics, The Pennsylvania State University, University Park, PA 16802, USA\\
$^{3}$ Institute for Cosmology and Gravity, The Pennsylvania State University, University Park, PA 16802, USA\\
$^{4}$ Dipartimento di Fisica, Sapienza, Universit$\grave{a}$ di Roma, Piazzale Aldo Moro 5, 00185, Roma, Italy  \\
$^{5}$ Department of Physics, University of California, Santa Barbara, CA 94720, USA \\
$^{6}$ Observatoire de Genève, Chemin des Maillettes, 51, 1290 Versoix, Switzerland \\
$^{7}$ Center for Computational Sciences University of Tsukuba, Tsukuba, Ibaraki 305-8577, Japan\\
$^{8}$ INFN, Sezione di Roma I, P.le Aldo Moro 2, 00185 Roma, Italy  \\
$^{9}$ Harvard-Smithsonian Center for Astrophysics, Harvard University, 60 Garden Street, Cambridge, MA 02138, USA
}

\date{Accepted XXX. Received YYY; in original form ZZZ}
\pubyear{2020}

\begin{document}
\pagerange{\pageref{firstpage}--\pageref{lastpage}} 
\label{firstpage}
\maketitle

\begin{abstract}
Supermassive black holes (SMBHs) of $\sim 10^9\, \Msun$ are generally believed to be the central engines of the luminous quasars observed at $z\gtrsim6$, but their astrophysical origin remains elusive. The $z\gtrsim6$ quasars reside in rare density peaks, which poses several challenges to uniform hydrodynamic simulations. To investigate the formation of these distant quasars, we perform a suite of zoom-in simulations on a favorable halo, with a mass of $\sim 10^{13}\, \Msun$ at $z = 6$ and a history of multiple major mergers, ideal for BH growth. We test BH seeds of $10 - 10^6\, \Msun$, and various accretion and feedback models, including thin-disk and slim-disk accretion. We find, contrary to previous studies, that light seeds of $\lesssim 10^3\, \Msun$ fail to grow to $10^8\, \Msun$ by $z\sim 6$ even with super-critical accretion; that the hyper-Eddington mode leads to lower accretion rates than the Eddington-limited case due to stronger feedback, resulting in significantly smaller BHs by two orders of magnitude; and that while the super-critical model boosts the growth of low-spin BHs, for high-spin BHs the mass may be reduced due to increased radiative feedback. Our simulations show that the first $10^8 - 10^9\, \Msun$ SMBHs may grow from heavy seeds of $\gtrsim 10^4\, \Msun$ via Eddington-limited or mild super-critical accretion facilitated by gas-rich mergers and self-regulated by feedback, and they co-evolve with their host galaxies, producing bright quasars such as those at $z\sim$6 and ULAS J1342+0928, currently the most distant quasar at z = 7.54.
\end{abstract}


\section{Introduction}

Over the last two decades, more than 200 quasars have been discovered at high redshift $z \gtrsim 6$ \citep[e.g.,][]{Fan2001,  Willott2010, Venemans2015, Jiang2016, Banados2016, Reed2017, Matsuoka2018, Shen2019}, among them a handful are at $z>7$ \citep[e.g.,][]{Mortlock2011, Matsuoka2019, Wang2019, Yang2019}, with the record-holder of the most distant quasar being at $z= 7.54$ \citep{Banados2018, Yang2020} when the age of Universe is less than $700$ Myr (e.g., see a recent review by \citealt{Inayoshi2020}). 

Multi-band  observations suggest that these quasars are powered by supermassive black holes (SMBHs) of mass $\sim 10^8 - 10^9\, \Msun$, with the most massive of all, SDSS J0100+2802, having an estimated mass of $1.2\times10^{10}\, \Msun$ at $z=6.3$  \citep{Wu2015}, while the two most distant ones, ULAS J1342+0928 and the recently reported {\it P$\bar{o}$niu$\bar{a}$'ena} (J1007+2115, \citealt{Yang2020}), having a mass of $7.8\times 10^8\, \Msun$ at $z=7.54$ \citep{Banados2018}, and $1.5 \times 10^9 \rm  M_\odot$ at $z = 7.52$ \citep{Yang2020}, respectively.

Furthermore, the observations suggest that these SMBHs accrete at sub- or near-Eddington rates \citep[e.g.,][]{Willott2010, Mazzucchelli2017, Onoue2019, Shen2019, Vito2019}. \citet{Shen2019} reported a median value of the Eddington ratio $\lambda_{\rm Edd} \sim 0.3$ for a sample of 50 $z>5.7$ quasars, while \citet{Mazzucchelli2017} found an average $\lambda_{\rm Edd} \sim 0.4$ at $z \gtrsim 6.5$, and \citet{Vito2019} reported the a range of $0.2 - 1.76$ for a sample of 21 $z > 6$ quasars. More recently, \citet{Davies2019} suggested that the $z >7$ quasars may have high accretion rates without violating the Eddington limit due to low radiative efficiency. 

These extraordinary observations raise fundamental questions about the origin and the rapid formation of $\sim 10^8 - 10^{10}\, \Msun$ SMBHs within the first billion yeas after the Big Bang: what were the BH seeds?  And, how did they grow? 

Unlike stellar mass black holes, the origin of SMBHs is much less well understood. In recent years, three major seed scenarios have been proposed (see reviews by \citealt{Volonteri2010, Volonteri2012a, Latif2016, Woods2019}, and more recently by \citealt{Greene2020}): (1) light seeds  $\sim 10^{1-2}\, \Msun$ from Pop~III stars \citep[e.g.,][]{Madau2014, Lupi2016, Ryu2016, Valiante2016, Valiante2017, Pezzulli2016, Pezzulli2017};  (2) intermediate seeds of $\sim 10^{3 - 4}\, \Msun$ from the collapse of super-massive stars \citep[e.g.,][]{Ferrara2014, Woods2017, Woods2019} or stellar collisions \citep[e.g.,][]{Katz2015, Yajima2016, Sakurai2017, Tagawa2019}; and (3) heavy seeds of $\sim 10^{5 - 6}\, \Msun$ from the  collapse of hot and dense gas clumps, so-called direct collapse black holes (DCBHs) \citep[e.g.,][]{Agarwal2012, Glover2015a, Glover2015b, Pacucci2015a, Inayoshi2016, Valiante2016, Regan2017, Chon2018, Regan2019, Wise2019, Luo2020}.

The rapid growth of the BH seeds, on the other hand, poses yet another long-standing challenge. Assuming the seeds grow at Eddington-limited accretion, the timescale for a seed to grow from $\Mseed$ to a given BH mass $\Mbh$ is given by \citep[e.g.,][]{Soltan1982, Woods2019, Inayoshi2020}:
\begin{equation}\\
t_{\rm grow} \approx \frac{0.45}{f_{\rm duty}}\frac{\epsilon_{\rm r}}{(1-\epsilon_{\rm r})}
~\ln \left(\frac{M_{\rm BH}}{M_{\rm seed}}\right)~{\rm Gyr,} 
\label{eq:tgrow}
\end{equation}
where $f_{\rm duty}$ is the duty cycle of Eddington accretion, and $\epsilon_{\rm r}$ is the radiative efficiency. Assuming a radiatively-efficient, thin-disk accretion model \citep{Shakura1973} with $\epsilon_{\rm r}=0.1$, a seed will grow exponentially with an e-folding time of $\sim 50\,  {\rm Myr}$. So for $100\, \Msun$ BH seeds from Pop~III stars, constant accretion at the Eddington rate with a full duty cycle ($f_{\rm duty} =1$), $\sim 810\,  {\rm Myr}$ are required to grow to $10^9\, \Msun$.  

In reality, it may be difficult to achieve such an ideal growth scenario as many simulations have shown inefficient accretion onto Pop~III remnants due to strong radiative feedback \citep[e.g.,][]{Alvarez2009, Jeon2012, Tanaka2012, Smith2018}.  To decrease the growth timescale, a number of solutions have been proposed: (1) super- or hyper-Eddington accretion for light seeds \citep[e.g.,][]{Pacucci2015a, Inayoshi2016, Ryu2016}; (2) jump-starting from heavy seeds with a radiatively-efficient, thin-disk accretion model \citep[e.g.,][]{Valiante2016, Pezzulli2017, Pacucci2017, Lupi2019}; and (3) super-critical accretion through a radiatively-inefficient ``slim disk" solution in which the radiative efficiency $\epsilon_{\rm r}$ drops quickly as the accretion rate approaches super-Eddington \citep[e.g.,][]{Sadowski2009, Madau2014, Volonteri2015, Lupi2016, Pezzulli2016}.

Indeed, numerous semi-analytical works have suggested that light seeds from Pop~III stars can grow to $10^9\, \Msun$ by $z \sim 6$ with super- or hyper-Eddington accretion \citep[e.g.,][]{Tanaka2009, Agarwal2013,Pezzulli2016,Ricarte2018}. To test these theories in detail, a number of groups have used hydrodynamical simulations to model the growth of SMBHs in the context of galaxy formation, taking into account gas dynamics, star formation, black hole accretion, and feedback processes. In particular, hydrodynamics simulations are essential for capturing the non-linear and intricate interplay between BH feedback and galaxy formation.

In an early study, \citet{Li2007} presented the first hydrodynamical simulation of the formation of $z \sim 6$ quasars, following the hierarchical mergers of protogalaxies of a $\sim 10^{13}\, \Msun$ halo at $z \sim 6$, the most massive one in a $\sim 2.5~{\rm Gpc}^3$ volume, with $\sim 10^5\, \Msun$ BH seeds and a self-regulated BH growth model, as well as other important physical processes of star formation and feedback. They found that the seed BHs grow rapidly via efficient gas accretion facilitated by gas-rich mergers, producing a bright quasar at $z \sim 6.5$ with a $\sim 10^9\, \Msun$ SMBH.  Subsequently, post processing the simulation with radiative transfer calculations and a supernova dust model, \citet{Li2008} found that the quasar reproduced a number of observational properties of many luminous quasars at $z  \sim 6$.

Since then, BHs were included in direct cosmological simulations \citep{DiMatteo2008}, and more recently in a number of impressive large-scale simulations ($box size \sim 100 - 500\, {\rm Mpc}$), such as Illustris \citep[][]{Vogelsberger2014, Genel2014}, EAGLE \citep{Schaye2015, Crain2015}, Horizon-AGN \citep{Dubois2014}, MassiveBlack \citep{Khandai2015}, BlueTides  \citep{Feng2016}, Magneticum \citep{Dolag2016}, IllustrisTNG~\citep[][]{Pillepich2018b, Springel2018, Weinberger2018, DeGraf2020}, and Simba \citep{Dave2019}. However, most of these simulations used similar BH recipes such as heavy seeds of $10^{4-6}\, \Msun$ in $10^{9} - 10^{10}\, \Msun$ halos and Eddington-limited Bondi accretion, and due to the limitation in volume, or in resolution, or in the redshift range of the simulation, none of them produced $10^9\,  \Msun$ SMBHs at $z > 6$ (the BlueTides Phase II was recently extended to $z \sim 7.4$ and it formed a single SMBH of $\sim 7\times 10^8\, \Msun$ in the end, \citealt{Tenneti2019}). It is a common feature shared by the new generation of simulations that much stronger feedback is introduced, in particular a large mass loading in the outflow than models prior to 2014, to reproduce the stellar mass functions. It is interesting to ask whether the rare density peaks at high-$z$ are then still able to produce luminous quasars. If not, what additional processes are required to reconcile discrepancies with observations? What do the quasar hosts look like? To address these questions, hydrodynamic simulations again prove to be a powerful tool.

It appears that high density peaks in the primeval initial conditions and high mass resolution are both required to study high-$z$ quasars, which makes uniform cosmological simulations inefficient. To overcome the limitation set by the box size and resolution, zoom-in re-simulations of targeted regions, while retaining the large-scale tidal field using lower resolution particles, offer a viable route to model high-$z$ quasars.  \citet{Sijacki2009} performed a set of 6 zoom-in re-simulations of a massive halo with a mass $10^{13}\, \Msun$ at $z=6$ from the Millennium simulation with various resolutions, using conventional BH recipes ($10^{5-6}\, \Msun$ seed in $10^{9} - 10^{10}\, \Msun$ halo) but also including BH gravitational recoil. They produced a $10^9\, \Msun$ SMBH at $z \sim 6$. Recently, several groups have performed various zoom-in simulations of massive halos using different codes, BH models and at various resolutions, and they achieved similar success in producing $\sim 10^9\, \Msun$ SMBHs at $z \gtrsim 6$ \citep[e.g.,][]{Feng2014, Costa2014, Curtis2016, Smidt2018, Huang2019, Lupi2019}. 

Despite the impressive progress, most of the simulations have used similar models for BH seeds, accretion, and feedback (e.g., heavy seeds, Eddington-limited Bondi accretion, thin-disk with $\epsilon_{\rm r}=0.1$), with little investigation of other scenarios. It is therefore important to  comprehensively study and compare various seed schemes, BH growth models and feedback processes.  To this end, we have performed a suite of 15 zoom-in cosmological simulations of a $10^{13}\, \Msun$ halo at $z \sim 6$ similar to that of \citet{Li2007}, with different BH seeds in the mass range of $10^1 - 10^6\, \Msun$, accretion modes from near-Eddington to super-, hyper- and unlimited-Eddington, and feedback models including both thin- and slim-disk accretion and radiation.   

This paper is organized as follows. In Section~\ref{sec:methods}, we describe the cosmological simulations and initial conditions in Section~\ref{sec:ic}, the physical processes of star formation and feedback in Section~\ref{sec:sfr}, the BH seeds in Section~\ref{sec:seed}, BH accretion in Section~\ref{sec:acc} and feedback in Section~\ref{sec:feedback}, and mergers in Section~\ref{sec:merger}. In Section~\ref{sec:ref}, we present the formation of the first quasars with our fiducial simulation, including the emergence of the first galaxies in Section~\ref{sec:firstgal}, the growth history of a $10^9\, \Msun$ SMBH at $z \sim 6$ in Section~\ref{sec:smbh}, and the host galaxy properties in Section~\ref{sec:host}. In Section~\ref{sec:tests}, we test different models with 15 zoom-in simulations, including BH seeds (Section~\ref{sec:testseed}), Eddington limits (Section~\ref{sec:testel}), super-critical accretion and feedback (Section~\ref{sec:testfb}),  and Bondi accretion variations and other parameters (Section~\ref{sec:testba}), and we highlight the viable models for $z \sim 6$ quasars in Section~\ref{sec:viablemodels}. We discuss the limitations of our simulations and comparison with previous works in Section~\ref{sec:discussion}, and summarize our findings in Section~\ref{sec:summary}.

\section{Methods}
\label{sec:methods}
To study the formation of $z \gtrsim 6$ quasars with $10^9\, \Msun$ SMBHs, we have performed a a suite of 15 zoom-in cosmological simulations of the same massive halo with different BH seeds of $10^1 - 10^6\, \Msun$, accretion modes from limited- to unlimited-Eddington, and feedback models including both thin- and slim-disk accretion and radiation. We describe the simulations, the physical processes, and the models in this Section.

\subsection{The Cosmological Zoom-in Simulations}

\subsubsection{Hydrodynamics Code}

The simulations were carried out with the Lagrangian particle code \Gizmo \citep{Hopkins2015} with the implementation of physical processes by \citet{Zhu2016}. \Gizmo was developed to overcome  problems encountered by smooth particle hydrodynamics (SPH) methods \citep{Agertz2007, Bauer2012, Zhu2015, Zhu2016}. It solves the Riemann problem at the interface of two overlapping fluid elements, embodying the advantages of both SPH and grid-based methods. As a result, \Gizmo captures the instabilities of fluid mixing well, and greatly reduces numerical noise and artificial viscosity. 

As is the case with many cosmological simulations, sub-grid recipes are used to describe physical processes due to insufficient resolution to resolve individual stars or BHs \citep[e.g.,][]{Vogelsberger2014, Schaye2015, Khandai2015, Springel2018, Pillepich2018b, Dave2019}. In \citet{Zhu2016}, we have implemented in \Gizmo a sub-grid recipe for star formation, multi-phase interstellar medium (ISM) and stellar feedback based on \citet{Springel2003}. Similarly, we have implemented a sub-grid recipe for BHs and related physical processes based on \citet{Springel2005model} and \citet{DiMatteo2005}, in which BH accretion is calculated using a spherical Bondi model \citep{Bondi1944}.

\subsubsection{Initial Conditions}
\label{sec:ic}
The target halo for the 15 zoom-in simulations is the most massive one within a $\sim{\rm Gpc}^3$ volume, with a total mass of $1.16 \times 10^{13}\, \Msun$ at $z=6.1$. Note that while the halo is selected using the same technique as that in \citet{Li2007}, the subsequent re-simulations are different in that the former performed a high-resolution hydrodynamical simulation of 6 major mergers from $z=14.4$ to $z=6$ extracted from the merger tree of the halo, but all the simulations here are full cosmological zoom-in simulations with hydrodynamics. This galaxy was shown to have a number of physical and panchromatic properties similar to some host galaxies of $z \sim 6$ quasars \citep{Li2007, Li2008}.

To obtain the initial conditions for the zoom-in simulations, we first ran a dark matter - only parent simulation with $1024^3$ particles in a 1~Gpc/$h$ box. We then selected the largest halo at $z=6$ with the friend-of-friend (FOF) group-finding algorithm \citep{Davis1985}, and regenerated the initial conditions, which contain both dark matter and baryons, with a nested higher-resolution region of this halo using the multi-scale initial conditions code {\Music}\citep{Hahn2011}. This high-resolution region has a size of $\sim 10\ {\rm Mpc}/h$ (comoving) for the selected halo. Finally, the zoom-in simulations with the new initial conditions were run with a full list of comprehensive physical processes and models from $z=99$ to $6$.

\subsubsection{Numerical Resolution and Cosmological Parameters}
\label{sec:resolution}
To determine the resolution for the zoom-in region, we had to consider a balance between the number of simulations and the cost of computations due to limited resources. In order to test the various models of BH seeding, accretion and feedback, we need at least 15 simulations, in addition to another set of 30 simulations to statistically study the properties of quasar hosts which will be presented in an upcoming paper. It is desired to have the same resolution for all these simulations for consistent comparisons, so we chose a reasonable resolution based on previous simulations \citep[e.g.,][]{Sijacki2009, Feng2014, Costa2014, Curtis2016}. Both \citep{Sijacki2009} and \citep{Feng2014} performed extensive resolution studies for their zoom-in simulations of early quasars, and found no significant effect of resolution on the final BH mass even when the resolution differed by a factor of 64, thanks to the numerically well behaved feedback model. Therefore, we chose a cost-effective resolution close to those of \citet{Sijacki2009} and \citet{Feng2014} with $m_{\rm gas} = 6.5\times10^5\, \Msun$/$h$,  $m_{\rm DM} = 3.5\times10^6\,  \Msun$/$h$, $\epsilon = 0.5\, {\rm kpc}/h$, for particle mass resolution and gravitational softening length, respectively. 

It is a significant challenge to include small BHs in cosmological simulations as we cannot achieve sufficient numerical resolution to resolve these objects due to limited computational resources. However, a large mass ratio between particles may cause two-body heating and other numerical problems. To alleviate these potential numerical artifacts, we adopted a numerical technique used by \citet{Springel2003} to treat star formation, in which a fraction of a gas particle becomes a new star, to treat BH seeds of $10 - 100\, \Msun$ from remnants of Pop~III stars in our simulations. The newly formed BH is treated as a fraction of the parent Pop~III star particle from which the BH formed, while the parent star particle becomes a ``ghost particle" associated with the new BH. The ``ghost particle" interacts with other particles only via gravity; the BH particle has the same dynamics as its ``ghost" while maintaining its own mass and accretion activity. This technique reduces two-body relaxation as the star particles have comparable mass as other particles, and it enables small BHs to robustly evolve and grow consistent with other massive BHs.

The simulations were run with a set of cosmological parametera consistent with the WMPA 1- and 5-year results \citep{Spergel2003, Komatsu2009}: $\Omega_{\rm M} = 0.25$, $\Omega_{\rm b} = 0.04$, $\Omega_{\Lambda} = 0.75$,  $h = 0.73$ and $\sigma_8 = 0.9$. We have also re-run the fiducial simulation with the cosmological parameters from \citet{Planck2016}: $\Omega_{\rm M} = 0.308$, $\Omega_{\rm b} = 0.048$ $\Omega_{\Lambda} = 0.692$,  $h = 0.678$ and $\sigma_8 = 0.82$, and found no significant differences between the two simulations, as shown in Section~\ref{sec:tests}.  

\subsubsection{List of Simulations}
\label{sec:simu_list}

As listed in Table~\ref{table:sims}, the 15 zoom-in simulations of the same halo aim to test various models and parameters, with different BH seeds of $\Mseed = 10^1 - 10^6\, \Msun$/$h$, Eddington limits of  $\max(\lambda_{\rm Edd}) = 1 - 10^4$,  and feedback from both constant-$\epsilon_{\rm r}$ radiatively-efficient accretion and varying-$\epsilon_{\rm r}$ from super-critical accretion models, as well as variations in Bondi accretion and cosmological parameters. Note in the simulations, we use units with the dimensionless Hubble parameter $h$ such as $\Msun$/$h$ and ${\rm kpc}/h$, but we present the results with $h$-less units for direct comparison with observations. The physical processes will be described in detail in the next sections.

\begin{table*}
\begin{center}
\caption{Summary of the 15 cosmological zoom-in simulations and model parameters.}
\label{table:sims}
\begin{tabular}{l|l|l|l|l|l|l|l}
\hline
Simulation & Model Description & $\rm{M}_{\rm seed}$  & Seeding                           & Accretion \& Feedback & $\epsilon_{\rm r}$  & $\max(\lambda_{\rm edd})$ & $\rm {M_{BH}}$ \\ 
                  &           			 & $[\Msun/h]$                &  Scheme                          &                                      &                               &                                              & $[\Msun/h]$\\ 
\hline
\multicolumn{8}{l}{\bf Seed Models} \\
\hline
S1		& Light Seed        & $10^1$ 	 	& PopIII   	     & Super-critical, $a=0$  &  Eq.~(\ref{eq:sc}) & $10^4$  & $7.3\times10^4$ \\  
S2		& Light Seed        & $10^2$ 	 	& PopIII         & Super-critical, $a=0$  &  Eq.~(\ref{eq:sc}) & $10^4$  & $2.5\times10^6$ \\
S3 		& Intermediate Seed   & $10^3$  	& Halo           &  Thin disk   		  &  0.1                      & 1	        & $5.1\times10^6$ \\ 
S4 		& Intermediate Seed   & $10^4$        & Halo           &  Thin disk   		  &  0.1                      & 1           & $2.4\times10^8$  \\ 
S5-REF 	& Heavy Seed, Fiducial  & $10^5$  	& Halo           &  Thin disk   		  &  0.1                      & 1           & $2.7\times10^9$ \\ 
S6          	& Heavy Seed     & $10^6$  		& Halo           &  Thin disk   		  &  0.1                      	& 1           & $6.8\times10^9$ \\ 
\hline

\multicolumn{8}{l}{\bf Eddington Limits} \\
\hline
S5-EL1 & Super-Eddington 	& $10^5$  & Halo      & Thin disk   	&  0.1          & 2 		& $7.6\times10^8$ \\ 
S5-EL2 & Super-Eddington       & $10^5$  & Halo      & Thin disk    	&  0.1          & 5 		& $1.8\times10^8$ \\ 
S5-EL3 & Hyper-Eddington 	& $10^5$  & Halo      & Thin disk   	&  0.1          & $10^4$ 	& $7.2\times10^6$\\ 
\hline

\multicolumn{8}{l}{\bf Feedback Models} \\
\hline
S5-FB1 & Slim disk, no spin  		& $10^5$  & Halo      & Super-critical, $a=0$       &  Eq.~(\ref{eq:sc})      &   $10^4$	 & $5.8\times10^{9}$\\ 
S5-FB2 & Slim disk, maximal spin	& $10^5$  & Halo      & Super-critical, $a=0.99$  &  Eq.~(\ref{eq:sc})      &	  $10^4$	 &  $4.0\times10^8$\\
S5-FB3 & High radiative feedback    & $10^5$  & Halo       & Thin disk                         &  0.2         		     & 	   1    	 &  $2.7\times10^7$\\ 
\hline

\multicolumn{8}{l}{\bf Bondi Accretion Variations} \\
\hline
S5-BA1 & Bondi constant $\alpha$    & $10^5$ & Halo       & Thin disk, $\alpha=100 $                                  & 0.1 & 1 & $2.2\times10^9$\\ 
S5-BA2 & Bondi power-law $\alpha$ & $10^5$ & Halo       & Thin disk, $\alpha=({\rho}/{\rho_{\rm th}})^{2}$ & 0.1 & 1  & $6.9\times10^8$\\ 
\hline

\multicolumn{8}{l}{\bf Cosmological Parameters} \\
\hline
S5-CP & Planck parameters            		& $10^5$   & Halo       & Thin disk              						  & 0.1  & 1 & $2.6\times10^9$ \\ 
\hline
\multicolumn{8}{c}{
\begin{minipage}[t]{2.0\columnwidth}
Notes: (1) Simulation name. (2) Model description. (3) Black hole seed mass. (4) Black hole seeding scheme. Light seeds come from Pop~III stars with metallicity $Z\le10^{-4}$, while intermediate and heavy seeds are planted in the host halo when its  virial mass $M_{200}=10^{10}\, \Msun$/$h$ at redshift $z=18.8$. (5) Accretion and feedback model. The accretion rate is calculated using Eq.~(\ref{eq:bhar}), while the feedback is described by the radiative efficiency $\epsilon_{\rm r}$. For the thin disk model $\epsilon_{\rm r}=0.1$ , while for the slim disk model $\epsilon_{\rm r}$ depends on the BH mass, accretion rate and spin, as calculated using Eq.~(\ref{eq:sc}). (6) Radiative efficiency of BH accretion. (7) Maximum Eddington limit. The hyper-Eddington model is set with a high number $\max(\lambda_{\rm edd})=10^4$ in the simulation. (8) The final black hole mass in the simulations at redshift $z=6.16$. Note S5-REF  is the fiducial simulation for comparison reference in this study. 
\end{minipage}
}\\
\end{tabular}
\end{center}
\end{table*}

\subsection{The Physics of Star Formation and Feedback}
\label{sec:sfr}

\subsubsection{Star Formation and Multi-phase ISM}

We use the sub-grid multi-phase interstellar medium model of \citet{Springel2003} for star formation 
and thermal feedback. In this model, gas consists of cold and hot phases, so the total gas density $\rho$ is the sum of $\rho_c$ and $\rho_h$, $\rho = \rho_h + \rho_c$. In star-forming gas, the evolution of the cold /hot phases is governed by  star formation, supernovae explosions, and heating and cooling. At high density ($\sim 10^5 \rho_{\rm th}$), this two-phase model returns a close-to-unity mass fraction of cold phase gas fueling rapid star formation. 

For each star-forming gas particle, surrounding gas is pressurized according to the multi-phase 
model that balances the evaporation of cold gas and cooling of hot gas that is explicitly 
solved in coupled equations. This approach naturally gives additional pressure support against 
self-gravity. Moreover,  \citet{Springel2003} showed that this method has good numerical convergence 
properties. We adopt a Chabrier initial mass function (IMF) with the stellar mass ranging between 
$[0.1, 100] \,  \Msun$. Stars more massive than $8\,  \Msun$ explode as Type~II supernovae 
and return the mass and metals into the surrounding ISM. The number density threshold for star 
formation is $0.13\,  \rm{cm^{-3}}$, which is commonly used in such cosmological simulations. 

\subsubsection{Metal Enrichment and Cooling}

Radiative cooling of optically thin gas is modeled using the cooling table from \citet{Smith2017}. 
The cooling curves are generated with {\sc CLOUDY} for gas exposed to a redshift-dependent
UV background from \citet{Faucher2010}. We adopt an exact time integration scheme introduced by 
\citet{Townsend2009} to update the thermal energy due to radiative cooling \citep{Zhu2017}. 

To model stellar evolution and chemical enrichment, we treat each star particle as a simple stellar 
population with an initial total mass of $\rm{10^6\,  M_{\odot}}$, and a metallicity in the range of 
$Z=[0.0004, 0.004, 0.008, 0.02, 0.05]$, adopting a Chabrier initial mass function.  Then 
each stellar population is evolved with stellar population synthesis  (SPS) code {\sc Starburst}99 
\citep{Leitherer1999, Leitherer2011} with Padova models (including the contribution of AGB stars) to produce tables 
of enrichment from both Type {\small\sc{II}} SNe and AGB stars. At each time step, we tabulate the 
returned mass and chemical enrichment from active star particles by linear interpolation on a 2D grid of
stellar age and metallicity. 

\subsubsection{Feedback from Supernovae}

Even with the sub-grid multi-phase ISM model of \citet{Springel2003}, galactic outflows are difficult to generate since most of the thermal energy from SNe is in the form of the effective pressure. A model of  ``energy-driven" winds  \citep{Vogelsberger2013} is a phenomenological approach to address the observed low star formation efficiency in low mass halos by launching gas into a galactic wind. We implement this model into our codes, assuming $70\%$ of the supernovae energy is available to launch the wind in two polar directions. Once a particle is flagged as a wind particle, we temporarily decouple it from hydrodynamics as in \citet{Springel2003} and \citet{Vogelsberger2013} to allow the wind particles to travel up to a certain distance away from star-forming regions, but when the gas density around the wind particle is a factor of $0.1$ of the density threshold for star formation, the wind particles are re-coupled hydrodynamically. This model successfully produces a realistic $L^*$ galaxy with a well-defined stellar disk like the Milky Way, as demonstrated by \citet{Zhu2016}. 

\subsection{The Black Hole Seeds}
\label{sec:seed}

Currently, there are three competing theories for the seeds of SMBHs which grow quickly into bright quasars. The mass of the BH seeds
varies from 10--100 $\Msun$ for stellar BHs from the first stars to $10^5\Msun$ for the direct collapse of pre-galactic gas clouds. Unfortunately,
our understanding of the first stars and the processes of direct collapse are both limited due to the complexity of the physical mechanisms involved. 
Therefore, the seed BH mass is treated as a free parameter in current cosmological hydrodynamic simulations. 

\subsubsection{Light BH seeds}
\label{sec:p3bh}
The first stars that formed out of primordial gas between $z=20$--30 in $\sim$$10^6\,  \Msun$ halos provide a natural channel to form the first BHs after their brief lifetimes \citep[see a recent review by][]{Haemmerle2020}. The IMF of Pop~III stars is still poorly constrained due to the interplay between gas cooling, turbulence, radiation, the evolution of accretion disks,  and possibly magnetic fields \citep[e.g.,][]{Stacy2016,Hirano2017, Chon2019, Fukushima2020, Sugimura2020}. 

It has been suggested that the BHs from Pop~III stars have a mass in the range of $10 - 100\, \Msun$ \citep[e.g.,][]{Volonteri2012a, Madau2014, Latif2016, Valiante2016, Becerra2018, Woods2019, Inayoshi2020}. We implemented a simplified model of BH seeds from Pop~III stars. Since we directly follow the chemical enrichment from stellar evolution, we are able to track the gas metallicity throughout the simulation. So we use the stellar metallicity as a criterion to select BH seeds. When a star particle metallicity of $Z$ is below a critical value of $10^{-4} \, Z_\odot$, we convert this particle into a BH particle 2 Myr after the star is born. The critical metallically is similar to the value $10^{-3.8} \, Z_\odot$ used in \citet{Valiante2016, Pezzulli2016}. We test two seed masses of $M_{\rm seed} = 10$ and $100\,  \Msun$/$h$ for this model with simulations. 

It is obvious that the resolution of our simulations cannot resolve these small BHs, and that the mass ratio between other particles and the BHs is very large, which could cause two-body relaxation and other numerical artifacts. As described in Section~\ref{sec:methods}, in order to alleviate the problems, we have adopted a numerical technique to treat the BH as a fraction of the Pop~III star particle from which it was born, similar to that used by \citet{Springel2003} to treat star formation in which a fraction of the gas particle becomes a new star, while the parent star particle becomes a ``ghost" attached to the BH. Since the ``ghost  particles" have comparable mass as the gas and dark matter particles, this technique reduces two-body heating and allows the BH particle to have the same dynamics as its parent star particle while maintaining its own mass and accretion activity. As we will see in Section~\ref{sec:testseed}, this treatment enables the small BHs to obtain consistent growth paths as their massive counterparts.

\subsubsection{Intermediate BH Seeds}

BH seeds with mass $\sim$$10^3 \Msun$ may result from collisional dynamics within dense star clusters in the early Universe \citep{Zwart2000, Freitag2006a, Freitag2006b}. Using N-body simulations, \citet{Zwart2002} demonstrate that in a very dense star cluster a massive star of $10^3\,  \Msun$ could form due to stellar collisions. Recently, \citet{Katz2015} and \citet{Yajima2016} have studied this channel and found a very massive star of $\sim$$10^3 \Msun$ can appear at redshifts as high as $z\sim20$. Subsequently, the very massive stars can form BHs without going through a supernova stage. Alternatively, \citet{Woods2017, Woods2019} argued that $\sim$$10^3 \Msun$ can also form from super-massive stars. For this model, we test two seed masses of $M_{\rm seed} = 10^3$ and $10^4\,  \Msun$/$h$ with simulations. 

Since the formation conditions of the intermediate BHs are uncertain, it is difficult to determine when and where they form, so for direct comparison with the heavy seeds, we use the same scheme as the latter to plant the seeds of $10^3$ and $10^4\,  \Msun$/$h$ in halos with mass $M_{200}=10^{10}\, \Msun$/$h$. 

\subsubsection{Heavy BH Seeds}

BH seeds could also form in the isothermal collapse of gas clouds in atomic-cooling halos ($T_{\rm vir} \ge 10^4$ K) \citep{Rees1978, Bromm2003, Begelman2006, Mayer2010, Johnson2011, Inayoshi2015, Becerra2015, Regan2017, Wise2019}. The physical processes, as depicted by \citet{Begelman2006}, show that the high density gas may quickly develop runaway global gravitational instabilities in the form of ``bars within bars", efficiently transferring angular momentum outwards, to form a dense gas core. Then, a central black hole can form in this dense core due to catastrophic cooling via thermal neutrinos if very high core temperature is achieved. The formation of direct collapse seeds depends sensitively on the location of the cooling halo, where ${\rm H}_2$ formation and fragmentation is supposed to be suppressed by the UV radiation from nearby Pop~III and Pop~II stars. The early chemical enrichment and the intensity of ${\rm H}_2$ dissociating UV background are the among the top uncertainties in modeling direct collapse. At redshift $z=15$ -- 18, a seed BH of $10^5\,  \Msun$ could form in $10^7$--$10^8\,  \Msun$ halos based on the most recent studies \citep{Valiante2017}.  Alternatively, other routes have been proposed to form massive BHs via merger-driven direct collapse \citep[e.g.,][]{Mayer2010, Mayer2019}.

Our simulation does not include the necessary resolution and physical process to model direct collapse. Instead, we follow the method used in previous studies \citep[e.g.,][]{Li2007, DiMatteo2008, Vogelsberger2014, Schaye2015, Khandai2015, Springel2018, Pillepich2018b} to place a massive BH seed at the center of halos with masses above a threshold $M_{200}=10^{10}\,  \Msun$/$h$, identified by frequently running the {\sc FOF} algorithm on the fly. In this study, we test two seed masses of $M_{\rm seed} = 10^5$ and $10^6\,  \Msun$/$h$, respectively, for this model.

\subsection{Black Hole Accretion}
\label{sec:acc}

One of the most important channels of BH mass growth is the accretion of gas. 
\citet{Bondi1944} described the accretion rate $\dot{M}_{\rm BH}$ 
around a central mass $M_{\rm BH}$ embedded in isothermal gas 
without net angular momentum as:
\begin{equation}\\
\dot{M}_{\rm BH} =  \frac{4\pi G^2 M_{\rm BH}^2 \rho}{(c_s^2 + v^2)^{3/2}}, 
\label{eq:Bondi}
\end{equation}
where $\rho$, $c_s$ and $v$ are the gas density, sound speed and  
relative velocity between gas and central mass $M_{\rm BH}$, respectively. 

However, it is currently not feasible to use the above expression directly in 
the cosmological simulations, which do not resolve the 
physical scales of black accretion (but see recent work by \citealt{Curtis2015} 
and \citealt{Hopkins2016}). Instead, most of the simulations adopt a sub-grid description 
that is often optimized towards more efficient accretion than 
that based on the simple Bondi-Hoyle accretion rate. In this study, we compare the following models used in some recent simulations.

\subsubsection{Constant Boost Factor $\alpha$ of Bondi Accretion}

This model was first proposed by \citet{Springel2005model} to adjust the effects of numerical resolution on the accretion rate with a constant boost factor $\alpha$:
\begin{equation}\\
\dot{M}_{\rm BH} =  \frac{\alpha\,  4\pi  G^2 M_{\rm BH}^2 \rho}{(c_s^2 + v^2)^{3/2}}
\label{eq:SpringelBH}
\end{equation}

The numerical value $\alpha$ is set to $100$ \citep{Sijacki2009}, which is calibrated with local scaling relations using galaxy merger simulations \citep{Springel2005model, DiMatteo2005}.  This was the approach employed in the original Illustris simulations \citep{Vogelsberger2014a}.  In the IllustrisTNG Project, however, no boost factor was applied to the estimated accretion rate.

\subsubsection{Power-law Boost Factor $\alpha$ of Bondi Accretion}

\citet{Booth2009} argued that the above model tends to over-estimate the accretion rate 
around a BH when the physical scale of gas is well-resolved, which 
could occur for low-density high-temperature non-star forming gas. Instead, 
they propose a density-dependent $\alpha$ model:
\begin{equation}\\
\alpha = \begin{cases} 1, \,  \,\,\,\,\,\,\,\,\,\,\,\,\,\,\,\,\,\,\,\,\,\, \rho < \rho_{\rm th} \\
                  \Big(\frac{\rho}{\rho_{\rm th}} \Big)^{\beta}, \, \,\,\,\,\,\,\,\, \rho \ge \rho_{\rm th}
\end{cases}
\label{eq:SchayeBH}
\end{equation}

This model uses a power-law index $\beta$ to capture the unresolved 
ISM physics while returns to the nominal Bondi-Hoyle rate for non-star forming gas. 
This model is also adopted in the  {\sc RAMSES} simulations
\citep[][]{Dubois2012}. 

\subsubsection{Chaotic Gas Accretion From Cold Clouds}

The switch between non-star forming and star-forming gas is certainly not unique. 
For example, \citet{Pelupessy2007} suggested a model based on a two-phase 
ISM as an alternative to eliminate the $\alpha$ factor. Instead of using $\rho$ and $u$, 
the accretion rate of star-forming gas is split into the contribution from both cold and hot phases. 
Since the accretion rate scales as $\rho/(c_s^2+v^2)^{3/2}$, 
the accretion mostly comes from the cold gas phase:
\begin{equation}\\
\dot{M}_{\rm BH} =  \frac{x\,  4\pi  G^2 M_{\rm BH}^2 \rho}{(c_{s,{\rm cold}}^2 + v^2)^{3/2}} + 
	\frac{(1-x)\,  4\pi  G^2 M_{\rm BH}^2 \rho}{(c_{s,{\rm hot}}^2 + v^2)^{3/2}}
\label{eq:bhar}
\end{equation}
where the mass fractions of cold gas $c_{s,{\rm cold}}$ and hot gas $c_{s,{\rm hot}}$ around the BH are obtained using the cold -- hot ISM break down in \citet{Springel2003}. 

Physically, this model aims to capture the large amount of molecular clouds around BHs, 
as recently been observed by \citet{Tremblay2016}. This model resembles the chaotic accretion in \citet{Gaspari2013, Gaspari2015} and it will be our fiducial choice for the gas accretion rate estimate. 

This Bondi--based accretion model is different from that in the Simba simulation \citep{Dave2019}, in which the BH accretion rate estimator scales linearly with the  gas inflow rate via gravitational torque for the cold gas, while for the hot gas it uses the Bondi accretion. The Simba accretion model would produce an invariant $\Mbh - \Mstar$ correlation across time, because both BHAR and SFR scale with the gas inflow rate. However, observations suggest that the $z \sim 6$ quasars do not seem to follow the local $\Mbh - \Mstar$ correlation \citep[e.g.,][]{Wang2013, Venemans2018, Pensabene2020}.

In this study, we test the chaotic cold accretion variations in the Bondi estimate of the BH accretion rate. We use the Eddington ratio $\lambda_{\rm Edd}$ to describe the accretion mode, which is the ratio of the BH accretion rate to the Eddington rate: 
\begin{equation}\\
\lambda_{\rm Edd} = \frac{\dot{M}_{\rm BH}}{\dot{M}_{\rm edd}}
\end{equation}
where the Eddington accretion rate $\dot{M}_{\rm edd}$\footnote{In the case of 
super-critical accretion, we use the definition $\dot{M}_{\rm Edd} = 16L_{\rm Edd}/c^2$ 
as in \citet{Madau2014}.} is calculated as:
\begin{equation}\\
\dot{M}_{\rm edd} = \frac{4 \pi G M_{\rm BH} m_p}{\epsilon_{\rm r} \sigma_{\rm T} c} = \frac{L_{\rm Edd}}{c^2 \epsilon_{\rm r}},  
\end{equation}
where $m_{\rm p}$ is the proton mass, $\epsilon_{\rm r}$ is the radiative efficiency and $\sigma_{\rm T}$ is the Thompson cross-section. 

We define three modes of accretion based on past convention: near-Eddington for $ 0.6 \le \lambda_{\rm Edd} \le 1$,  super-Eddington for $1 < \lambda_{\rm Edd} \le  10$, and hyper-Eddington for $\lambda_{\rm Edd} > 10$. In the simulations, we use a free parameter $\max(\lambda_{\rm Edd})$ to limit the maximum accretion rate in order to investigate the effects of these different modes on the BH growth.

\subsection{Black Hole Feedback}
\label{sec:feedback}

As suggested by \citet{Springel2005model}, feedback from BHs generally falls in two modes depending on the accretion rate: the quasar and radio modes for high and low accretion rates, respectively. Since we focus on the quasar phase of early SMBHs in this study,  we only consider ``quasar feedback", which couples the radiation from BH accretion to surrounding gas in the form of thermal energy with a fixed fraction of $\epsilon_{\rm r}\epsilon_{\rm f}$, where $\epsilon_{\rm r}$ is the radiative efficiency, and $\epsilon_{\rm f}$ is the efficiency of converting radiation into thermal energy, which is typically fixed at 5\% as suggested by \citet{DiMatteo2005}. 

The thermal energy is then distributed isotropically to the surrounding gas. There are some subtleties related to the coupling of thermal feedback energy in the two-phase ISM model. For non-star forming gas, the thermal injection from BHs is a simple additional term to the gas thermal energy, but for star-forming gas, the thermal energy from BHs is coupled to that from SNe and distributed over a characteristic decay timescale, as described in \citet{Springel2005model}. Note even within this simple model, large theoretical uncertainties are present due to our current understanding of BH accretion disks (e.g., see reviews by \citealt{Abramowicz2013} and \citealt{Haemmerle2020}). 

For this study, we consider two feedback mechanisms: radiatively-efficient thin-disks with constant $\epsilon_{\rm r}$, and super-critical slim-disks with varying $\epsilon_{\rm r}$ depending on the BH mass, accretion and spin, as described below.  

\subsubsection{Thin-disk Radiative Accretion with Constant $\epsilon_{\rm r}$}

Our fiducial choice of $\epsilon_{\rm r} = 0.1$ is the same value used in \citet{Springel2005model}, which corresponds to the radiatively-efficient thin disk model of \citet{Shakura1973}. This model assumes local thermal equilibrium and is such that the heat generated by viscosity is radiated away immediately. This model is widely adopted in cosmological simulations. The numerical value of the radiative efficiency $\epsilon_{\rm r}$ ranges from $0.1$\citep{Springel2005model, DiMatteo2005}, to 0.15 \citep{Booth2009}, and to $0.2$ \citep{Vogelsberger2013}. Similar to the seed mass, $\epsilon_{\rm r}$ is another important parameter in cosmological simulations with BHs. 

As listed in Table~\ref{table:sims}, we also test this model with $\epsilon_{\rm r}=0.2$ to investigate the effects of radiation on BH growth.

\subsubsection{Slim-disk Super-critical Accretion with Variable $\epsilon_{\rm r}$}

Once the accretion luminosity is high ($L\ge L_{\rm Edd}$), the radiative efficiency 
drops due to the fact the heat generated by viscosity is insufficiently radiated away
and the radial velocity within the disk cannot be neglected. 
Advection within the accretion disks dominant.
As a result, a substantial amount of heat is transported inwards by gas advection thus 
reducing the overall radiative efficiencies. \citet{Abramowicz1988} first proposed a 
slim-disk model that describes a super-critical radiatively inefficient accretion flow. Recently \citet{Madau2014} have (also used by \citealt{Lupi2016}) fitted the radiative efficiency as a function of high accretion rate $\dot{M}_{\rm BH}$ and black hole spin parameter $a$ according to the solutions constructed by \citet{Scdowski2009} with the following functions:

\begin{equation}\\
\begin{aligned}
\epsilon_{\rm r} & = r A(a) \Big[ \frac{0.985}{r + B(a)} + \frac{0.015}{r + C(a)} \Big], \\
r & = \frac{L_{\rm Edd}}{\dot{M}_{\rm BH} c^2} = \frac{1}{\dot{M}_{\rm BH}} \frac{4 \pi G M_{\rm BH} \mu_{e} m_p c}{c^2 \sigma_{\rm T}}.
\label{eq:sc}
\end{aligned}
\end{equation}
where the functions $A(a), B(a)$ and $C(a)$ depend on BH spin $a$:
\begin{equation}\\
\begin{aligned}
 A(a) &= \big(0.9663 - 0.9292\,  a \big)^{-0.5639},\\
 B(a) &= \big(4.6270 - 4.4450\,  a \big)^{-0.5524},\\
 C(a) &= \big(827.30 - 718.10\,  a \big)^{-0.5639}.
\end{aligned}
\end{equation}

In this model, the radiative efficiency $\epsilon_{\rm r}$ quickly drops from $\sim$0.05 to $\sim$0.01 for non-spinning BHs accreting at super-Eddington rates. For BHs with a maximum spin of $a \sim 1$,  $\epsilon_{\rm r}$ is considerably higher. Despite the super-Eddington rate of BH mass accretion, because of low radiative efficiency the resultant luminosity may appear moderately super-Eddington with $\max(\lambda_{\rm Edd}) \sim 3$  \citep{Madau2014}. Recently, \citet{Sadowski2016a} and \citet{Narayan2017} have found that, while the radiative efficiency for the slim-disk accretion is low, large mechanical power in the form of jets and outflows can be released.

A proper treatment of super-critical accretion thus requires more inputs from small-scale simulations of accretion flows around BHs.

For this study, we test the super-critical accretion model for both light BH seeds of $10, 100\, \Msun$/$h$ from Pop~III stars and heavy seeds of $10^5\, \Msun$/$h$, as listed in Table~\ref{table:sims}. Although it is possible to calculate the spin of a BH in the simulation by tracking the angular momentum of all the gas particles accreted by the BH, it is time consuming to do so on-the-fly for all the BHs in the simulations, so in practice we use two spins $a=0$ and $a=0.99$ to bracket the spin range to investigate the effects of spin on BH growth.

\begin{figure*}
\includegraphics[width=0.32\linewidth]{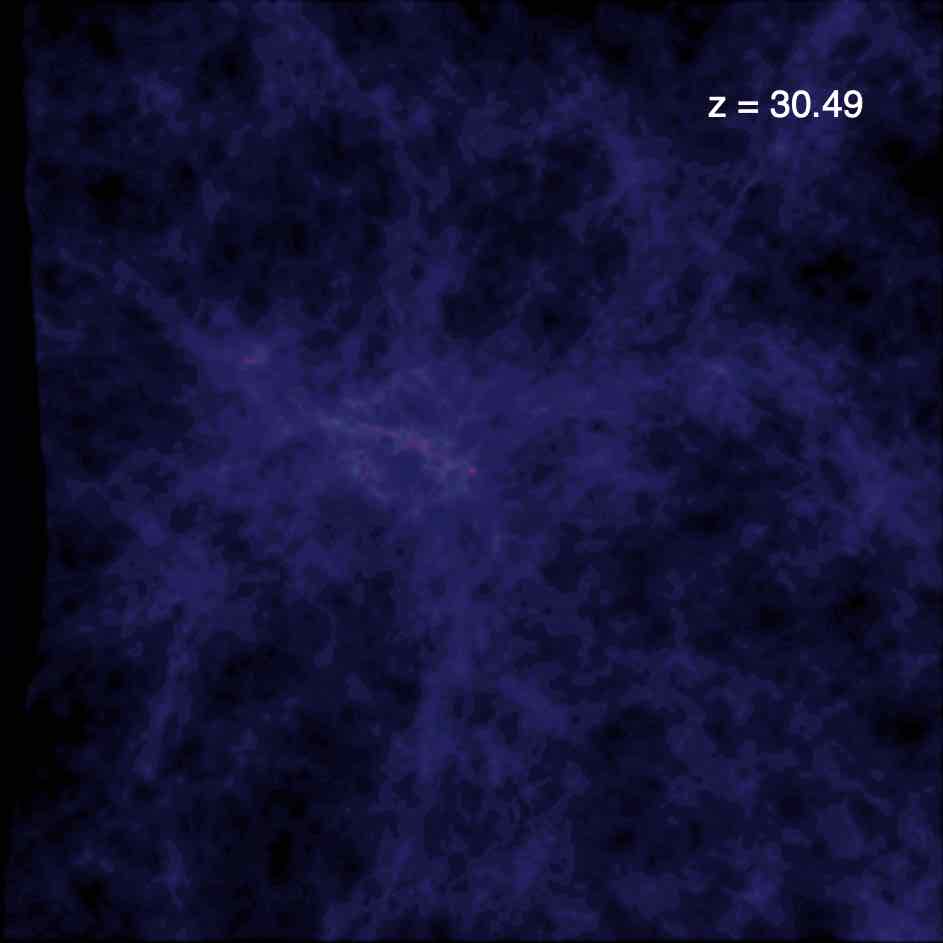}
\includegraphics[width=0.32\linewidth]{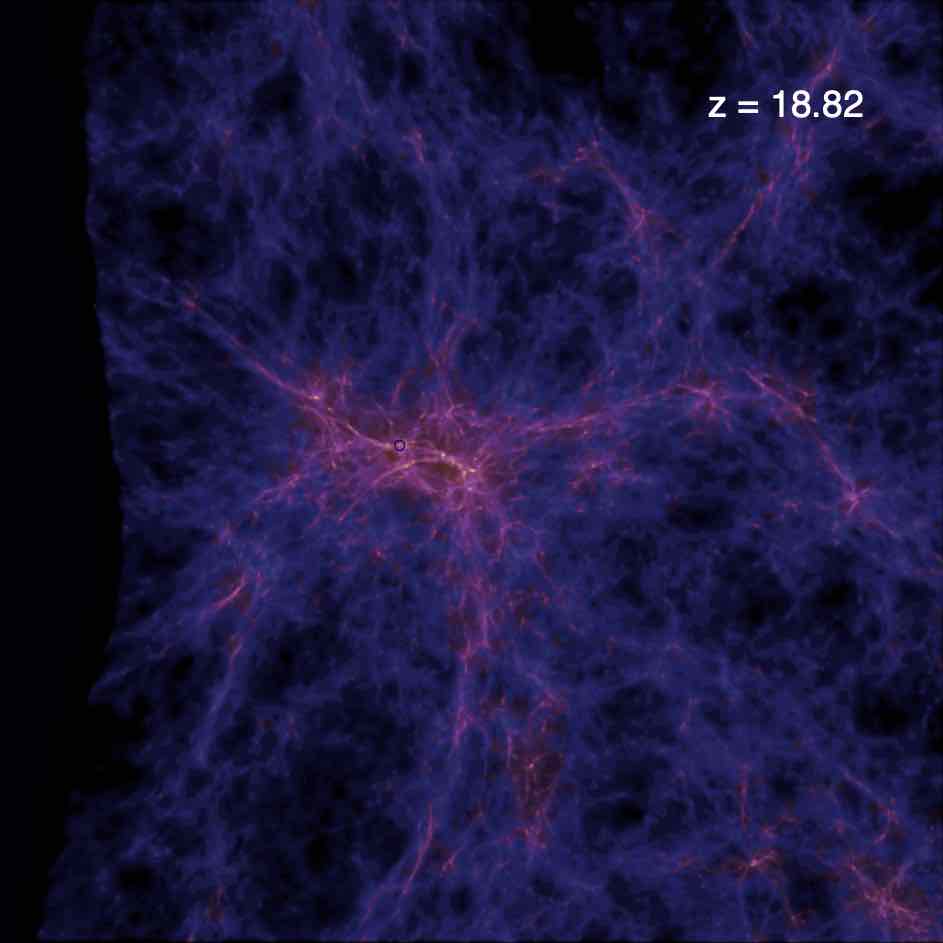}
\includegraphics[width=0.32\linewidth]{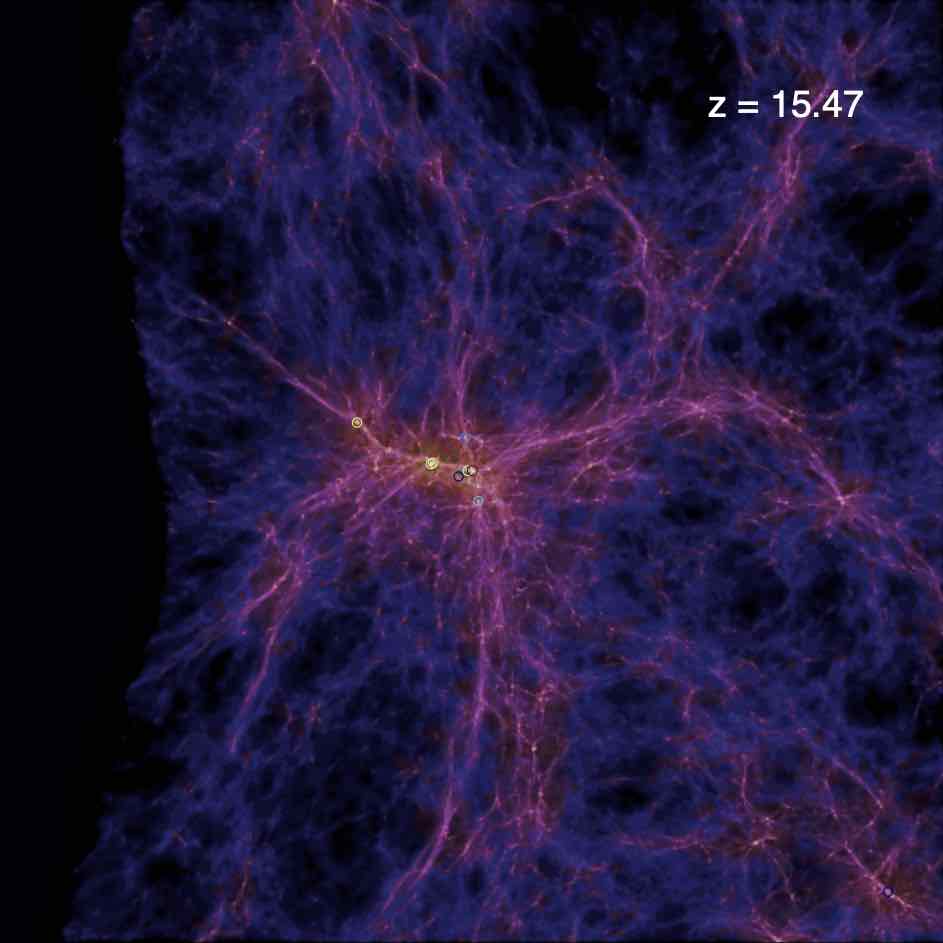}\\
\includegraphics[width=0.32\linewidth]{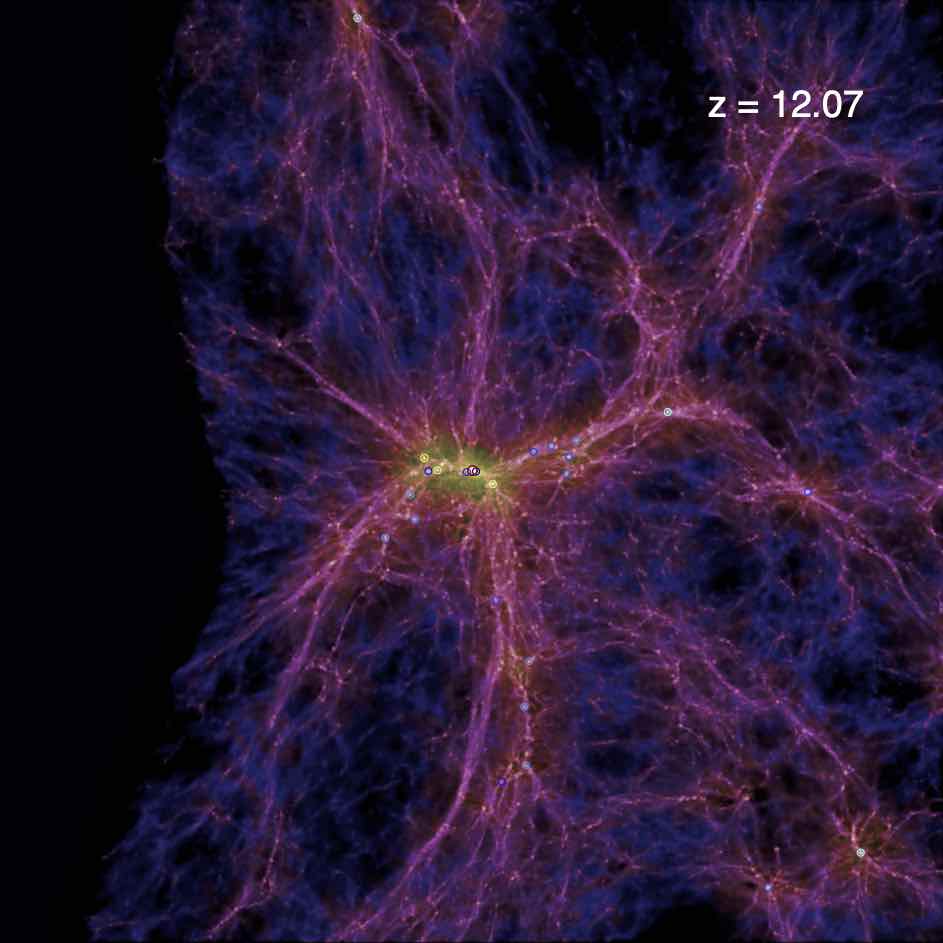}
\includegraphics[width=0.32\linewidth]{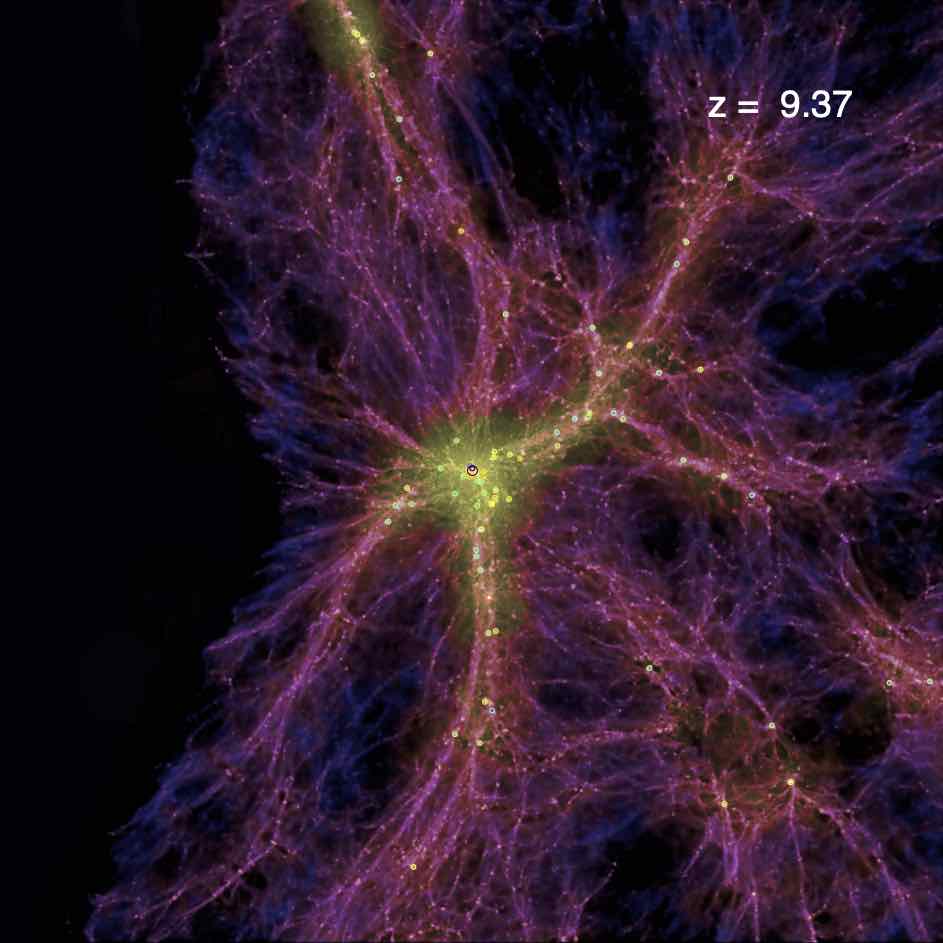}
\includegraphics[width=0.32\linewidth]{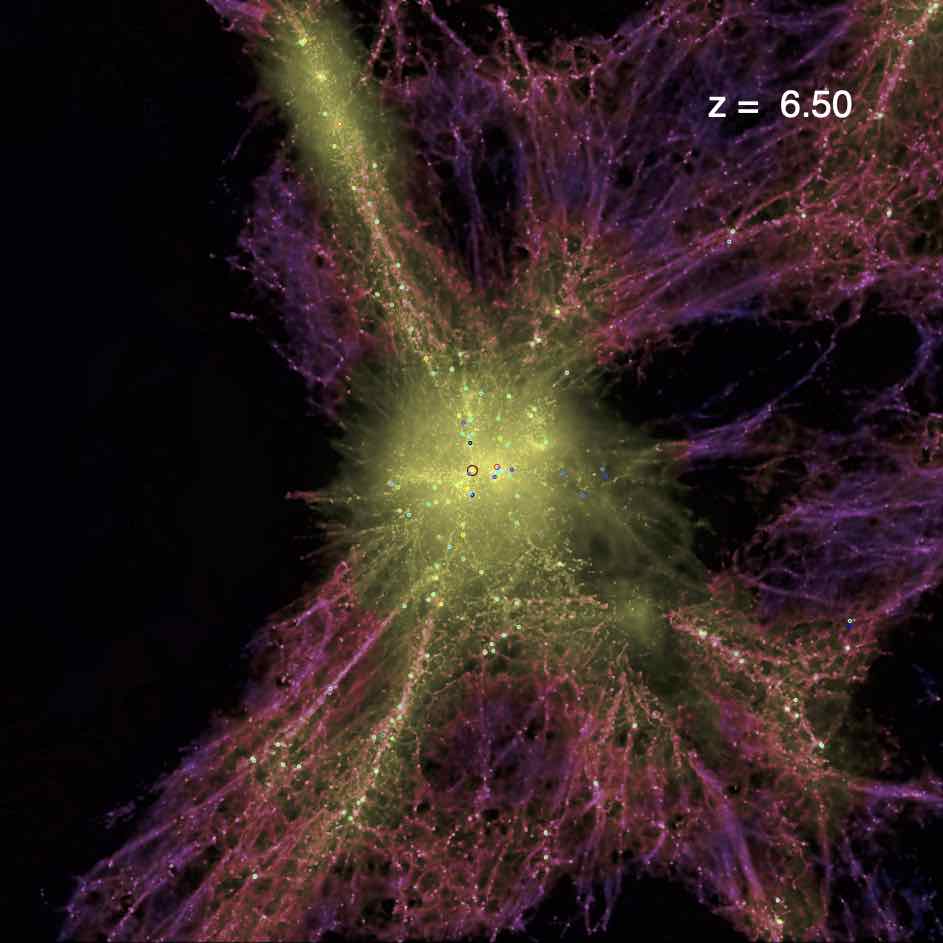}\\
\caption{The emergence of the first galaxies and BHs and their evolution from $z \sim 30$ to $z \sim 6.5$ in the fiducial S5-REF simulation. The images are the projected gas density, color-coded by temperature: blue represents cold, while yellow represents hot gas heated and ionized by feedback from supernovae and accreting BHs. The colored circles indicate locations of the BHs. The zoom-in region is $\sim 10$ comoving Mpc centered on a galaxy halo of $\sim 1.16 \times 10^{13}\, \Msun$ at $z = 6.1$, the most massive one in a ${\rm Gpc}^3$ volume.}
\label{fig:firstgal} 
\end{figure*}

\subsection{Black Hole Mergers}
\label{sec:merger}

In addition to gas accretion, BHs can grow through binary mergers. In our simulations, BH binary mergers happen whenever two BHs are within the smoothing radius of either BH and the relative velocity between the two BHs is below the circular velocity of the BH pair \citep{Booth2009}. 

During the merger of two BHs, a large velocity kick might be expected for the remnant because of gravitational recoil \citep{Bekenstein1973, Volonteri2006,  Blecha2008}. The recoil velocity, which depends upon the mass ratio of the BH binary and the spins of each BH, can temporarily kick the BH from the center of the galaxy and delay its mass accretion as shown by \citet{Blecha2011}. Gravitational recoil has been included in some simulations \citep[e.g.,][]{Sijacki2009, Kelley2017}. In particular, \citet{Kelley2017} presented a comprehensive modeling of the recoiling BH including processes such as dynamical friction, stellar loss-cone scattering, and gas drag. In our simulations, we do not consider BH recoil in the modeling, because as demonstrated in \citet{Li2007}, the escape velocity of this massive host halo we selected is large enough even for the most extreme recoil velocities, and our main targets are the most massive BHs in the galaxies, which are usually located in the deepest potential well of the hosts.

\section{Formation of the First Quasars}
\label{sec:ref}

\begin{figure*}
\includegraphics[width=0.45\linewidth]{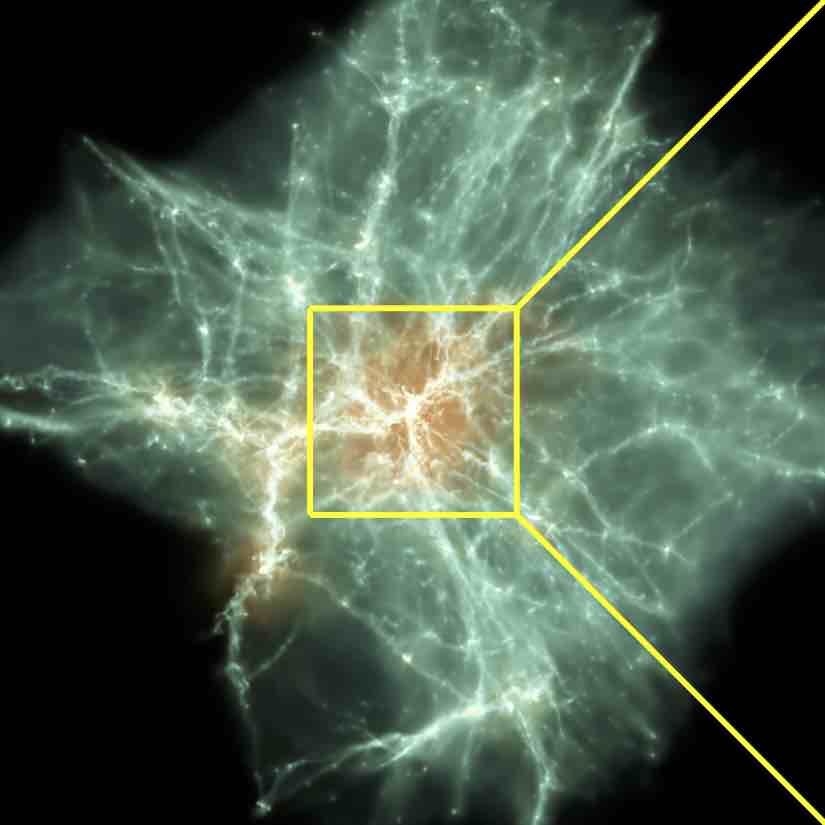}
\includegraphics[width=0.45\linewidth]{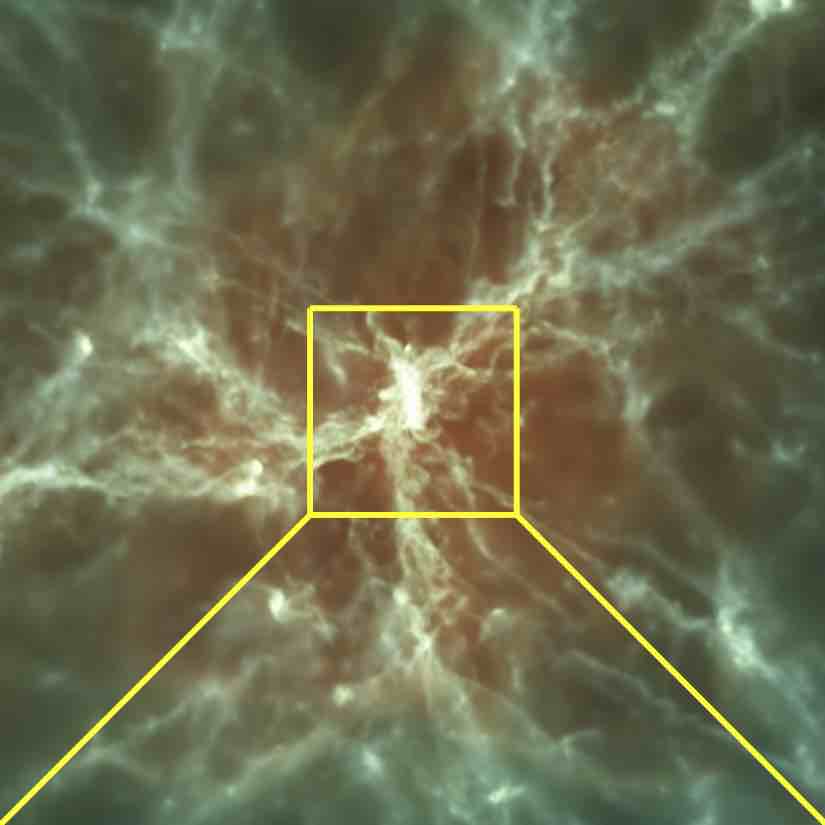}\\
\includegraphics[width=0.45\linewidth]{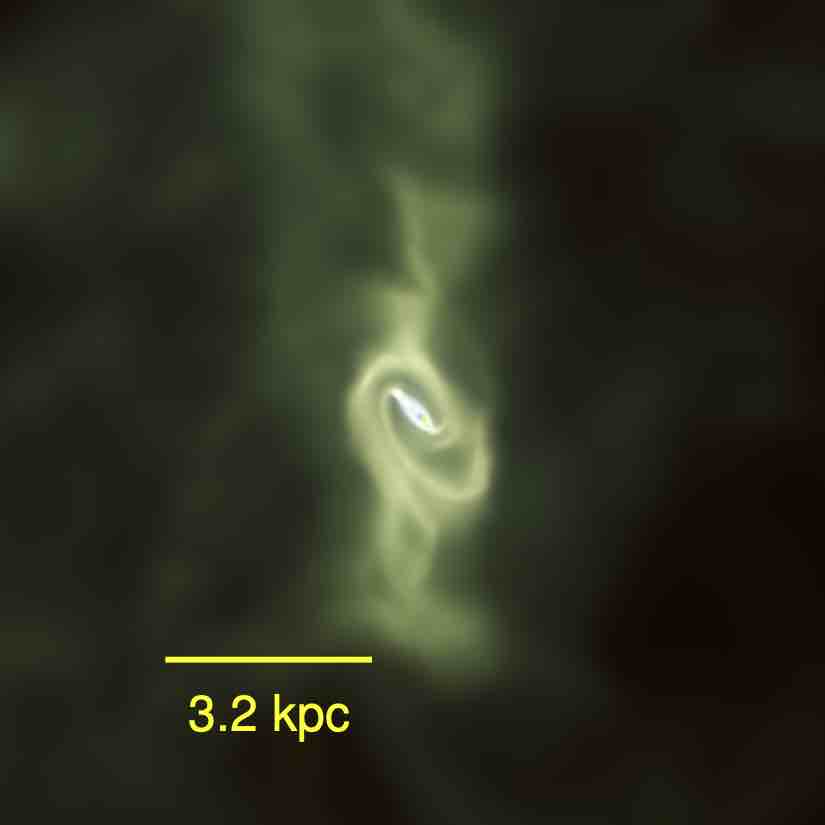}
\includegraphics[width=0.45\linewidth]{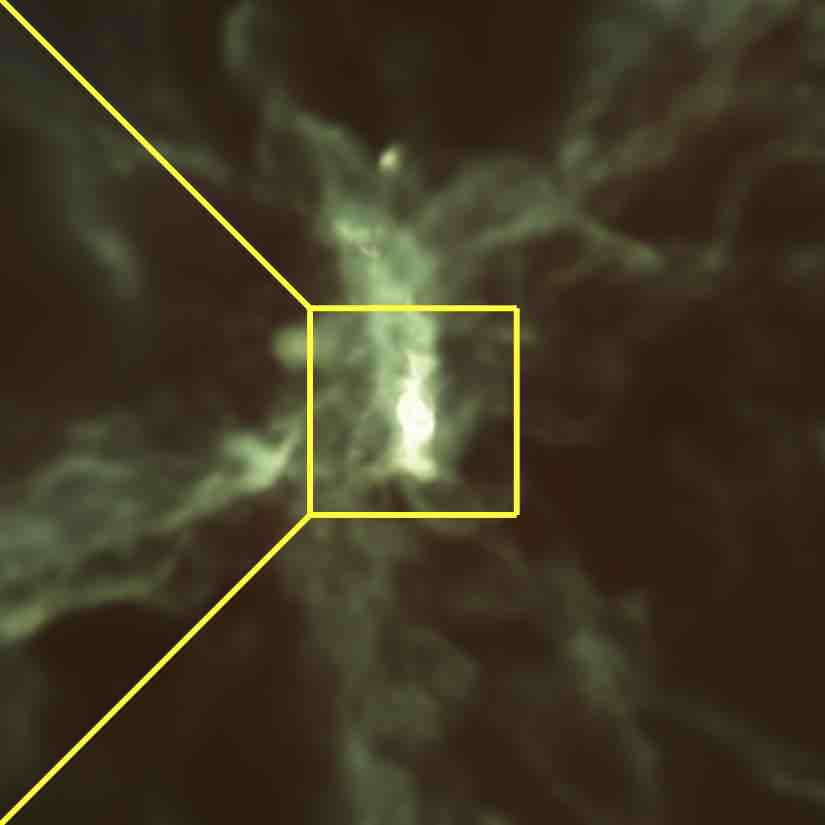}
\caption{The gas distribution of the central galaxy at $z = 7.5$ from the S5-REF simulation. The images are the projected gas density at different scales. In the central 3 kpc region, the gas forms a disk with spiral arms connected to the central bar-like structure. }
\label{fig:gas} 
\end{figure*}

In this Section, we present the formation and evolution of the first galaxies and black holes from the S5-REF simulation, which covers the region of the largest halo at $z \sim 6$ in a $(1\, {\rm Gpc}/h)^3$ volume. As listed in Table~\ref{table:sims}, the S5-REF simulation is our fiducial model with a heavy seed of $10^5\, \Msun$/$h$ BHs are planted in a halo when its mass reaches $10^{10}\, \Msun$/$h$, and the BH grows with the canonical Eddington-limited, thin-disk chaotic accretion with a constant $\epsilon_{\rm r}=0.1$. It also includes all the physics of a multi-phase ISM, star formation, and feedback processes, as described in Section~\ref{sec:methods}.

\subsection{Emergence of the First Galaxies}
\label{sec:firstgal}

\begin{figure*}
\includegraphics[width=\linewidth]{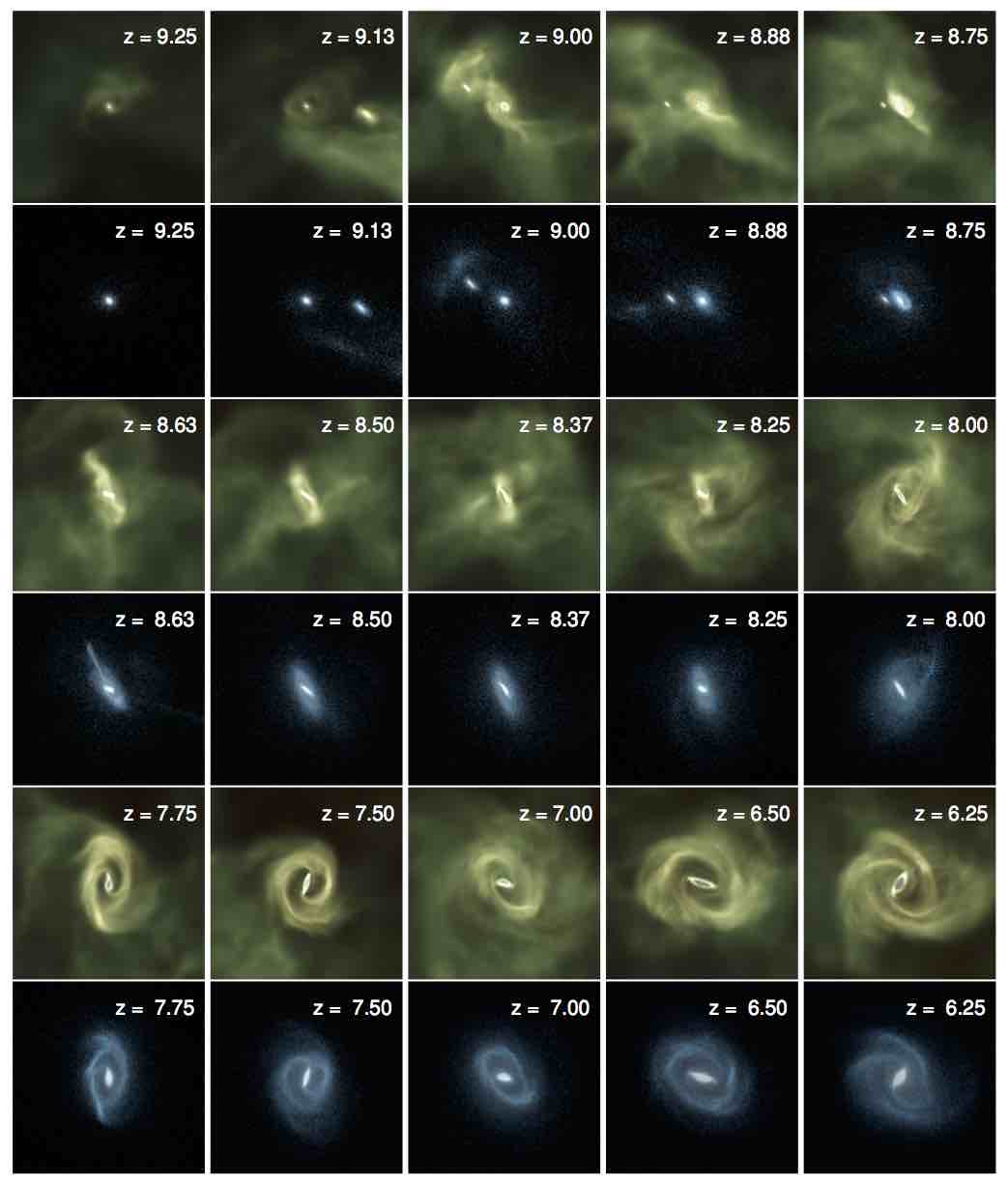}
\caption{A time sequence of the merger event between $z = 9.13$ and $z = 6.1$ from the S5-REF simulation. The images are the projected gas density at different redshifts and corresponding projected stellar density. The size of each panel is 10 kpc in physical coordinates. The stellar map is color-coded using stellar photometry from Starburst99 \citep{Leitherer1999, Leitherer2011} based on the metallicity and stellar age such that old stellar population appears in red while the young one is in blue. }
\label{fig:gas-star} 
\end{figure*}

\begin{figure*}
\includegraphics[width=0.32\linewidth]{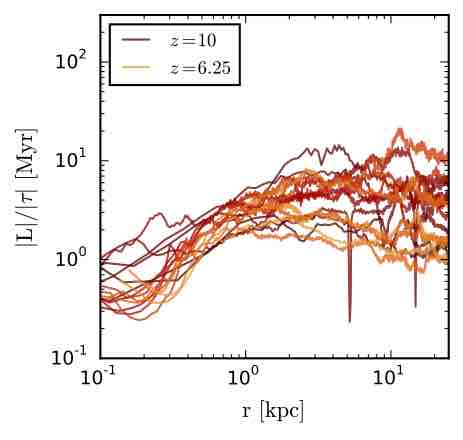} 
\includegraphics[width=0.32\linewidth]{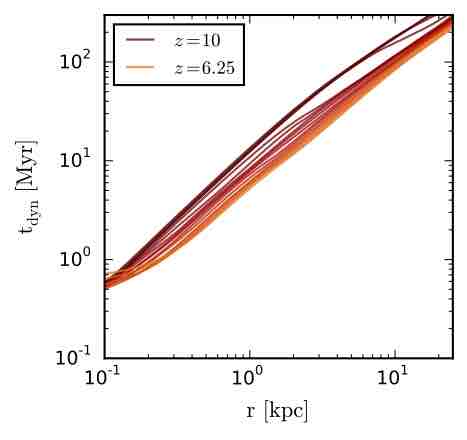}
\includegraphics[width=0.32\linewidth]{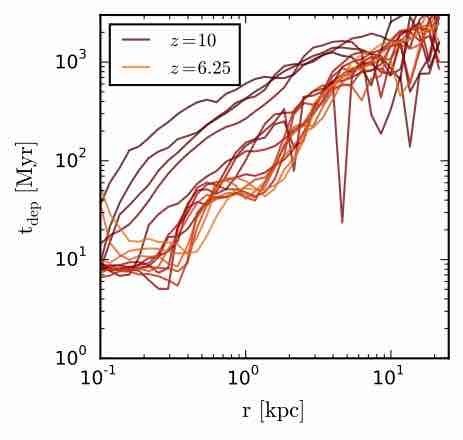}
\caption{A comparison of the angular momentum transport timescale (left panel), the local dynamical timescale (middle), and the gas depletion timescale (right) at different radii from $z\sim 10$ to $z \sim 6$ in the S5-REF simulation. Since there is no well-defined orientation of the gas plane in the simulation, we use the magnitude ratio of vectors $|\textbf{L}|/|\tau|$ as estimates of the timescale for angular momentum change.}
\label{fig:timescale_evolution} 
\end{figure*}

The S5-REF simulation is a highly overdense region where the first galaxies are expected to form. Figure~\ref{fig:firstgal} shows the cosmic web of this region and its evolution from $z= 30.45$ to $z=6.16$. The filamentary structure formed under gravity from perturbations, and the first mini halos collapsed in the dense intersections of the filaments and continued to grow through accretion and mergers with other mini halos. When these atomic-cooling halos reach mass $\gtrsim 10^6\, \Msun$ and temperature $T_{\rm vir} \gtrsim 10^4$~K at $z \sim 30$,  Pop~III stars form in the metal-free gas. These first stars evolved rapidly, and the feedback, metal enrichment and remnants from the stars shaped the subsequent formation of galaxies. In the simulation, the first BH seed emerged in the largest halo at $z \sim 18.8$ when it reached $10^{10}\, \Msun$/$h$.

 The largest halo has undergone more than two dozen major mergers from $z \sim 24$ to $z \sim 6$ when it assembled a mass of $6.86 \times 10^{12}\, \Msun$/$h$. These gas-rich galaxy mergers provide an abundant gas supply and compress the gas to high density in the central galaxy. Figure~\ref{fig:gas} shows the gas distribution of the central galaxy at $z=7.5$ from the simulation. A large amount of cold gas is continuously channelled into the central galaxy from the filaments and frequent mergers. In the central region of the galaxy, a disky gas distribution with a bar-like structure and spiral arms are present. Such a structure is highly unstable and it facilitates efficient angular momentum transport of the gas to fuel a vigorous circumnuclear starburst and rapid black hole accretion in the center. 

Figure~\ref{fig:gas-star} shows the distribution of gas and stars of a merger sequence from $z = 9.13$ to $z = 6.1$. During the merger event, a nuclear gas disk is formed in the central region. The axis of the nuclear gas disk is misaligned with the outer gas distribution, as can be seen between $z = 8.63$ and $z = 8.37$. This behavior is similar to the nuclear disk formed in the merger simulations by \citet{Barnes1996}, who showed that the remnant of the nuclear disk becomes part of the central stellar bar later on. 

The tidal torques are an efficient means to transfer gas angular momentum \citep{Hopkins2010}, which is manifested in the forms of density waves, in particular in cases involving $m=1, 2$ (number of spiral arms). The strong features of spiral arms, bars and misaligned gas planes in multiple snapshots in Figure~\ref{fig:gas-star} indicate that a strong gas inflow induced by torques is taking place at these redshifts. 

In order to quantify the effects of tidal torques on angular momentum transport and gas inflow, we compare three timescales: angular momentum transport timescale $t_{\rm ang}$, local dynamical timescale $t_{\rm dyn}$, and gas depletion timescale $t_{\rm dep}$. 

The angular momentum transport timescale $t_{\rm ang}$ is defined as the ratio between the angular momentum $\textbf{L} = \textbf{r} \times \textbf{v}$ and tidal torques ${\tau}= \textbf{r} \times \textbf{F} / m$:
\begin{equation}\\
t_{\rm ang} = \frac{\tau}{\textbf{L}}=\frac{\textbf{r} \times \textbf{F} / m}{\textbf{r} \times \textbf{v}}. 
\end{equation}
where $\textbf{r}$, $m$, $\textbf{v(r)}$ and $\textbf{F(r)}$ are the radius, mass, velocity of the gas and force on the gas at the location, respectively.

The local dynamical timescale $t_{\rm dyn}$ is defined as:
\begin{equation}\\
t_{\rm dyn} = \frac{2\pi r}{v_{\rm cir}(r)}.
\end{equation}
where $v_{\rm cir}(r)$ is the circular velocity at the radius $r$. 

The gas depletion timescale $t_{\rm dep}$ by star formation is defined as:
\begin{equation}\\
t_{\rm dep} = M_{\rm gas} / \rm{SFR}, 
\end{equation}
where $M_{\rm gas}$ and  \rm{SFR} are the total mass of star-forming gas and total star formation rate at each $r$, respectively. 

The difference between the timescale for angular momentum exchange and the dynamical timescale is key to understanding the origin of the angular momentum transport. \citet{Tohline1982} showed that gravitational torques cause misaligned gas to settle into a preferred plane and the timescale of this settling process is proportional to the local dynamical timescale. Recently, \citet{vandeVoort2015}  reported that the timescale of this settling can be much longer than the local dynamical timescale if a substantial amount of angular momentum is accreted such as that in mergers, and the gas depletion timescale from observations ranges from 2 Gyr in local spiral galaxies \citep[][]{Leroy2008} to 10--100 Myr in starburst galaxies at low- and high-$z$ \citep[][]{Genzel2010}. 

A comparison of the three timescales at different locations and redshifts from the simulation is shown in Figure~\ref{fig:timescale_evolution}. It is clear that all three timescales show a strong dependence on the distance from the galactic center. While both dynamical and depletion timescales generally show rapid increase with radius, the angular momentum transport timescale increases gradually with radius, reaching a plateau to several Myr at $\sim 3$~kpc where $t_{\rm dyn} > 10$~Myr and $t_{\rm dep} > 100$~Myr, while in the central sub-kpc region it is typically less than 1 Myr, shorter than both dynamical and depletion timescales.  The gas depletion timescale quickly increases from $\sim$10 Myr at 1~kpc to $\sim10^3$~Myr at 10~kpc, in good agreement with observations of starburst galaxies \citep[][]{Genzel2010}, suggesting that star formation in the circumnuclear region is the most efficient. 

Figure~\ref{fig:timescale_evolution} also shows that all three timescales change with redshift and they become shorter as the galaxy evolves. In particular, the angular momentum timescale decreases across all radii from $z = 10$ and $z = 6$, due to the galaxy interaction and merger during this period as shown in Figure~ref{fig:gas-star}. This trend suggests that galaxy interactions can boost efficient angular momentum removal from gas. \citet{Dubois2012} have argued that the low angular momentum gas can be accreted through almost radial cold inflow for high-$z$ galaxies. In our simulation, angular momentum is quickly transported outwards as characterized with the short timescale of several Myrs from the cosmic web to feed the central galaxy. Moreover, the angular momentum transport happens before the net cancellation of angular momentum due to multiple streams as outlined in \citet{Dubois2012}. It is likely that the frictional force between the cold inflow and the hot halo gas plays an important role to remove angular momentum in the cold gas streams.  Any residual angular momentum is further transported outwards once the gas settles into a rotationally supported disk in the central several kpc region. Since the gas depletion timescale is larger than the other two timescales, the central gas disk is not immediately consumed by star formation but is able to maintain its rotation instead. 

\begin{figure*}
\includegraphics[width=0.92\linewidth]{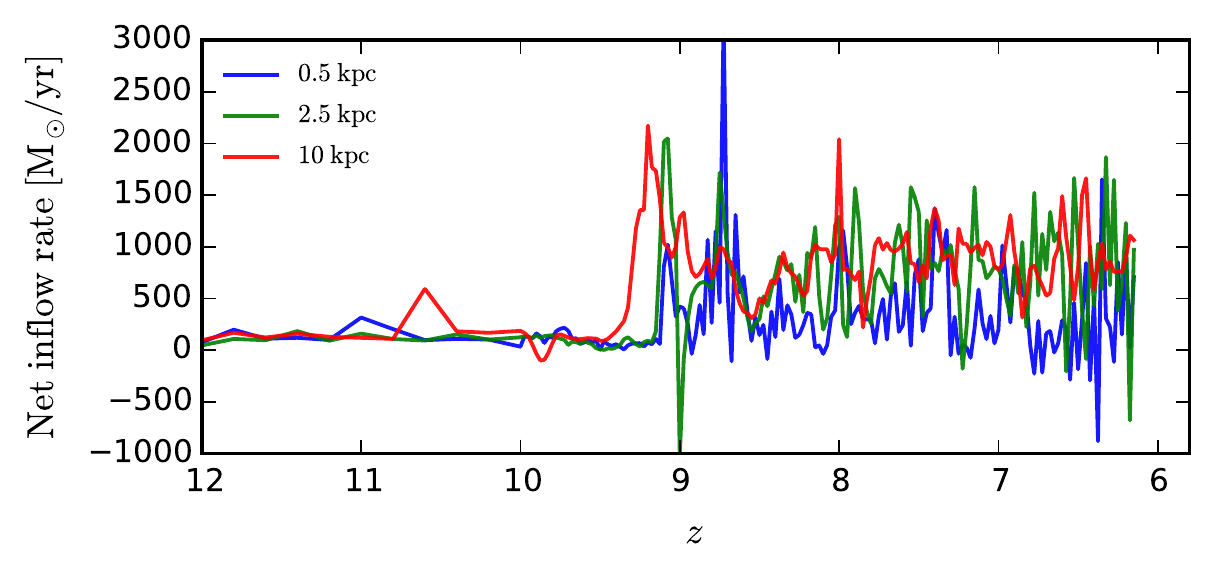}\\
\vspace{-0.5cm}
\includegraphics[width=0.92\linewidth]{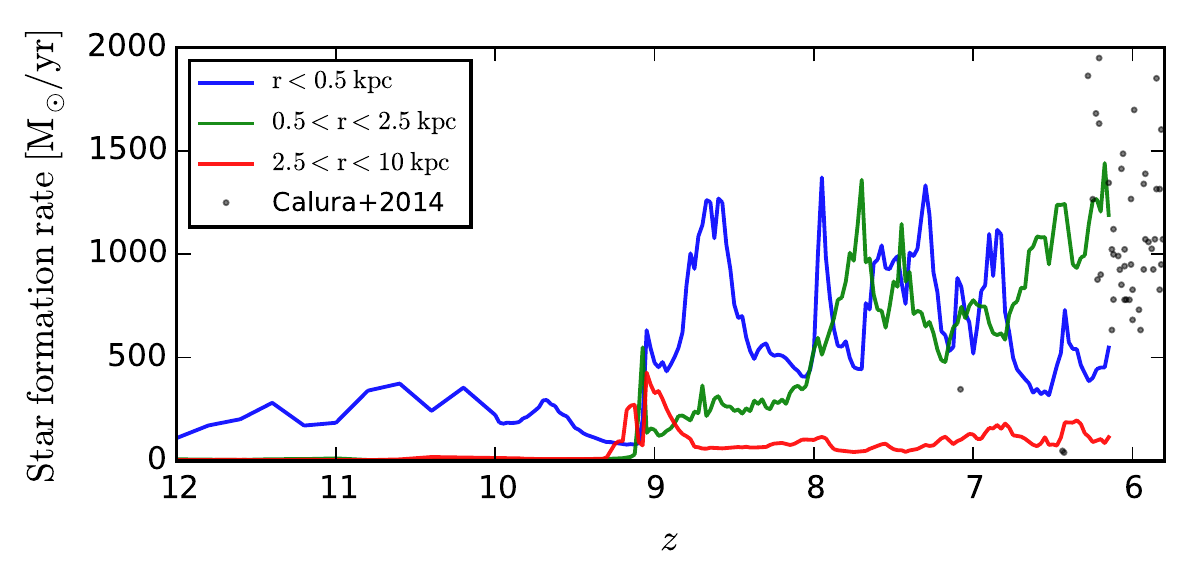}\\
\vspace{-0.5cm}
\caption{A comparison of the net gas inflow rate (top panel) with the star formation rate (bottom) at different redshifts within three regions at 0.5, 2.5 and 10~kpc from the galaxy center, respectively, from the S5-REF simulation. The grey circles represent observed SFRs of quasar host galaxies at $z \sim 6 - 7$ by \citet{Calura2014}. }
\label{fig:gas_inflow_and_sfr} 
\end{figure*}

The short timescale of angular momentum change facilitates strong gas inflow to the center, as can be seen from the gas distribution in Figure~\ref{fig:gas-star}. Following \citet{Muratov2015},  the net gas inflow rate can be computed as:
\begin{equation}\\
\dot{M}_{\rm inf} = \sum m \frac{ \mathbf{v} \cdot \mathbf{r} } {|r|} \frac{1}{{\rm d} L}.
\label{eq:inflow_rate}
\end{equation}

We computed the net gas inflow rate at different redshifts for a shell between $0.9r$ and $1.1r$ for $r=0.5$, 2.5, and 10 kpc, respectively, and show the result in Figure~\ref{fig:gas_inflow_and_sfr}, in comparison with the star formation rate at each corresponding redshift. Both the inflow rate and SFR increase rapidly during the merger event starting at $z \sim 9.5$, and the gas inflow precedes the star formation. Large inflow rates ($\sim10^3 \Msun/{\rm yr}$) continue from $r = 10$ kpc all the way down to $r = 0.5$ kpc owing to efficient angular momentum transport. Meanwhile, the large inflow of cold gas leads to strong star formation with rates $> 10^3 \Msun/{\rm yr}$, in particular within the central 2.5 kpc region. The quasar host in the simulation grows rapidly through intense star formation, and it assembles a stellar population of $\sim 5\times10^{11}\Msun$ by $z \sim 6$ as shown in Section~\ref{sec:host}.

The vigorous circumnuclear starbursts in the simulation agree well with the observations of $z \sim 6 - 7$ quasar hosts, and the extremely high SFRs fall within the  measurement range of \citet{Calura2014}. This suggests that the extreme starbursts observed in the quasar host galaxies may be triggered by violent galaxy interactions and mergers.

\subsection{Growth of the Supermassive Black Holes}
\label{sec:smbh}

\begin{figure*}
\includegraphics[width=0.85\linewidth]{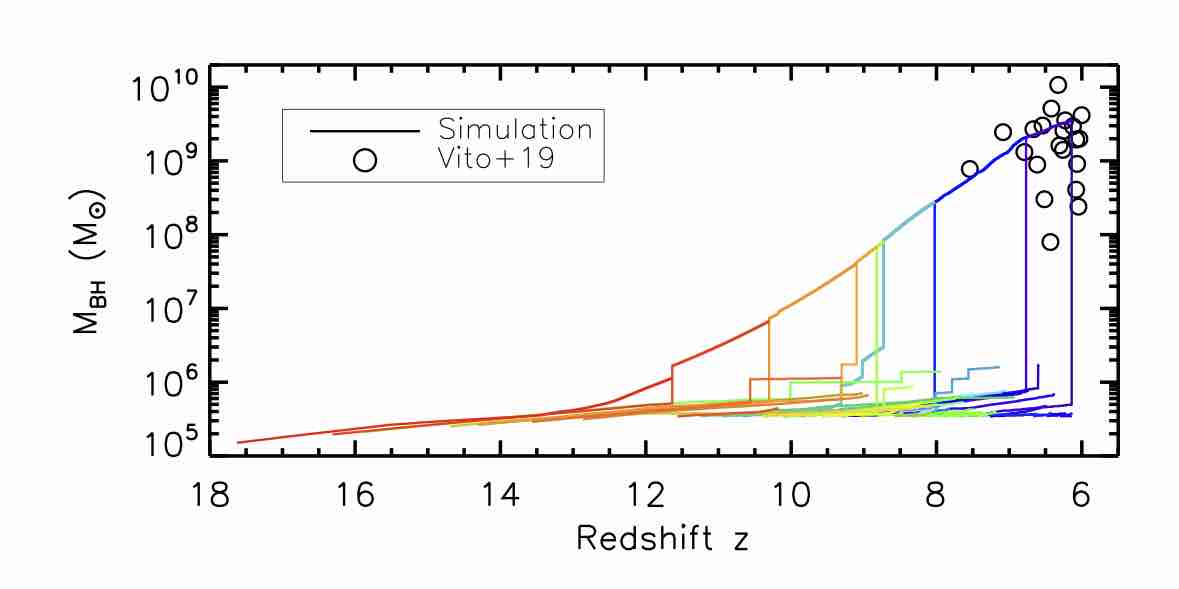}\\
\vspace{-0.5cm}
\includegraphics[width=0.85\linewidth]{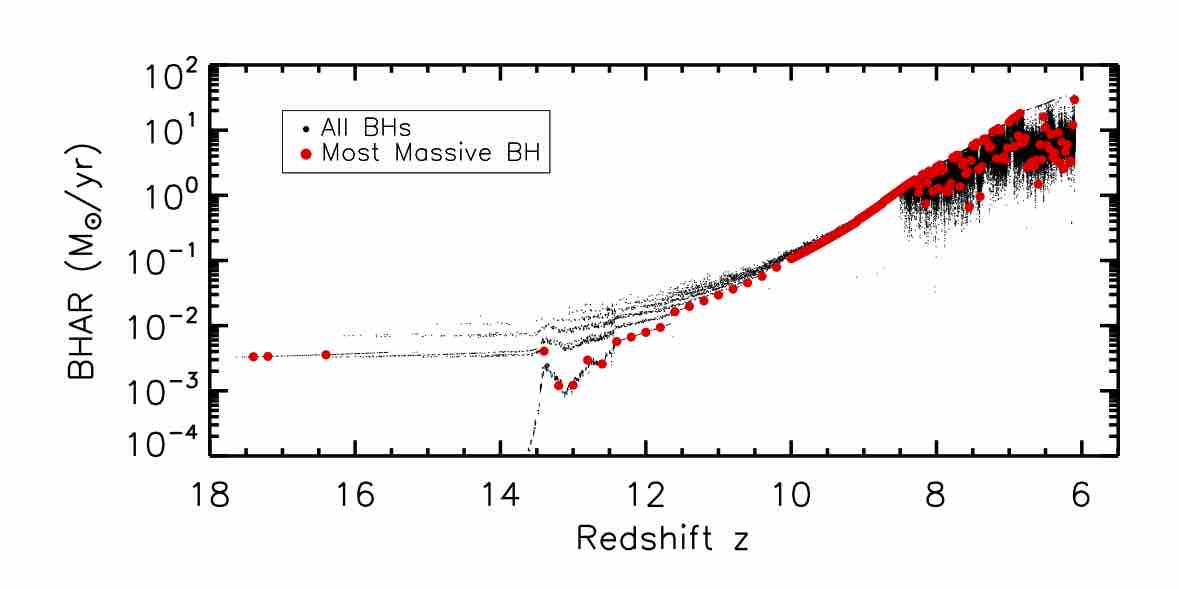}\\
\vspace{-0.5cm}
\includegraphics[width=0.85\linewidth]{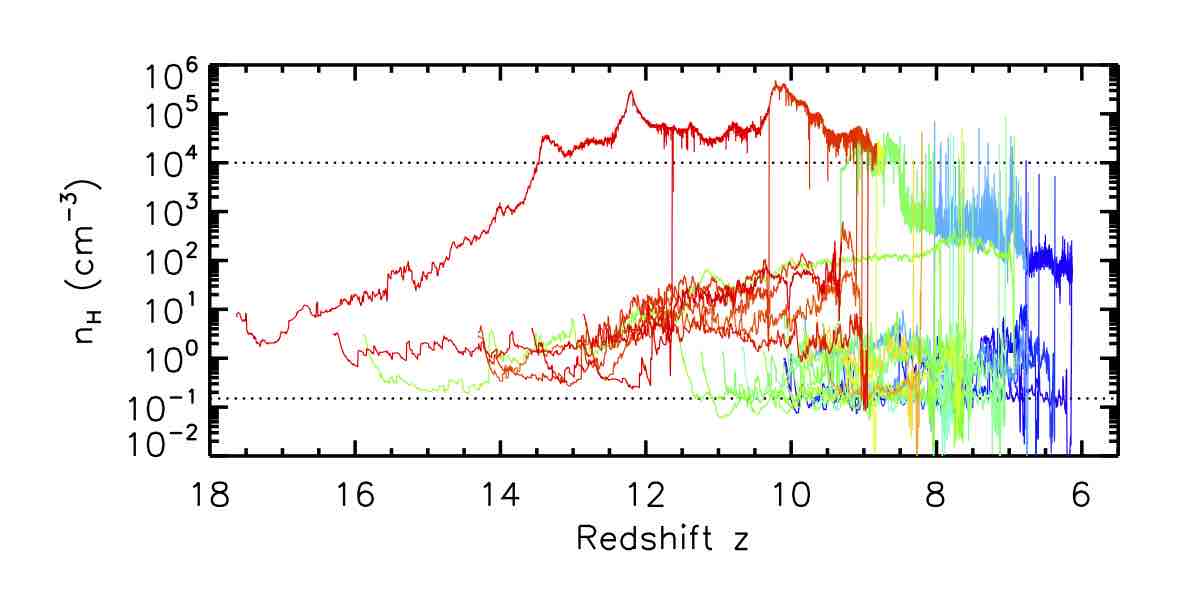}\\
\vspace{-0.5cm}
\caption{The growth history of the most massive BH from the S5-REF simulation through gas accretion and a series of mergers, in comparison with observations of $z \sim 6$ quasars by \citet{Vito2019} (top panel),  the accretion rates of all BHs and the most massive one at different redshifts (middle), and the gas density around each BH before it merged (bottom). The colored lines represent individual BHs merged into the most massive one. In the bottom panel, the upper dashed line indicates the density threshold above which the effective sound speed drops, while the lower one indicates the critical density for star formation. This SMBH assembled a mass of $6.9 \times 10^8\, \Msun$ at $z \sim 7.5$, similar to the most distant quasar ULAS J1342+0928 at z = 7.54, and $3.9\times 10^9\, \Msun$ at $z \sim 6.1$, matching a number of luminous $z > 6$ quasars such as SDSS J1148+5251 and SDSS J2310+1855 \citep{Vito2019}.}
\label{fig:bh_history} 
\end{figure*}

Similar to the strong star formation in the host galaxy, the seed BHs grow quickly owing to the abundant gas supply from the filaments in the highly biased region and the frequent interaction and mergers with other protogalaxies, which brings in large gas inflows to the galactic center through efficient angular momentum transport.

Figure~\ref{fig:bh_history} shows the growth history of the most massive BH from the simulation. It grows rapidly from $z \sim 13$ to  $z \sim 6$ through near-Eddington accretion, boosted by multiple gas-rich mergers during this period. The SMBH assembled a mass of $\sim 6.9 \times 10^8\, \Msun$ at $z \sim 7.5$, and $3.9\times 10^9\, \Msun$ at $z = 6.1$, matching the SMBH of the most distant quasar ULAS J1342+0928 at $z = 7.54$, and many bright $z \sim 6$ quasars.

The fact that our simulation produced a SMBH at $z \sim 6$ similar to that \citet{Li2007}, who only followed six major mergers from $z=14.4$ to $z=6$, highlights the significant role of galaxy interaction in BH growth by providing large gas inflow to feed the BHs. We note that although this SMBH underwent multiple mergers, the mass assembled from merged BHs is insignificant compared to its total mass at $z \sim 6$.  The most massive BH merger occurred at $z \sim 8.7$ and it acquired a BH of only $ 1.3 \times 10^6\, \Msun$, about $\sim 1\%$ of its total mass at that time,  while most of the BH mergers had large mass ratios $> 100$. Our main target is the most massive BH, but the minor mergers may result in the much smaller BHs being kicked out due to large gravitational recoil \citep{Tanaka2009}. In total, for this SMBH from the simulation, only $\sim 4\times 10^6\,  \Msun$ came from mergers of BHs, which is $\sim 0.1\%$ of the total BH mass at $z =6.1$. Therefore, the most important channel for the BH growth in this simulation is gas accretion rather than BH mergers, supporting the previous claim by  \citet{Li2007}. 

The efficient gas accretion by the most massive BH is fueled by dense gas surrounding the BH. As shown in the bottom panel of Figure~\ref{fig:bh_history}, the gas density around the first progenitor increased rapidly by 4 orders of magnitude from $n_{\rm H} = 10^1\,  \rm{cm^{-3}}$ at  $z \sim 18$ to $n_{\rm H}  = 10^5\,  \rm{cm^{-3}}$ at $z = 12$. Once the gas density rises above $n_{\rm H} = 10^4\,  \rm{cm^{-3}}$, the density threshold above which the effective sound speed drops, the Bondi accretion rate would increase \citep{Hopkins2010}. Therefore, as demonstrated in the bottom panel of Fig.~\ref{fig:bh_history}, a combination of a high gas density and low sound speed significantly enhanced efficient gas accretion. 

\subsection{Host Galaxy of the First Quasar}
\label{sec:host}

\begin{figure}
\includegraphics[width=1\linewidth]{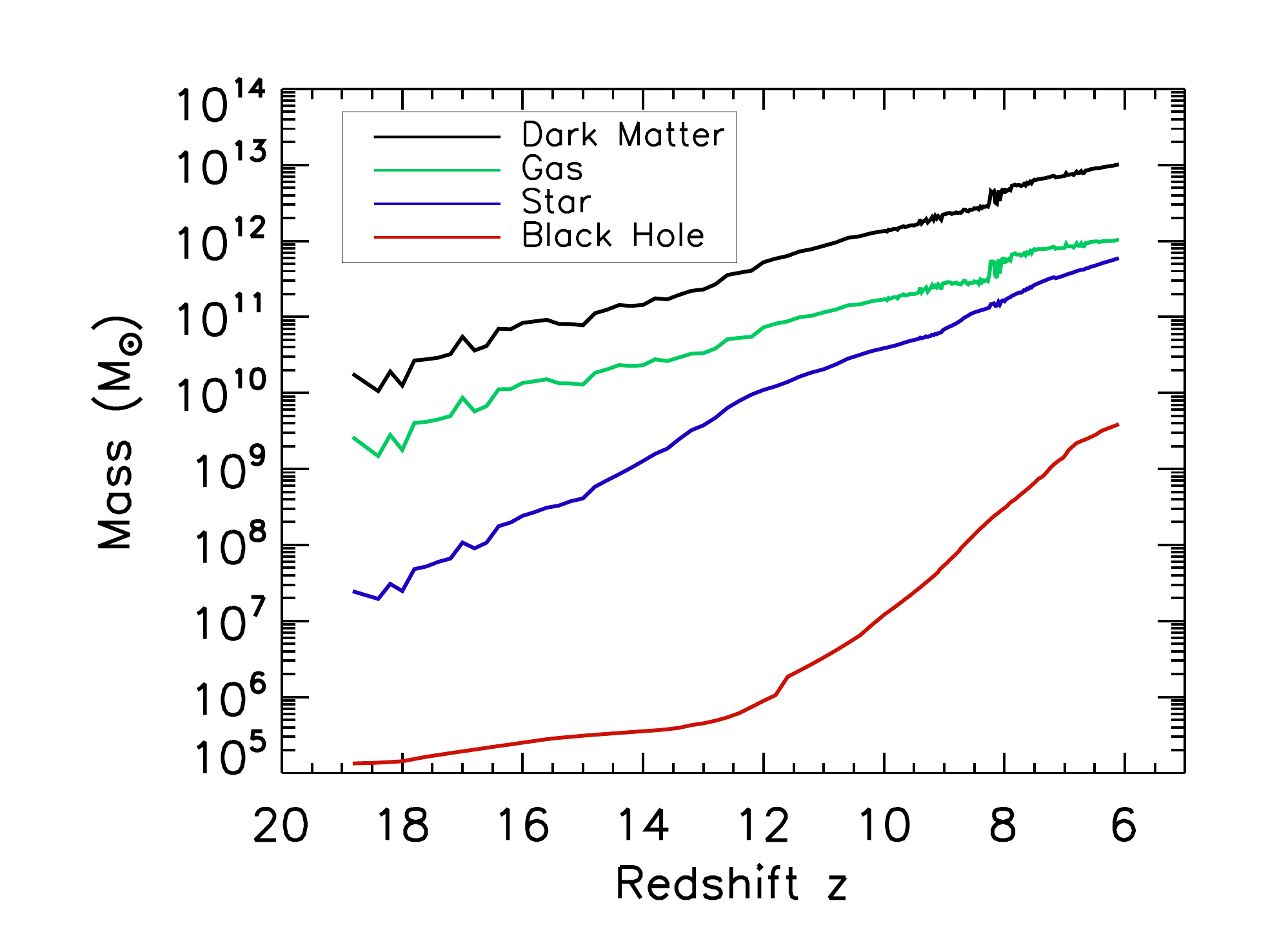}\\
\vspace{-0.5cm}
\caption{The assembly history of the most massive galaxy in the S5-REF simulation, which hosts the most massive BH at $z \sim 6$.  The wiggles show the mass changes during galaxy interactions. }
\label{fig:galaxy_history} 
\end{figure}

\begin{figure}
\includegraphics[width=1\linewidth]{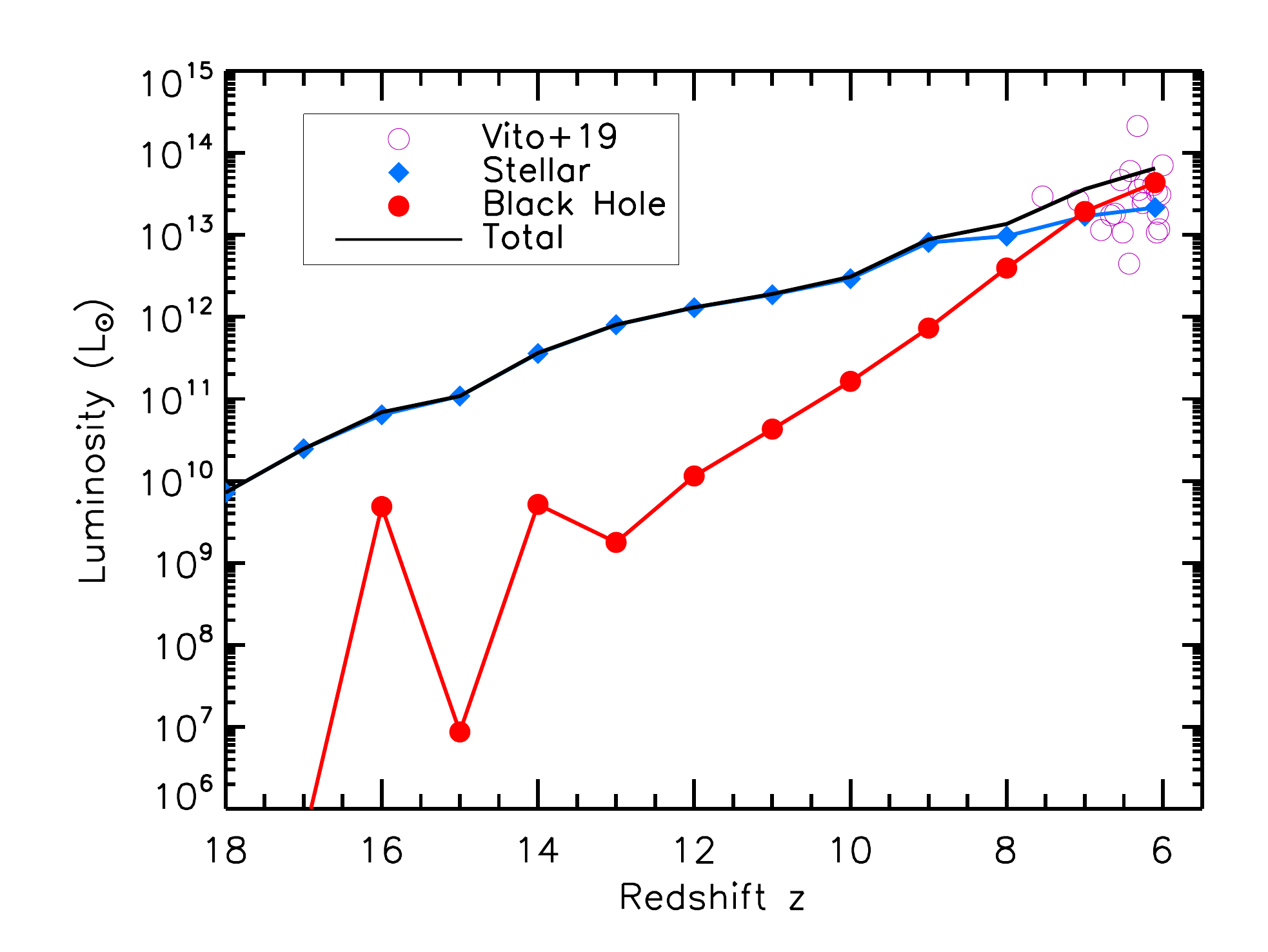}\\
\vspace{-0.5cm}
\caption{The evolution of luminosities from the stars and the most massive BH, as well as the total of both, for the quasar host galaxy in the S5-REF simulation. The luminosity from stars is calculated using Starburst99 \citep{Leitherer1999, Leitherer2011}, while that from the accreting BH is the bolometric luminosity $L_{\rm bol} = \epsilon_{\rm r} \Mdot c^2$. The open circles are bolometric luminosities of $z > 6$ quasars from \citet{Vito2019}.}
\label{fig:lum_z} 
\end{figure}

\begin{figure*}
\includegraphics[width=0.33\linewidth]{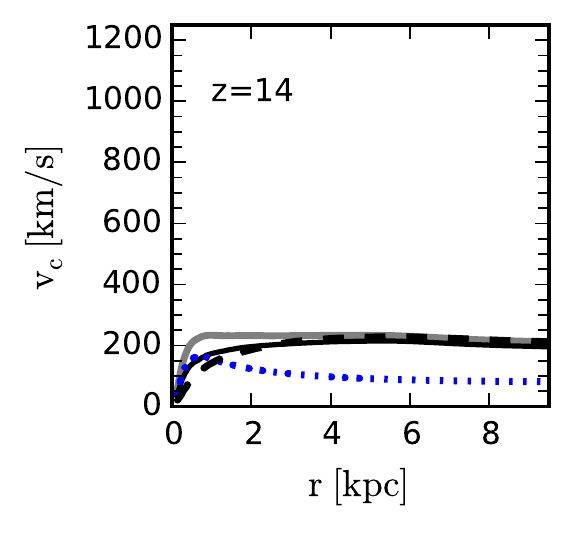}
\includegraphics[width=0.33\linewidth]{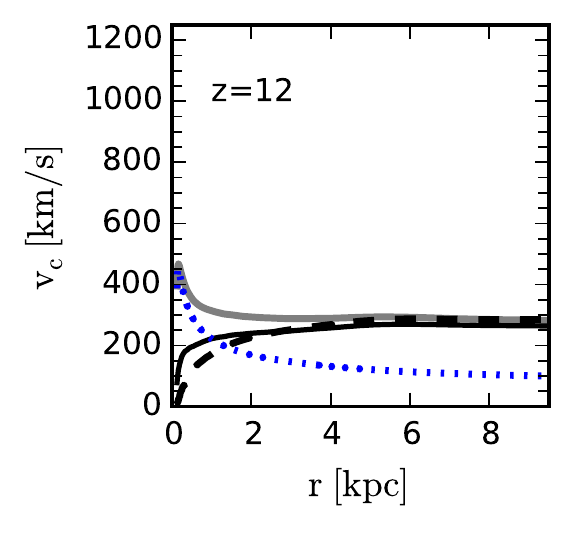}
\includegraphics[width=0.33\linewidth]{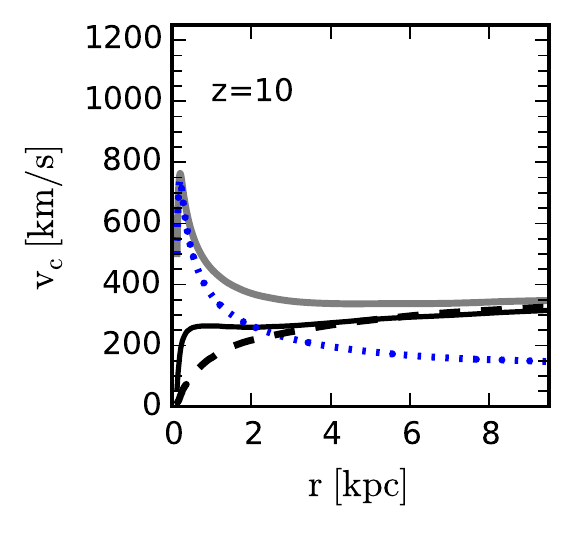}\\
\includegraphics[width=0.33\linewidth]{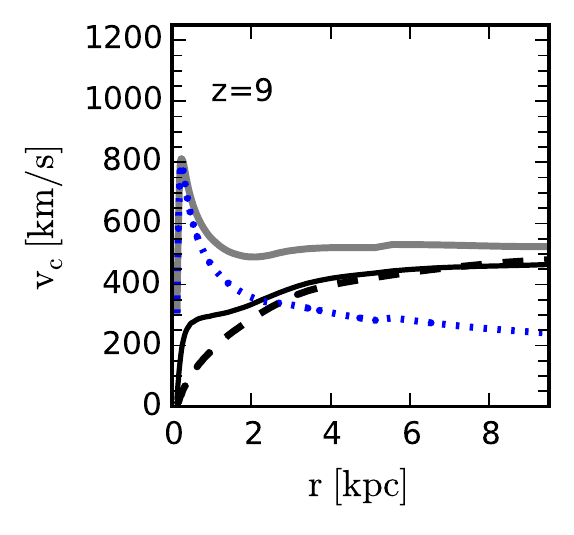}
\includegraphics[width=0.33\linewidth]{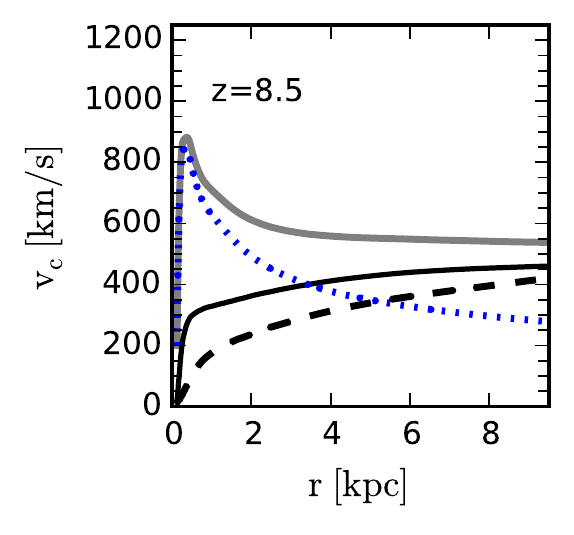}
\includegraphics[width=0.33\linewidth]{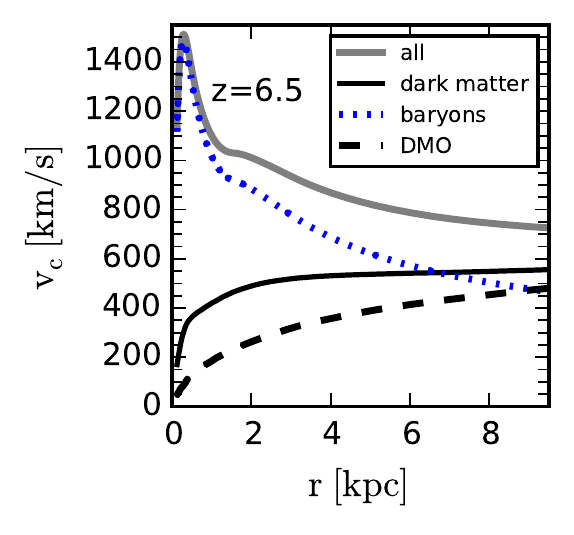}\\
\vspace{-0.2cm}
\caption{The circular velocity curves of the quasar host galaxy from $z = 14$ to $z = 6.5$ in the S5-REF simulation. For comparison, contributions from dark matter (black line) and baryons (blue dotted line) are shown, as well as the circular velocity curves of the same halo from a dark matter - only (DMO) simulation (black dashed line). 
$v_{\rm c}(r)$ from $z = 14$ to $z = 6.5$. }
\label{fig:vc}
\end{figure*}

Along with the intense star formation and BH growth, the host galaxy assembled its mass rapidly through accretion from the dense filaments and a series of mergers with other halos. Figure~\ref{fig:galaxy_history} shows the mass growth history of the main components of the galaxy, including the dark matter, gas, stars and the most massive BH. All the components appear to grow coevally as the host galaxy experiences a series of collisions and mergers with others in the highly overdense environment. These interactions lead to dramatic mass losses and gains, in particular when the galaxy was small at $z >15$, as evidenced in the wiggles of the mass curves of the gas and stars. After that, the galaxy became more massive and was able to maintain steady growth. The most massive BH grew slowly in the beginning, but starting at $z \sim 13$, its mass growth was accelerated exponentially by rapid accretion fueled by gas-rich mergers. By $z=6.1$,  the galaxy has accumulated $M_{\rm DM}=1\times 10^{13}\, \Msun$ for dark matter, $M_{\rm gas}=9.7\times 10^{11}\, \Msun$ for gas, $\Mstar=5.9\times 10^{11}\, \Msun$ for stars, and $\Mbh=3.9\times 10^{9}\, \Msun$ for the most massive BH, with a total mass of $M_{\rm tot} = 1.16\times 10^{13}\, \Msun$. 

The stellar population grew much faster than the BHs because the star formation rate was much higher than the BH accretion rate during the period $z=20 - 6$, as shown in  Figure~\ref{fig:gas_inflow_and_sfr}  and Figure~\ref{fig:bh_history}. At $z = 6.1$, the SMBH and the host galaxy fall on the local $\Mbh - \Mstar$ correlation suggested by \citet{Kormendy2013}, as shown in Section~\ref{sec:viablemodels}.

Figure~\ref{fig:lum_z} shows the evolution of the luminosity of the quasar host galaxy, with contributions from both the stars and the most massive BH. From early on, the luminosity of the host is dominated by stellar radiation until $z \sim 7$, when the BH catches up and outshines the stars. At $z=6.1$, the SMBH accretes at a rate of $\dot{M}_{\rm BH} = 29\, \Msun/{\rm yr}$ and produces a luminosity of $L_{\rm AGN} = 4.3\times 10^{13}\, \Lsun$ assuming a radiative efficiency $\epsilon_{\rm r} = 0.1$.  By taking into account the emission of $L_{*} = 2.2\times 10^{13}\, \Lsun$ from young stars calculated using Starburst99 \citep{Leitherer1999, Leitherer2011}, this galaxy would shine brightly with a total luminosity of $L_{\rm tot} = 6.5\, \times 10^{13}\, \Lsun$ at $z \sim 6$, making it a highly luminous quasar like SDSS J1148+5251 and SDSS J2310+1855 \citep{Vito2019}. Moreover, this galaxy reaches  $L_{\rm tot} \sim 10^{12}\, \Lsun$ by $z \sim 12$, which would represent a starbursting galaxy with a buried AGN in the early stage of the cosmic reionization. Therefore, this simulation offers valuable insights into the formation and evolution of the first galaxies and quasars. In a companion study (Li et al. in prep), we perform multi-wavelength radiative transfer post-processing on the S5-REF and other simulations to derive the panchromatic properties of $z > 6$ quasar hosts for direct comparison with multi-band surveys.
 
After the completion of the last major merger at $z \sim 7$, the host galaxy remains isolated for a period of time. A disk of rotating gas formed within the center of the galaxy potential. Newly born stars trace closely the prominent spiral arms within the gas disk. At $z = 6.1$, the host galaxy exhibits a bright concentrated bulge with old stars and a more extended (5 kpc in size) young stellar population. 

To study the structural properties of the host galaxy, we compute the circular velocity, $v_{\rm c} = (GM(<r)/r)^{1/2}$, of the galaxy and the contribution from each component, as shown in Figure~\ref{fig:vc}. We also include the circular velocity curves of the same halo from a dark matter - only (DMO) simulation for comparison. There is already a noticeable difference in $v_{\rm c}$ at redshifts as early as $z = 14$ where the maximum circular velocity $v_{\rm max}$ in the hydrodynamic simulation is larger than that in the DMO run, the distribution of dark matter in the hydrodynamic simulation shows clear enhancement in the central region, and the gap of $v_{\rm c}$ in the central region between the two simulations continues to increase with time.  

Starting from $z = 12$, the mass distribution in the central regions is dominated by baryons (stars and gas) rather than dark matter. The size of the baryon-dominated region increases from 1 kpc at $z = 12$ to 6 kpc at $z = 6.5$. Instead of a flat rotation curve as in the DMO simulation, the galaxy in the S5-REF simulation shows a strong Keplerian falloff from the central kpc to the outer radius. The rotational curves at $z=6.5$ suggest that the galaxy is compact, with a stellar mass of $\sim 4\times10^{11}\, \Msun$ within the central 2.5 kpc. 

The stellar mass of the galaxy at $z = 6.1$ in the S5-REF simulation is similar to that of previous hydrodynamic simulations \citep[][]{Li2007, Khandai2012, Feng2014, Lupi2019} and to semi-analytical modeling \citet{Valiante2014, Valiante2016, Pezzulli2016}. However, the compact size of the galaxy only signifies a mass discrepancy discussed in \citet{Valiante2014}. Recent measurements of gas kinematics of quasar hosts suggest that the inferred dynamical mass is $\sim 10^{10} - 10^{11}\,  \Msun$ \citep{Wang2010,  Wang2013}. Besides the effects of inclination of a $\sin^2(i)$ factor or the ``Lauer bias" \citep{Lauer2007}, another way to reduce this tension is to postulate that the stellar distribution is extended into tens of kpc as seen in \citet{Khandai2012}. However, an extended stellar distribution contradicts with the concentrated distribution in our simulation. It is thus crucial to determine whether the majority of high-$z$ quasar hosts are viewed face-on \citep[][]{Ho2007}. Moreover, recent simulations by \citet{Lupi2019} suggest that the galaxy mass in high-$z$ quasars observations is based on gas tracers that underestimate the true mass, which may explain the tension between models and observations. We will investigate this issue in a companion paper using the emission lines from radiative transfer calculations (Li et al. in prep).

The presence of massive galaxies at high-$z$ has already been reported in the literature \citep[e.g.][]{Mobasher2005, Wiklind2008, Marchesini2010, Stefanon2015}. It was argued that the majority of the stellar mass should be formed very rapidly in a manner similar to the monolithic collapse model \citep{ELS1962}, and that the rapid assembly of large stellar masses poses considerable challenges to the hierarchical structure formation model \citep[][]{Steinhardt2016, Glazebrook2017}. However, our results show that massive galaxies and hosts of luminous quasars can grow rapidly through accretion from filaments in a highly biased region and frequent mergers in a standard cosmological model of hierarchical structure formation. We will study the host galaxies of $z > 6$ quasars with a sample of 30 cosmological zoom-in simulations of different halos (Zhu et al. in prep).

\section{Test of  Models and Parameters}
\label{sec:tests}

In this Section, we present results of the 15 simulations listed in Table~\ref{table:sims} to test a number of models and parameters, including BH seeds with $\Mseed = 10^1 - 10^6\, \Msun$, Eddington limits with $\max(\lambda_{\rm Edd}) = 1 - 10^4$,  feedback from both constant-$\epsilon_{\rm r}$, radiative-efficient accretion and varying-$\epsilon_{\rm r}$ from super-critical accretion models, as well as Bondi accretion variations, and cosmological parameters. We will identify the viable models for $z \sim 6$ quasars with $10^9\, \Msun$ SMBHs. All these simulations use the same galaxy halo as S5-REF, as described in the previous Section~\ref{sec:ref}.

\subsection{The BH Seed Models}
\label{sec:testseed}

\begin{figure*}
\includegraphics[width=0.34\linewidth]{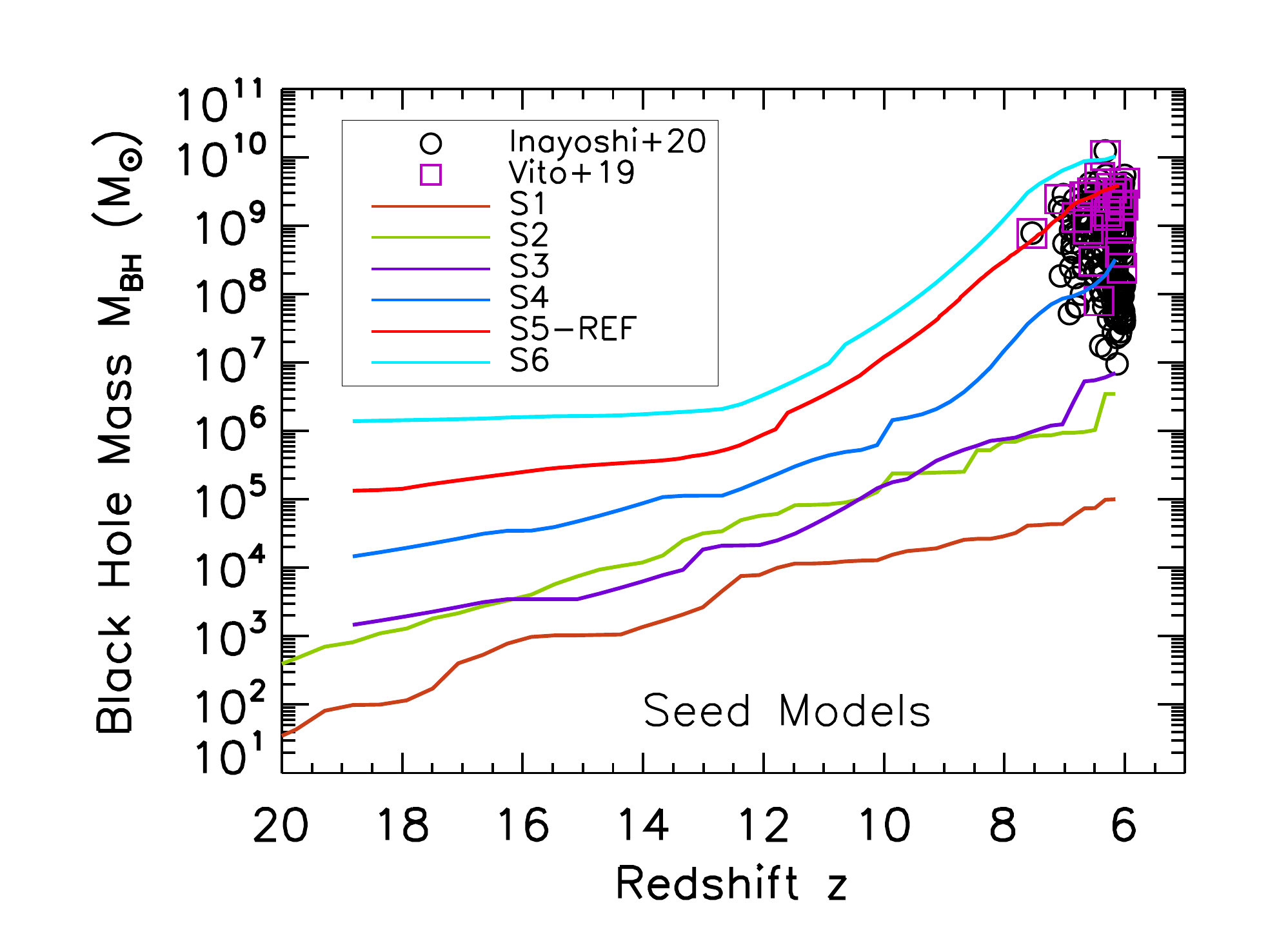}
\hspace{-0.5cm}
\includegraphics[width=0.34\linewidth]{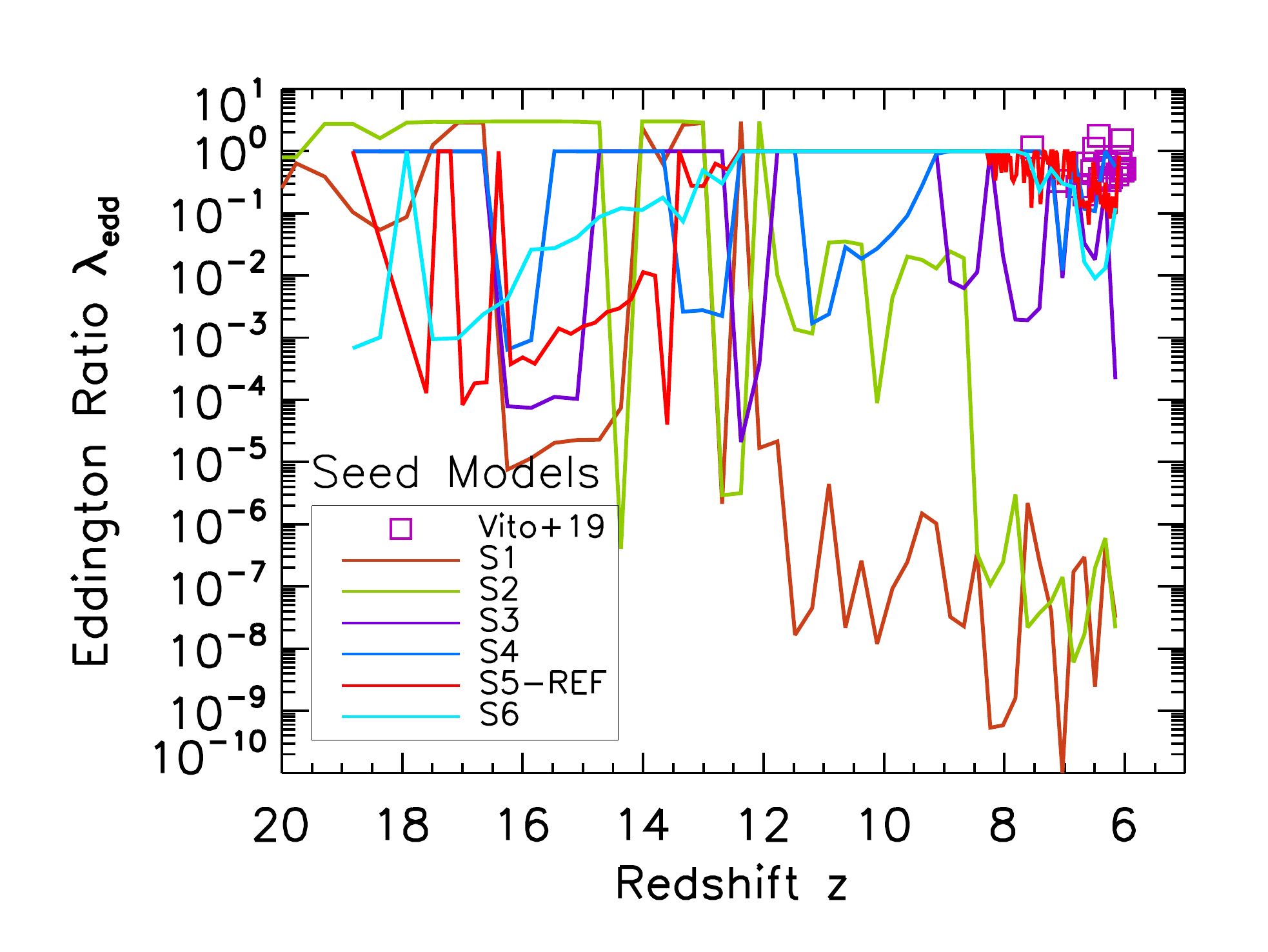} 
\hspace{-0.5cm}
\includegraphics[width=0.34\linewidth]{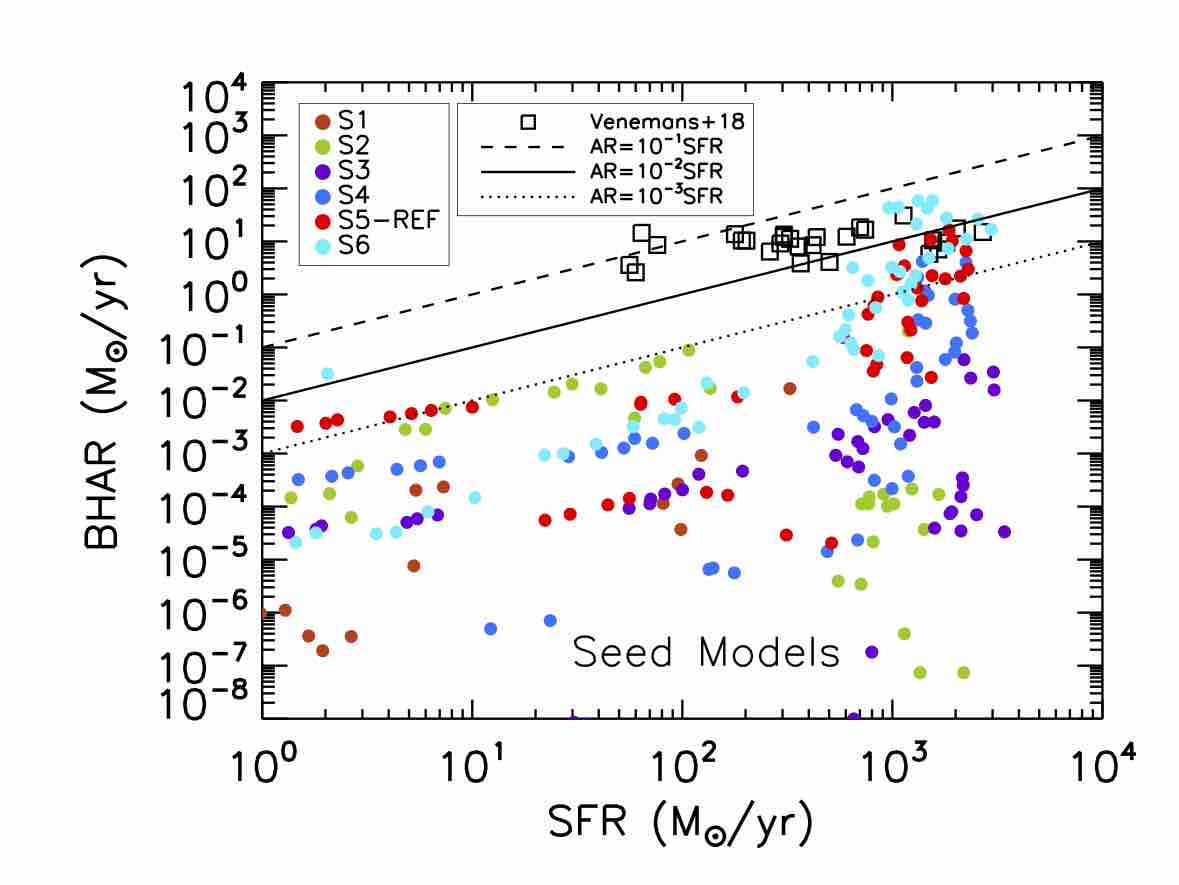}
\caption{A comparison of the Seed Model simulations with BH seed masses of $10, 10^2, 10^3, 10^4, 10^5$ and $10^6\, \Msun$/$h$, corresponding to S1, S2, S3, S4, S5-REF and S6, respectively, as listed in Table~\ref{table:sims}. Note the BH properties in the plots refer to the most massive BH in each simulation, and each colored line or circle represents a specific simulation as indicated in the legend. {\it Left panel:}  The BH mass growth history from each simulation in comparison with measurements of 25 $z > 6$ quasars from \citet{Vito2019}, and those converted from UV luminosity $\rm M_{1450}$ of the complete sample of 203 $z \gtrsim 6$ quasars from \citet{Inayoshi2020}.  {\it Middle panel:} Evolution of Eddington ratio of the BH accretion from each simulation in comparison with observations from \citet{Vito2019}.  {\it Right panel:}  Relation between the BH accretion rate and the star formation rate of the host galaxy from each simulation, in comparison with observations from \citet{Venemans2018}, and scaling relations of ${\rm AR  =10^{-1}, 10^{-2}, 10^{-3} SFR}$, as represented by the black dashed, solid and dotted lines, respectively. }  
\label{fig:seed} 
\end{figure*}

\begin{figure*}
\includegraphics[width=0.34\linewidth]{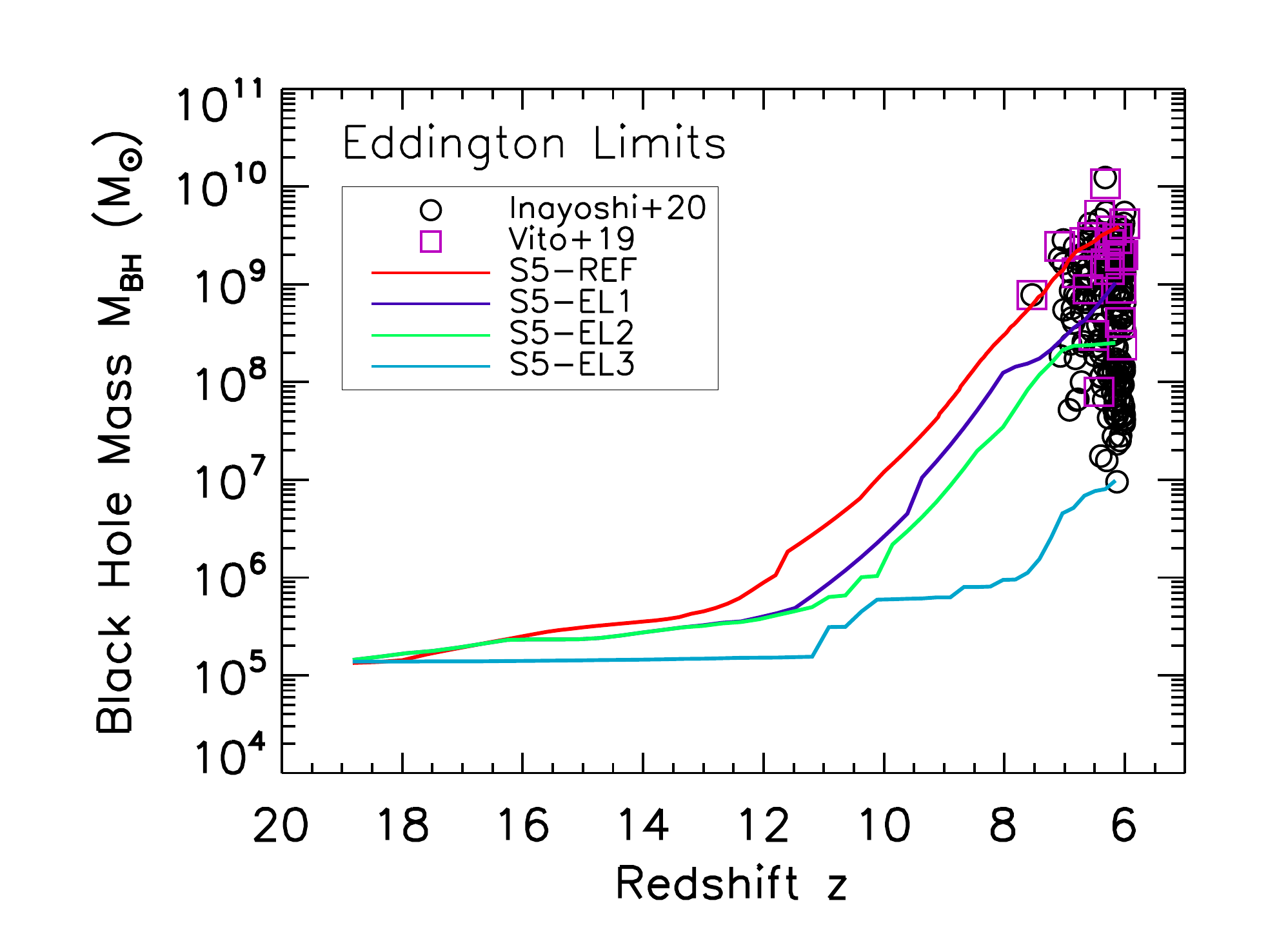}
\hspace{-0.5cm}
\includegraphics[width=0.34\linewidth]{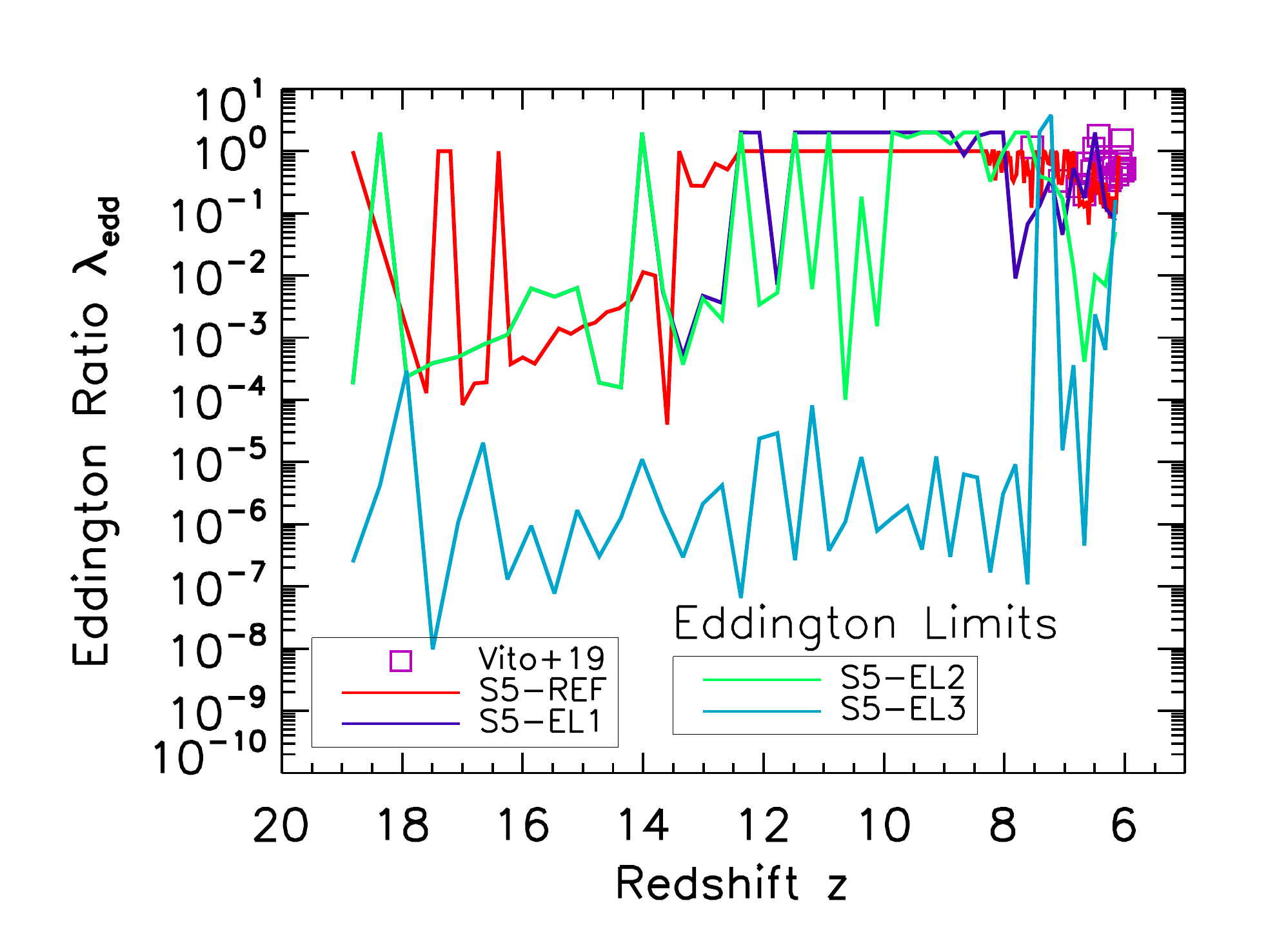} 
\hspace{-0.5cm}
\includegraphics[width=0.34\linewidth]{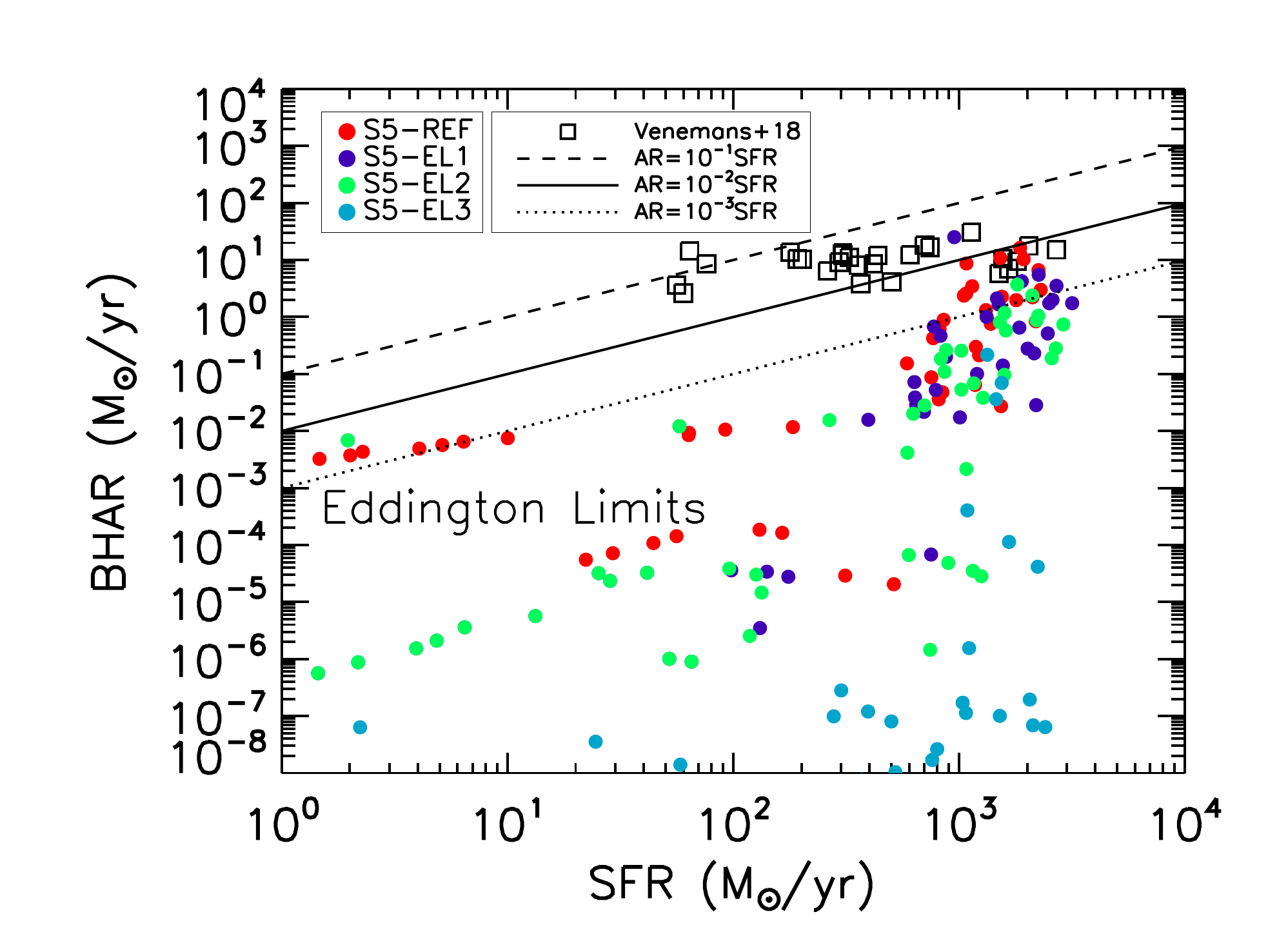}
\caption{Same as Figure~\ref{fig:seed}, but for simulations with different maximum Eddington limits: $\max(\lambda_{\rm Edd})$=1 (Eddington limited, S5-REF), 2 (super-Eddington, S5-EL1), 5 (super-Eddington, S5-EL2), and $10^4$ (hyper-Eddington, S5-EL3), respectively. } 
\label{fig:el} 
\end{figure*}

We test all three seed scenarios described in Section~\ref{sec:seed}: light seeds with $M_{\rm seed}=10$ and $10^2\, \Msun$/$h$, respectively; intermediate seeds of $10^3$ and $10^4\, \Msun$/$h$, respectively; and heavy seeds of $10^5$ and $10^6\, \Msun$/$h$, respectively. The light seeds are assumed to be remnants of Pop~III stars formed from metal poor gas of $Z < 10^{-4} \, Z_\odot$, so in the simulations they were created  2~Myrs after the Pop~III stars formed. For the intermediate and heavy seeds, since the initial conditions of their formation are not well constrained, we adopt the same scheme to plant the seed in $10^{10}\, \Msun$/$h$ halos so we can compare the results directly. In the simulation, the first Pop~III star formed at redshift $z \sim 26$, and the first halo reached $10^{10}\, \Msun$/$h$ at $z \sim 19$. For the simulations in this test, the light seeds were modeled with super-critical accretion with spin $a=0$, while the heavy seeds used the conventional Eddington-limited, thin-disk model with radiative efficiency $\epsilon_{\rm r}=0.1$ (a comparison between thin-disk and super-critical accretion for the heavy seeds is presented in Section~\ref{sec:testfb}). 

A comparison of the seed models is shown in Figure~\ref{fig:seed} with the six simulations S1, S2, S3, S4, S5-REF and S6, which correspond to the BH seed mass of  $10, 10^2, 10^3, 10^4, 10^5, 10^6\, \Msun$/$h$, respectively. The left panel shows the BH mass growth history of each seed model in comparison with BH masses of  $z > 6$ quasars measured by \citet{Vito2019}, as well as those converted from UV luminosity $\rm M_{1450}$ by \citet{Inayoshi2020} for the complete sample of 203 $z \gtrsim 6$ quasars.  As the curves demonstrate, small seeds S1, S2 and S3 grow slowly with a shallow slope, while big seeds S4, S5-REF and S6 grow exponentially from $z \sim 12$ on after an initial slow phase. At  $z =6.1$, the most massive BH of the S1 - S6 simulation grows to $\sim 10^5,  3.6\times 10^6, 7\times 10^6, 3.3\times 10^8, 5.3\times 10^9,  1\times 10^{10}\, \Msun$, respectively. The small seeds S1 - S3 fail to reach the BH mass threshold of $10^7\, \Msun$ of the quasars detected at $z > 6$, while the bigger seeds S4 - S6 produce SMBHs in the mass range of the observed $z > 6$ quasars, with S4 at the lower end and S6 on the high end above all but one, nearly matching the SMBH of SDSS J0100+2802 at $z=6.31$, the most massive SMBH in the early Universe \citep{Wu2015}. The S5-REF appears to be the most successful model of the six in the test, producing a SMBH close to that of the most distant quasar ULAS J1342+0928 at z = 7.54 \citep{Banados2018}, and numerous luminous quasars at $z \sim 6$ \citep{Vito2019}.

The middle panel of Figure~\ref{fig:seed} shows the Eddington ratio of each model as a function of time. All models have highly fluctuating accretion rates, although bigger seeds are able to maintain higher rates than the smaller ones. Even though S1 and S2 were run with the super-critical accretion model, both had only brief periods of super-Eddington accretion in the beginning, then dropped to sub-Eddington after $z \sim 12$ when multiple galaxy mergers took place; especially S1 dropped by nearly 9 orders of magnitude. The big seeds typically had $\lambda_{\rm Edd} \gtrsim 10^{-3}$, in particular S5-REF and S6 had near-Eddington accretion for a significant fraction of the time, and the Eddington ratios are in good agreement within the range of $\lambda_{\rm Edd} \sim 0.1 - 2$ of quasars at $z \sim 6 - 7.5$ \citep{Vito2019}. 

The right panel of Figure~\ref{fig:seed} shows the relation between the accretion rate of the most massive BH and the SFR of the host galaxy from simulations S1 - S6, in comparison with measurements of $z \gtrsim 6$ quasars from \citet{Venemans2018} in which the SFR was derived from the far-infrared luminosity while the BH accretion rate was estimated from the bolometric luminosity assuming $\epsilon_{\rm r}=0.1$  \citep{Decarli2018}. The hosts of the observed $z \gtrsim 6$ quasars are strong starbursts with SFR $\sim 50 - 3000\, \Msun/{\rm yr}$, and the BH accretion rates (BHARs) reach $\sim 5 - 40\, \Msun/{\rm yr}$. These observed quasars fall in the range of ${\rm  10^{-3} SFR \lesssim  BHAR \lesssim 10^{-1}SFR}$.  In our modeling, the host halo is the same for all the simulations, and it is a strong starbursting galaxy as shown in Section~\ref{sec:ref} with SFR $\sim 300 - 1400\, \Msun/{\rm yr}$ at $z \sim 6-7$, agreeing with observed $z \gtrsim 6$ quasar hosts. The BH accretion, however, differs significantly from model to model. S1 - S3 models have low BH accretion rates $<10^{-2}\, \Msun/{\rm yr}$ most of the time, while S4 - S6 have higher rates than the lighter seeds, in particular during $z \sim 6-7$,  S5-REF and S-6 have rates $\sim 1 - 30\, \Msun/{\rm yr}$, and they fall in the range of ${\rm  10^{-3} SFR \lesssim  BHAR  \lesssim 10^{-1}SFR}$ like the observed $z \gtrsim 6$ quasars. 

We note that the Simba simulation \citep{Dave2019} produced the same linear $BHAR - SFR$  correlation from $z \sim 5$ to $z \sim 0$ \citep{Thomas2019}. This is because in their model, the BHAR estimator scales linearly with the gravitational torque-limited gas inflow rate, with which the SFR also scales linearly. In our simulations, as we show in Figure~\ref{fig:gas_inflow_and_sfr}, the SFR follows closely with the gas inflow rate, however, as shown in Figure~\ref{fig:seed}, the BHAR does not scale linearly with SFR, and the $z \sim 6$ quasar observations by \citet{Venemans2018} do not show a linear $BHAR$--$SFR$  correlation. Furthermore, analysis by \citet{Pensabene2020} suggest that $z \sim 6$ quasars do not follow the same $\Mbh - \Mstar$  correlation as local galaxies.

The two light seed simulations ($\Mseed$=10, 100 $\Msun$) were indeed run with the super-critical slim-disk model, and they did have super-Eddington accretion, although for some short periods only, as shown in the middle panel of Fig 10. The accretion history is rather chaotic, and the radiative efficiency fluctuates quite widely with the BH mass and accretion rate, so the growth is not steady as the semi-analytical calculations even with the super-critical mode. Resolution is certainly a big concern here as we cannot resolve these light seeds and the gas density around them, so it is possible that the accretion rate is underestimated.

Overall, it is clear from Figure~\ref{fig:seed} that the BH growth path and final mass depend strongly on the seed BH mass. In contrast to previous studies \citep[e.g.,][]{Madau2014, Volonteri2015}, our simulations show that light seeds of $10, 10^2\, \Msun$ from Pop~III remnants fail to grow to $10^7\, \Msun$ even with the super-critical accretion model due to strong feedback. Both models reach super-Eddington accretion, but for some short periods and with large fluctuations. The chaotic growth is not steady as is typically assumed in the semi-analytical calculations. The accretion rates may be underestimated due to our limited resolution as we cannot resolve the gas density around the BHs, but similar results were also reported by other studies with different resolutions \citep[e.g.,][]{Lupi2016, Smith2018}.  In particular, \citet{Smith2018} followed the growth of over 15000 BHs from Pop~III stars in the Renaissance simulations with ultra-high resolutions down to ${\rm m_{DM}} = 1.27\, \Msun$/$h$, and they found inefficient growth for most of the BHs, with an average mass increase less than 10\%. On the other hand, the bigger seeds of $10^4 - 10^6\, \Msun$ were able to achieve higher rates than the smaller ones and maintain near-Eddington accretion for a substantial period of time, enabling them to grow to $10^8 - 10^9\, \Msun$ by $z \sim 6$, in agreement with other simulations \citep[e.g.,][]{Sijacki2009, Dubois2013, Costa2014, Feng2014, Curtis2016, Smidt2018, Huang2019}. The $10^3\, \Msun$ seed was planted using the same scheme as the heavy ones in our simulations for a head-to-head comparison, and it grew to $\sim 10^7\, \Msun$  at $z \sim 6$, bridging the gap between the Pop~III and direct collapse BHs, although \citet{Huang2019} suggested that they could grow to $10^8 - 10^9\, \Msun$ if they were seeded in smaller halos formed at earlier times. 

The finding of the BH seed tests highlights the need of heavy seeds for the formation of the luminous quasars observed at $z \sim 6$ in the standard structure formation model, motivating more theoretical and observational investigations on the formation of heavy BH seeds in the early Universe. 

\begin{figure*}
\includegraphics[width=0.34\linewidth]{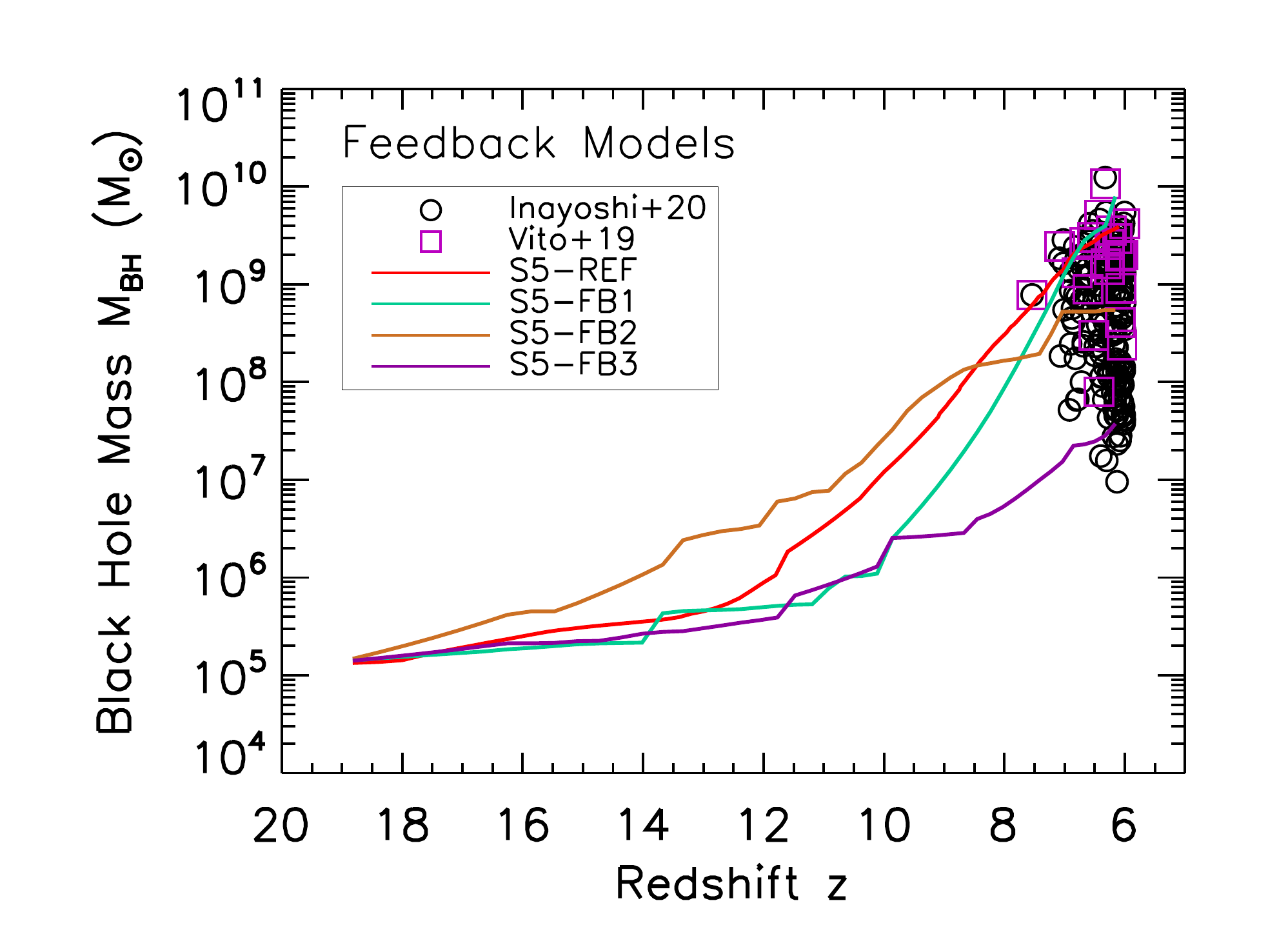}
\hspace{-0.5cm}
\includegraphics[width=0.34\linewidth]{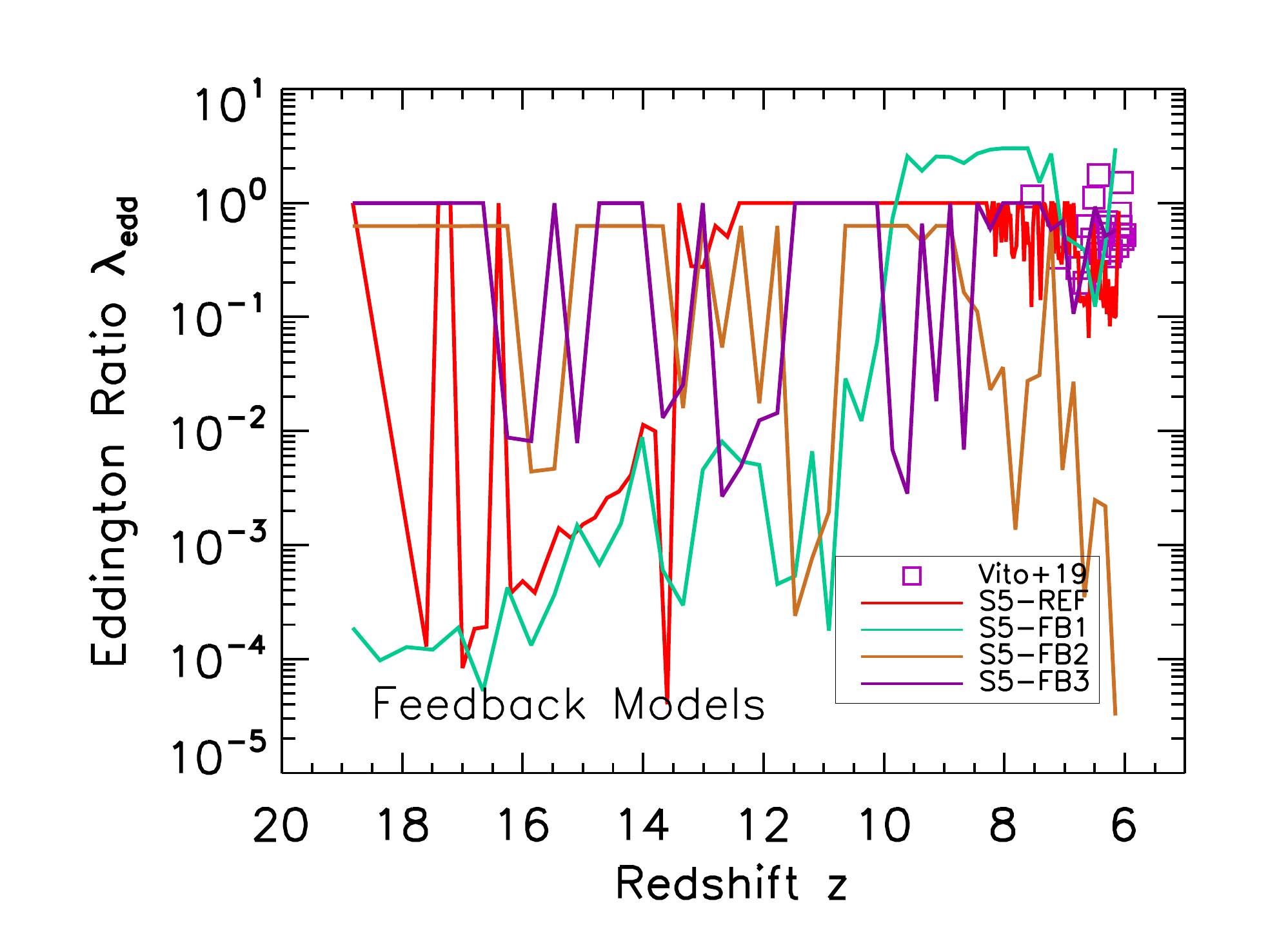} 
\hspace{-0.5cm}
\includegraphics[width=0.34\linewidth]{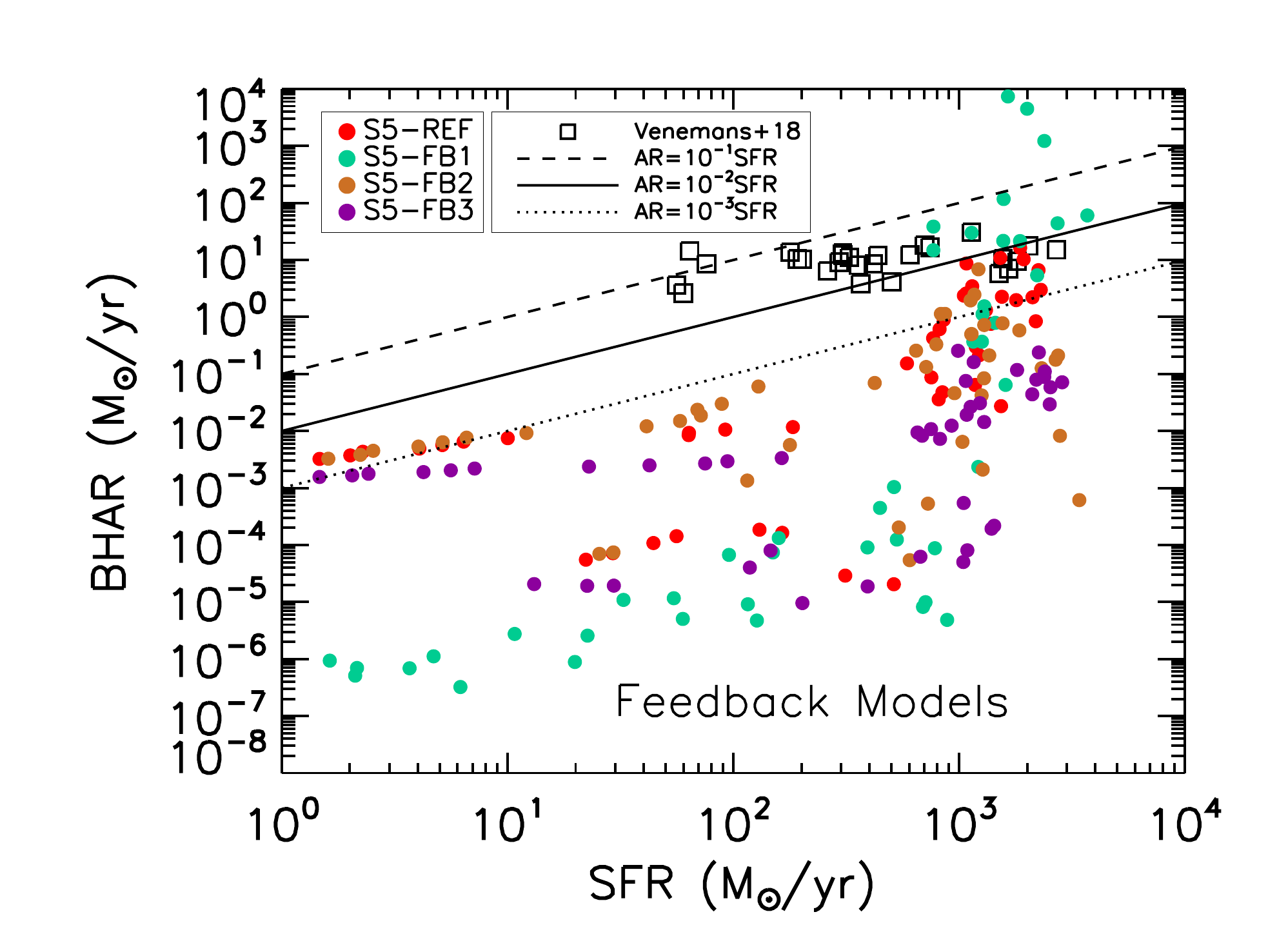}
\caption{Same as Figure~\ref{fig:seed}, but for simulations with different accretion and feedback models: thin disk accretion with $\epsilon_{\rm r}=0.1$ (S5-REF), super-critical accretion with varying $\epsilon_{\rm r}$ for spin $a=0$ (S5-FB1), $a=0.99$ (S5-FB2), and thin disk $\epsilon_{\rm r}=0.2$ (S5-FB3), respectively. }
\label{fig:fb} 
\end{figure*}

\begin{figure*}
\includegraphics[width=0.34\linewidth]{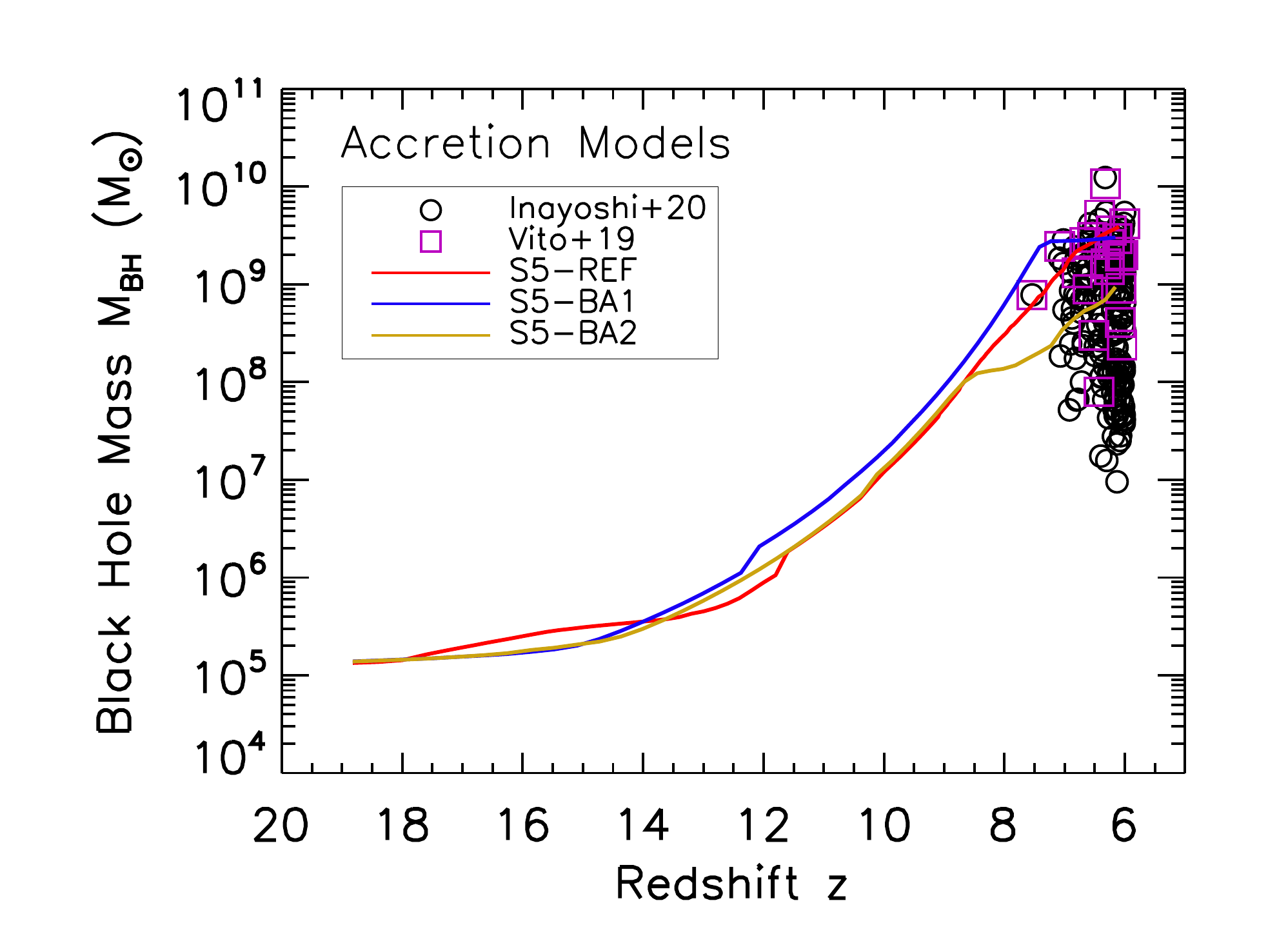}
\hspace{-0.5cm}
\includegraphics[width=0.34\linewidth]{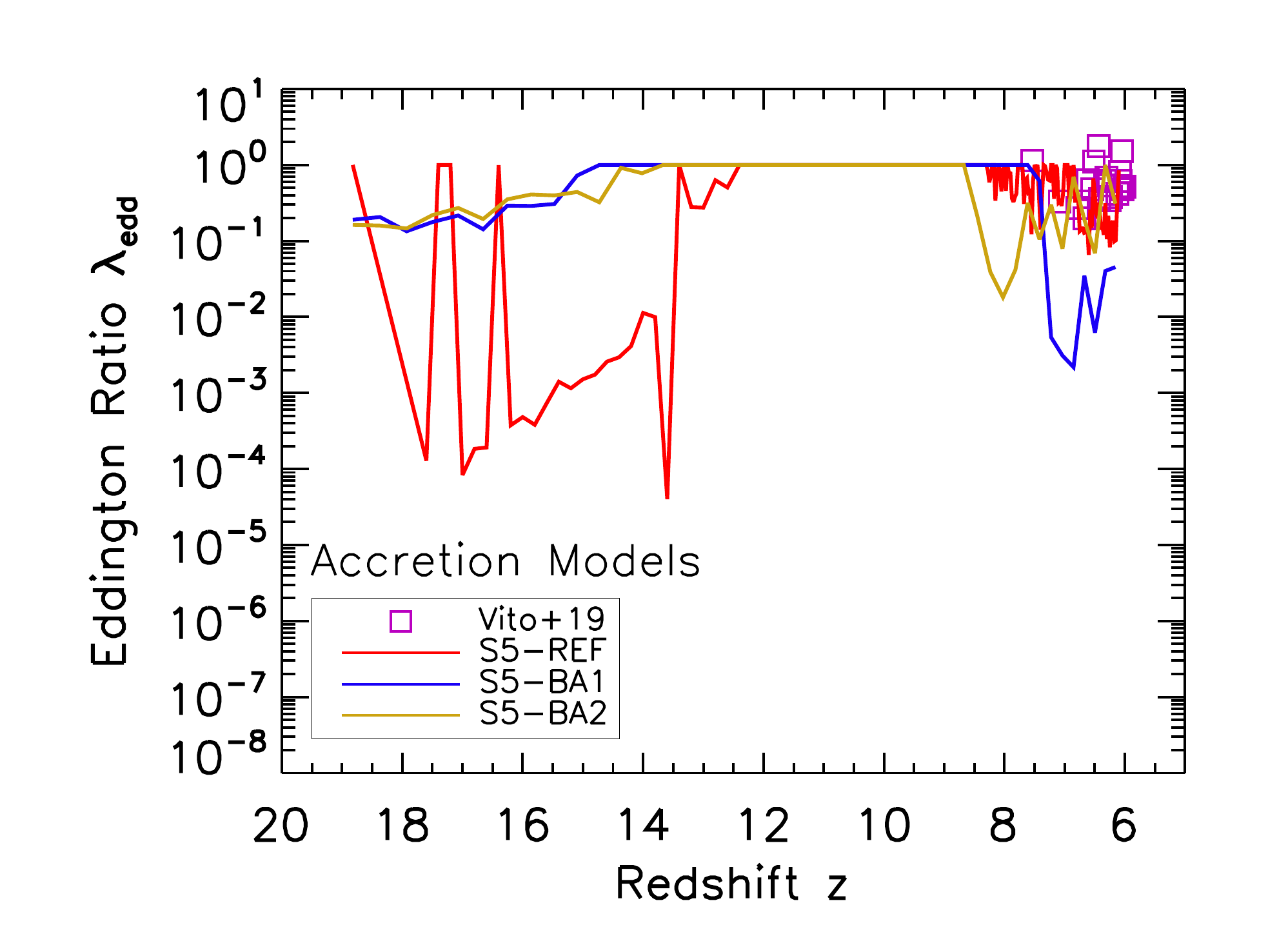} 
\hspace{-0.5cm}
\includegraphics[width=0.34\linewidth]{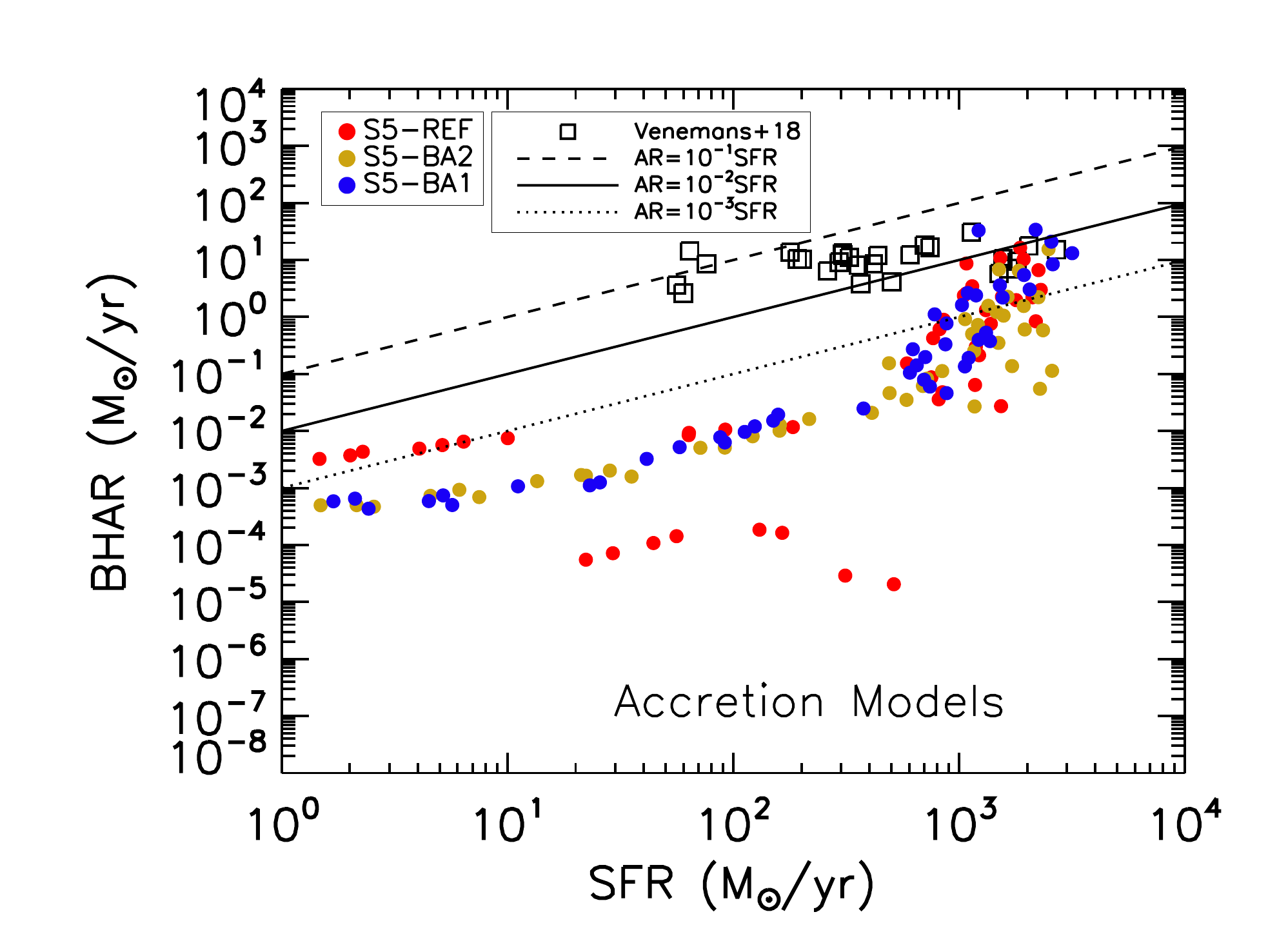}
\caption{Same as Figure~\ref{fig:seed}, but for simulations with different parameters for the Bondi accretion:  chaotic cold accretion (S5-REF), and conventional Bondi accretion with constant $\alpha=100$ (S5-BA1), and power-law $\alpha$ (S5-BA2).}
\label{fig:ba} 
\end{figure*}

\subsection{The Eddington Limit}
\label{sec:testel}

In order to explain the rapid formation of $\sim 10^9\, \Msun$ SMBHs within the first billion years, super- or hyper-Eddington accretion has been proposed \citep[e.g.,][]{Pacucci2015a, Inayoshi2016, Lupi2016, Ryu2016, Pezzulli2016, Sugimura2017, Becerra2018, Takeo2019}. However, some recent simulations suggested that feedback from super-Eddington accretion would suppress BH growth \citep[e.g.,][]{Smith2018, Regan2019}. Here we test the effects of the Eddington limit on the growth of BHs by varying the free parameter $\max(\lambda_{\rm edd})$. 

Figure~\ref{fig:el} shows a comparison of the four simulations S5-REF, S5-EL1, S5-EL2, and S5-EL3, which correspond to $\max(\lambda_{\rm edd})$=1, 2, 5, and $10^4$, respectively. Similar to Figure~\ref{fig:seed}, the left panel shows the BH mass growth history of each simulation in comparison with BH masses of  $z > 6$ quasars from \citet{Vito2019} and \citet{Inayoshi2020}. 

Counter-intuitively, Figure~\ref{fig:el} shows that a higher maximum Eddington rate leads to a lower BH mass growth; while the fiducial model with $\max(\lambda_{\rm edd})$=1, S5-REF produced a SMBH of $5.3\times 10^9\, \Msun$ at $z =6.1$, the Hyper-Eddington model with $\max(\lambda_{\rm edd}) = 10^4$ produced the least massive BH of $\sim$$10^7\,  \Msun$, more than two orders of magnitude lower than that of S5-REF $\max(\lambda_{\rm edd}) = 1$, due to strong feedback. This test of Eddington limits demonstrates the self-regulating of the growth rate of BHs by radiation feedback.

\subsection{The Radiative Feedback Models}
\label{sec:testfb}

To study the impact of feedback from an accreting BH on its own growth, we test two widely considered feedback mechanisms: the radiative-efficient thin-disk model with constant $\epsilon_{\rm r}$, and a super-critical slim-disk model with varying $\epsilon_{\rm r}$ depending on the BH mass, accretion and spin, as described in Section~\ref{sec:feedback}.

Figure~\ref{fig:fb} shows a comparison of the four simulations S5-REF, S5-FB1, S5-FB2, and S5-FB3, which correspond to simulations run with thin disk accretion with $\epsilon_{\rm r}=0.1$, super-critical accretion with varying $\epsilon_{\rm r}$ for spin $a=0$, $a=0.99$, and thin disk $\epsilon_{\rm r}=0.2$, respectively. Similar to Figure~\ref{fig:seed}, the left panel shows the BH mass growth history of each simulation in comparison with BH masses of  $z > 6$ quasars from \citet{Vito2019} and \citet{Inayoshi2020}. 

Clearly, the radiative efficiency $\epsilon_{\rm r}$ has a significant impact on the BH growth. For the thin-disk model, the final mass of S5-FB3 with $\epsilon_{\rm r} = 0.2$ is only $\sim 3.9 \times 10^7\, \Msun$, which is lower than that of the fiducial model S5-REF with $\epsilon_{\rm r} = 0.1$ by nearly two orders of magnitude. For the slim-disk model, S5-FB1 ($a=0$) reaches $\sim 8.3\times 10^9\, \Msun$, which is more than one order of magnitude larger than the $\sim 5.7\times 10^8\, \Msun$ of S5-FB  ($a=0.99$), as the latter has higher radiative efficiency due to higher spin. Owing to low radiative efficiency ($\epsilon_{\rm r}< 0.01$), the BH in S5-FB1 grows rapidly with super-critial accretion with an Eddington ratio $\max(\lambda_{\rm edd}) = 2.6$, as shown in the middle panel of Figure~\ref{fig:fb}. During the super-Eddington accretion phase around $z \sim 7 - 10$, the most massive BH in S5-FB1 reaches extremely high accretion rates $BHAR \sim 10^4\, \Msun/yr$, which is significantly larger than those of $z \sim 6 -7$ quasars estimated by \citet{Venemans2018}, as shown in the right panel of Figure~\ref{fig:fb}. As a result of super-Eddington accretion with low radiative efficiency, S5-FB1 produces the most massive BH among all the S5 simulations with the same BH seed, with $\Mbh=8.3\times 10^{9}\, \Msun$.         

\subsection{Bondi Accretion Variations} 
\label{sec:testba}

To compare the different Bondi accretion models employed in various cosmological simulations, Figure~\ref{fig:ba} shows the results from simulations using three models: our chaotic cold accretion model as in the fiducial run (S5-REF), the conventional model of \citet{DiMatteo2005} with constant $\alpha=100$ (S5-BA1), and the model of \citet{Booth2009} with a power-law $\alpha$ (S5-BA2). The overall growth of BH mass in these three models is in good agreement with each other. As shown in the left panel of Figure~\ref{fig:ba}, the S5-REF produces the most massive BH of the three, with $\Mbh \sim 3.9 \times 10^{9}\, \Msun$, while that of S5-BA1 is slightly smaller with $\Mbh \sim 3 \times 10^{9}\, \Msun$, followed by S5-BA2 with $\Mbh \simeq 10^{9}\, \Msun$. The detailed accretion of the BHs differs, however, as shown in the middle panel of Figure~\ref{fig:ba}, where S5-REF shows a more chaotic growth history, in particular in the early phase at $z > 14$, which results in lower accretion rates, as shown in the right panel of Figure~\ref{fig:ba}, but it becomes more stable at later times. All three simulations have a substantial period of near-Eddington accretion, which is critical to produce $\sim 10^{9}\, \Msun$ SMBHs at $z >6$.

We have also run the simulation with different cosmological parameters: WMAP 5-year results and Planck 2016 cosmology, respectively, but there is no significant difference in the BH growth histories and other results, as we will show in the next Section.

\subsection[Viable Models]{Viable Models for the $z \sim 6$ Quasars with $10^9\, \Msun$ SMBHs}
\label{sec:viablemodels}

\begin{figure}
\includegraphics[width=\linewidth]{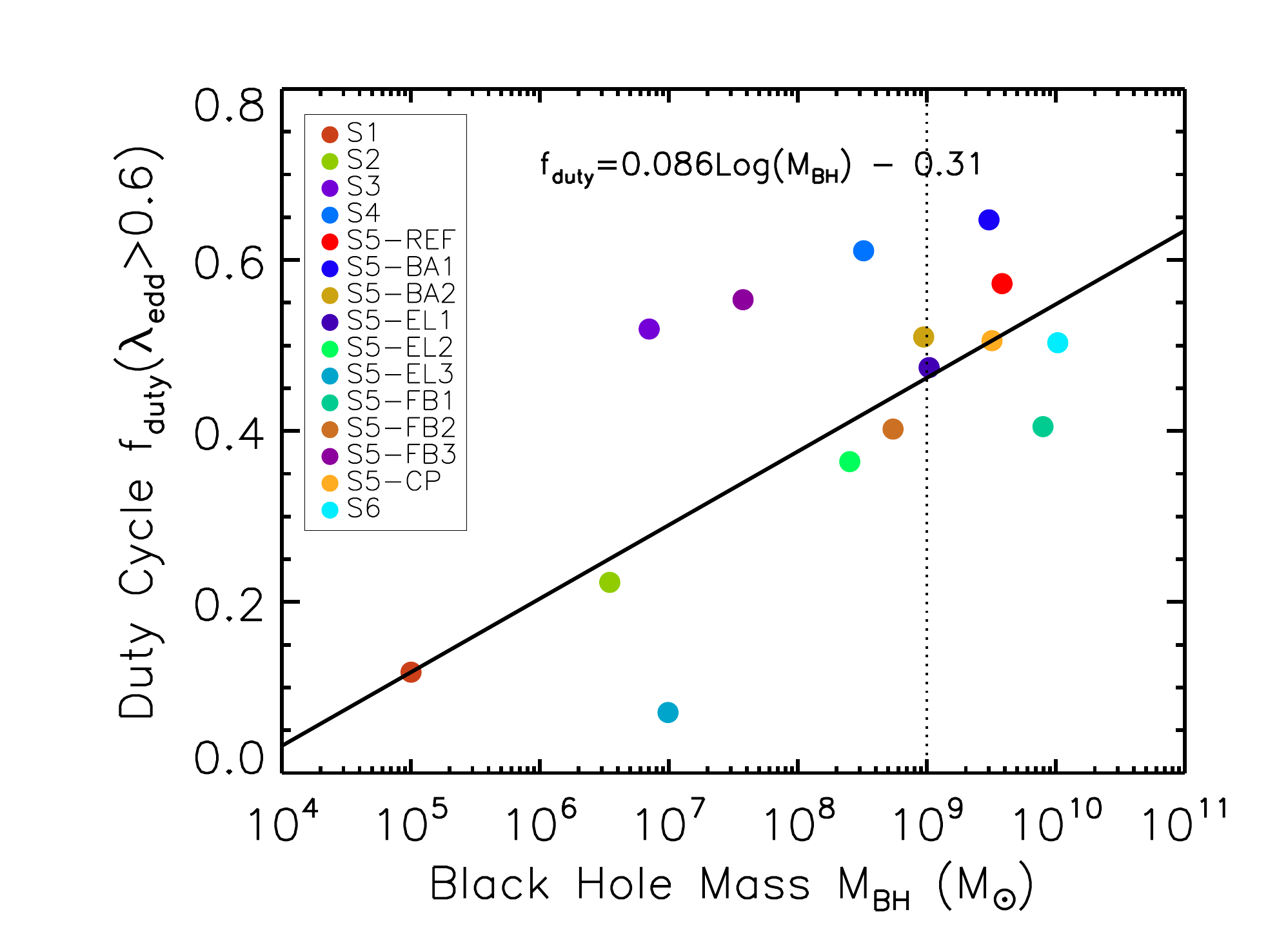}
\caption{The correlation between the duty cycle of near-Eddington accretion and the final BH mass of all simulations in this study.}
\label{fig:fduty} 
\end{figure}

\begin{figure}
\includegraphics[width=\linewidth]{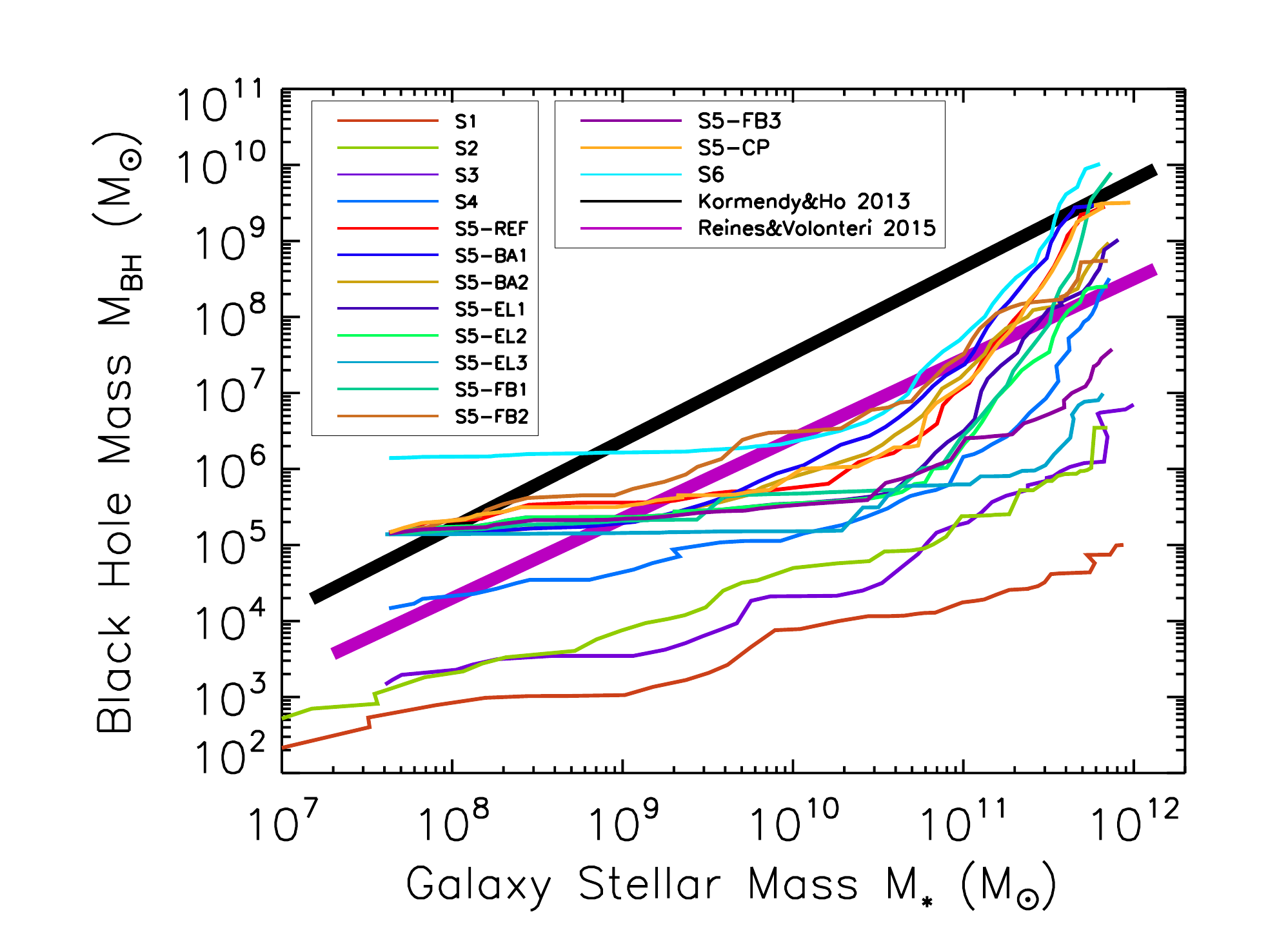} 
\caption{The correlation between BH mass $\Mbh$ and galaxy stellar mass $\Mstar$ across the growth history of all simulations in this study, in comparison with observed correlations of local ellipticals by \citet{Kormendy2013} and local AGNs by \citet{Reines2015}.}
\label{fig:mbh-ms} 
\end{figure}

To explore the viable paths to produce $\sim 10^{9}\, \Msun$ SMBHs at $z \gtrsim 6$, we examine the duty cycle of Eddington accretion of the BHs, $f_{\rm duty}$, as in Equation~(\ref{eq:tgrow}). Figure~\ref{fig:fduty} shows the correlation between the final BH mass from the simulations and the duty cycle $f_{\rm duty}$, with a fitting function $f_{\rm duty} = 0.086 \rm{log}(\Mbh) - 0.31$. This correlation suggests that the BH mass growth scales exponentially with $f_{\rm duty}$, so the longer a BH stays in the near- or super-Eddington accretion phase, the more rapidly it accumulates its mass. Figure~\ref{fig:fduty} also shows that heavy BH seeds of $\gtrsim 10^5\, \Msun$ can grow to $10^9\, \Msun$ by $z \sim 6$ in both thin- and slim-disk models, provided that the duty cycle of near-Eddington accretion with an Eddington ratio $\lambda_{\rm Edd} \gtrsim  0.6$ is maintained at $\gtrsim 40\%$.

Figure~\ref{fig:mbh-ms} shows the correlation between the final BH mass and the stellar mass of the host galaxy, throughout the assembly history until $z = 6.1$ for all simulations in this work.  In most cases, BHs start to grow more efficiently when the galaxy stellar mass exceeds $3\times 10^{10}\, \Msun$, corresponding to $z \sim 12$ when the host progenitor experiences major galaxy collisions. Compared to the local observed correlations, $\Mbh - \Mstar$, of ellipticals \citep{Kormendy2013} and AGNs \citep{Reines2015}, the growth curves of these early quasars do not follow the local $\Mbh - \Mstar$ relations, although at $z=6.1$, several models (S5-REF, S5-BA1, S5-CP, S5-FB1, S6) appear to fall on that of the local ellipticals, while S4, S5-EL1, S5-BA2, S5-EL2, and S5-FB2 fall on that of the local AGNs. This plot demonstrates that even with the same host galaxy, the BH growth history varies significantly with different  seeds, therefore they may have different $\Mbh - \Mstar$ relations at different stages of growth.  

As demonstrated in Figure~\ref{fig:mbh-ms}, the BH masses always catch up with the host galaxy mass, a feature that is not that surprising for our model. Recall that in the previous sections we show that most BH growth occurs between $z\sim12$ and $7$, determined by efficient gas inflow. It is only when the BH is massive enough so that its energetic feedback can effectively halt the gas accretion. By then, much of the gas has fueled intense star formation in the inner 2.5 kpc. Despite the differences in the seed mass or radiative efficiencies, there is no alternative route directly to feed the central BH without making stars along the way.

\section{Discussion}
\label{sec:discussion}

While our simulations provide a number of new insights on the formation of the first SMBHs and quasars, there are caveats as in other simulations.  In this section we discuss the limitations of our modeling and comparisons with previous studies.

\subsection{Limitations of Numerical Methods}

We stress that the models of star formation, BH accretion, and feedback from both supernovae and AGNs are all {\it phenomenological} descriptions, that the sub-grid recipes we employ to follow these processes in cosmological simulations are {\it simplified} approaches, and that the numerical methods and resolution are {\it limited} by the computational resources. 

It is computationally prohibitive to achieve sufficient numerical resolution to resolve individual stars or small BHs in cosmological simulations on $\sim 10$~Mpc scales, and so our resolution is far from resolving the BH seeds in the simulations. \cite{Smith2018} compared the growth of BHs from Pop~III remnants in the Renaissance simulations, which have a dark matter mass resolution of ${\rm m_{DM}} = 2.9\times10^4\,  \Msun$/$h$, to that from {\it Pop2Prime} simulations \citep{Smith2015} which have extremely high resolution (${\rm m_{DM}} = 1.27\, \Msun$/$h$), and found that the average accretion rate increased from $\sim 10^{-15}\,\Msun/yr$ to $\sim 10^{-13}\,\Msun/yr$ when the resolution increased by 4 orders of magnitudes. However, such an extreme regime with extremely low accretion is beyond the scope of our study, because the frequent hierarchical mergers of the halo in our simulations provide abundant gas supply and efficient angular momentum transport to fuel intense starbursts with star formation rates up to $10^3\, \Msun/yr$ and high BH accretion rates of $\sim 10^{-3} - 30\, \Msun/yr$, as shown in Section~\ref{sec:ref}. 

As mentioned in Section~\ref{sec:methods}, another significant challenge of this project is the treatment of small BHs of $10 - 100\, \Msun$ from Pop~III stars in simulations with mass resolution of $\sim 10^5 - 10^6\, \Msun$, because  the large mass ratio between particles would cause two-body heating and other numerical issues. To alleviate the problem, we adopted a numerical technique to treat the BH as a fraction of the parent Pop~III star particle from which the BH formed, such that the BH particle will have the same dynamics as its parent star particle while maintaining its own mass and accretion activity. This technique is similar to that used by \cite{Springel2003} to treat star formation in which a fraction of the gas particle becomes the new star, reducing two-body relaxation as the star particles have comparable mass to other particles, and it enables the small BHs to evolve robustly and with a growth history consistent with other massive BHs, as described in Section~\ref{sec:p3bh}. 

Recent developments, such as super-Lagrangian refinement \citep{Curtis2015} will be of great help to resolve the gas angular momentum before it reaches the accretion radius of the BH. In addition, quantities such as BH mass and accretion rate are less likely to be numerically converged than quantities such as the stellar mass \citep{Sijacki2015}. As the BH physics is always carried out on the resolution scale (grid size $\Delta x$ or smoothing length $h$), accurate modeling of BH growth in cosmological simulations may remain a significant challenge in the near future \citep{Wurster2013}.

\subsection{Caveats of Physical Models}

\begin{figure}
\includegraphics[width=\linewidth]{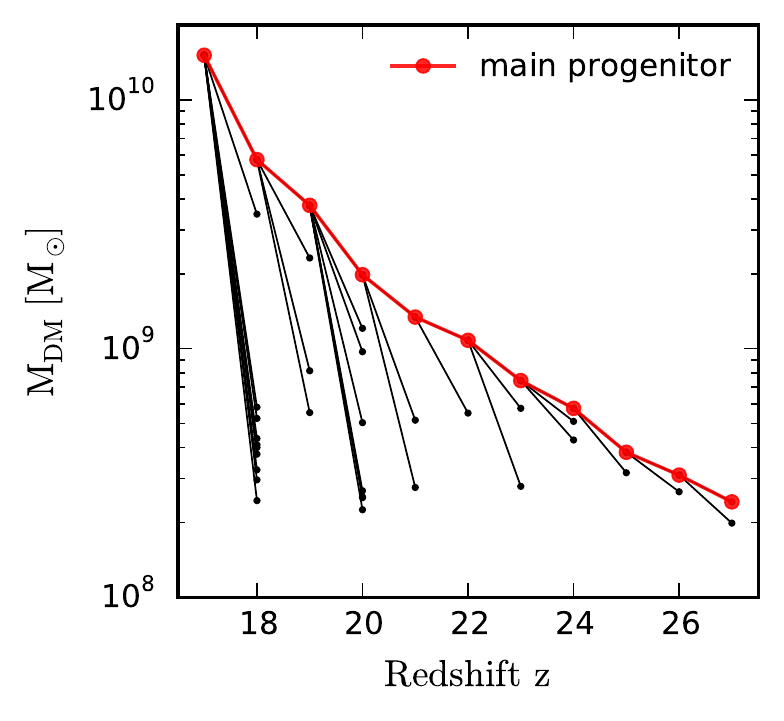}
\caption{Merger tree of the dark matter halo from redshift $z = 27$ to 17, when the first BH is seeded at $M_{200}=10^{10}\, \Msun/h$. The main progenitor is shown by the solid red symbols while the small symbols indicate the dark matter halos that merged with the main progenitor. There are 30 dark matter halos in total with mass above $10^8\, \Msun$ that merged between $z = 27$ and 18. }
\label{fig:merger_tree} 
\end{figure}

As we show in the model comparison section, BH mass is sensitive to model 
parameters such as the seed mass, seed halo mass, radiative efficiency, and 
the maximum Eddington ratio. Our model of BH accretion rate is based on the
Bondi-Hoyle formula with several different estimates of boost factors, although
the differences are found to be small due to the fact that most efficient mass
growth period is Eddington-limited. However doing this, we ignore the  
dependence on gas angular momentum \citep{Curtis2016b, Negri2016, 
AnglesAlcazar2017}. 

Since we focus on the bright quasar phase of the BHs in this study, only quasar mode feedback is included in our simulations.  We do not include radio feedback where jets or outflows from BHs at low state inflate giant radio lobes. Radio feedback model \citep{Sijacki2007, Dubois2011, Weinberger2017} is now routinely used in the current simulations \citep[][]{Vogelsberger2014b, Kaviraj2017, Pillepich2017}. We justify our choice of not modeling radio mode because the presence of radio-loud quasars at $z \ge 4$ is generally very rare \citep{Volonteri2011a, Yi2014}. Using numerical simulations, \citet{Sijacki2015} also found that for the most massive BHs, the mean Eddington ratio consistently drops from $z = 4$ to 0, a signature of cosmic downsizing. Therefore, the  radio mode feedback would be insignificant in the early quasars.

In our simulations, BH kick from velocity recoil during BH--BH mergers is not included. The circular velocity curve of the central galaxy has reached $\sim$800~\kms already at $z = 10$. In addition, the mass ratios of the BH--BH mergers are generally very low. The impact of any kick from gravitational recoil would be minimal. Furthermore, the strong gas drag force due to the presence of the gas disk also damps any kick substantially, so BH kick would not affect the growth of the most massive BHs we present here.

It is useful to examine how well the BH seed model for our fiducial choices agrees with the current direct collapse models. In Figure~\ref{fig:merger_tree}, we show
the dark matter merger tree from redshift $z = 27$ to 17 when the first BH
is seeded in our simulation.  There are 30 dark matter in total with mass above $10^8\,  \Msun$ that merged between $z = 27$ and 18. Since we only trace the major branch where the major progenitor is located, this number (30) is quite a conservative estimate. Assuming an occupation fraction of DCBHs between 5\% and 10\% \citep{DeGraf2020}, the seed BH mass at $z \sim 17$ at least should be $[1.5, 3] \times 10^5\,  \Msun$, which is slightly above our fiducial choice. Hence, our fiducial choice of BH seed mass is on the slightly conservative side. As a result, the BH mass at $z\sim6$ could be more massive than our calculation. 

However, the implementation of feedback and numerical resolution is likely to be
important.  It has been documented that BH mass also depends on how 
the feedback is coupled to the surrounding gas \citep{Wurster2013, Zubovas2016}. 
In addition, numerical resolution could potentially play an important role \citep{Bourne2015}. The BH mass in our high resolution is slightly lower than that in the low-resolution run. It is beyond the scope of this paper to study the impact of various implementations of feedback and resolution effects. 

In addition, the stellar mass of the host galaxy also contains some uncertainties. Essentially, we are extrapolating the wind scaling relations used in our previous work \citep{Zhu2016}, which works well for Milky Way-sized galaxies at $z = 0$, to this extremely massive system at high-$z$. We use a local dark matter velocity dispersion (around 60 dark matter particles) as a proxy to estimate the wind velocity as in \citet{Vogelsberger2013}. Recently, \citet{Pillepich2017} has shown that this estimate introduces a redshift dependence. If the wind injection velocity and the wind mass load factor are  increased similarly as in our simulation, one
might expect that the stellar mass of the host galaxy will be reduced. 

\subsection{Comparison with Previous Zoom-in Quasar Simulations}
\label{sec:compsims}

\citet{Sijacki2009} performed a set of 6 zoom-in re-simulations of the most massive halo with a mass $10^{13}\, \Msun$ at $z=6$ from the Millennium simulation with various resolutions, using the conventional BH recipes ($10^{5-6}\, \Msun$ seed in $10^{9} - 10^{10}\, \Msun$ halo) but also including BH gravitational recoil. They produced a $10^9\, \Msun$ SMBH at $z \sim 6$.  They also found that resolution did not have a significant effect on the final BH mass. For example, the final BH mass differed by only a factor of $\sim 1.5$ between the lowest- resolution simulation and the highest-resolution one when the resolution was increased by a factor of 8. 

\citet{Feng2014} used zoom-in SPH simulations of three high-redshift quasars from $10^{12}$--$10^{13}\,  \Msun$ halos at $z\sim6$ in the MassiveBlack simulation to examine cold flows as a viable feeding mechanism of high-$z$ quasars \citep{DiMatteo2012}. They show that the result of a $10^9\,  \Msun$ BH at $z\sim7$ seen in the uni-grid simulation, which was presented in \citet{DiMatteo2012}, is quite insensitive to numerical resolution, feedback prescriptions or SPH formulations (conventional SPH vs. pressure-entropy by \citealt{Hopkins2013}). At high-$z$, cold streams of gas from the cosmic web are able to directly feed the central BHs undisrupted by the feedback from accreting BHs. 

\citet{Dubois2013} simulated a slightly less massive halo ($5\times10^{11} \, \Msun$) at $z\sim6$ using the {\sc RAMSES} code. They found that cold filamentary gas fuels rapid Eddington-limited accretion in the early stages and the BH accretion is maintained both by smooth gas inflow and clump migration within gas disks once the feedback 
energy is sufficiently large to unbind and remove the cold gas. Also, the stellar-bulge mass content in the host galaxy is altered as the result of AGN feedback. The final BH mass at $z=6$ is $8 \times 10^{7} \, \Msun$, which is not consistent with billion-solar-mass BHs. However, this mass depends on the host halo mass as well as the adopted feedback efficiency. 

\citet{Costa2014} used the moving-mesh code {\sc Arepo} to simulate 18 dark matter halos from the Millennium simulation \citep{Springel2005qso} including the six most massive halos at $z = 6.1$. BH accretion in these regions proceeds at a much faster rate than the average region. At $z = 6.1$, the most massive BHs have already grown to $\sim$ $10^9 \, \Msun$ while the BHs in the average region are $\sim 10^6 \, \Msun$. In their model, they found that galactic winds from supernova feedback are also crucial to reduce the rapid stellar mass build-up,  in general, to reconcile with the observed galaxy numbers in the $z\sim6$ quasars' neighboring region \citep[e.g.,][]{Kim2009, Simpson2014}. Feedback from BHs is able to drive a large amount of cold gas with a velocity faster than 1000 \kms \citep{Costa2015}, which is similar to what is observed in velocity width distributions.  

\citet{Curtis2016} studied an $8\times10^{12}\Msun$ halo at $z = 5$ using the {\sc Arepo} code. Regions around BHs are refined using the super-Lagrangian refinement method in \citet{Curtis2015}. A massive rotationally supported gas disk is found in the presence of a $\sim 10^9\,  \Msun$ central BH. In comparison, the cold gas disk is otherwise not present without the refinement method. 

\citet{Smidt2018} implemented X-ray photon transportation into the hydrodynamics and chemistry network of {\sc ENZO} and used it to simulate a $10^{12}\Msun$ halo at $z = 7$ without prior major merger history. They reported that a black hole fed by cold gas streams, seeded at $z \sim 20$ with $10^5\,  \Msun$, is able to grow to a supermassive BH with $2.15\times10^9 \Msun$ at $z\sim7$.

\citet{Huang2019} carried out three zoom-in simulations of the same halo from the BlueTides simulation with different seed masses of $5\times10^3$, $5\times10^4$, and $5\times10^5\, \Msun$ while keeping the halo to BH mass ratio fixed. They found that the BH mass converges to $\sim 10^9\, \Msun$ by z=6 regardless of the seeding procedure while their early growth histories at $z>10$ differ. The Eddington-limited accretion between $z=10$ and $6$ within gas-rich environments, until the BH feedback sets in, is in good agreement with what we have found.  The constrained realization of the Gaussian field they employed appears to be a promising way to simulate rare density peaks at high $z$, within a small periodic box of $\sim 15$Mpc, similar to the zoom-in region in our study.

The role of cold accretion in feeding the first quasars was noted by  \citet{DiMatteo2012}, although the low resolution of the simulations and (possible)
SPH artifacts may affect the dynamics of cold accretion and its interaction between
hot halo gas and dense star-forming gas \citep[e.g.,][]{Nelson2013}. Our simulation offers a more detailed view of the gas distribution around the first quasar thanks to its higher spatial resolution and the use of {\sc Gizmo} for a better performance in fluid mixing than the SPH code \citep[see][]{Hopkins2015}. This is an improvement step forward to bridging the scale of cosmological gas accretion to the central BH scale.

\subsection{Comparison with Previous Works on Super-critical Accretion}
\label{sec:compsc}

Super-critical accretion through a slim disk provides a viable route for rapid mass growth of light BH seeds into high-$z$ quasars \citep{Madau2014, Volonteri2015}. \citet{Madau2014} have argued that a moderate super-Eddington ratio $\sim$3--4  is able to grow a $100\, \Msun$ stellar mass BHs to a $10^9\, \Msun$ BH before $z = 6$. Using simulations with the {\sc RAMSES} and {\sc Gizmo} codes,  \citet{Lupi2016} simulated the growth of BHs using a super-critical accretion model. They employ a star formation and feedback model with delayed cooling and the AGN thermal feedback model of \citet{Dubois2014}. In their low-resolution runs, super-critical 
accretion is able to increase BH masses from $100\, \Msun$ to $10^4\, \Msun$ within a rapid Myr timescale, which is much faster than the standard thin accretion disk model. In their high-resolution runs, the growth rates of the BH mass are both reduced significantly in the thin accretion disk and slim-disk accretion models. Note that, the accretion rate estimate in \citet{Lupi2016} estimates the mass flux rate within a prescribed accretion zone, which is different from Bondi-like accretion. 

Due to the different models and their implementations in \citet{Lupi2016}, it is difficult to directly compare their results with ours. In \citet{Lupi2016}, BH accretion is halted once the feedback energy from SNe and BHs creates a hot bubble. In their high resolution run with $\epsilon_{\rm r} = 0.1$, the feedback energy is strong enough such that the BH mass barely grows within 3 Myr. The most efficient growth of BHs occurs when they are embedded in cold gas, which could be the result of replenishment of cold gas from large scale filaments and mergers.  A maximum Eddington ratio of 500 in their simulation enables a brief episode of rapid mass growth. 

In our simulation, the accretion of cold gas is modeled within the two-phase ISM model. Any thermal feedback from BH accretion is immediately distributed around the neighboring gas particles.  As a result, the effect of BH feedback is not solved explicitly in the cold/hot gas phase. The effect of feedback appears to be strong as indicated by the large oscillations in the mass accretion rate before $z = 14$ shown in Figure~\ref{fig:bh_history}. This approach is less sensitive to numerical resolution, as discussed in the previous sections. As a result, the boost of mass growth rate with the super-critical accretion in our simulation is quite modest. There is one advantageous feature of the super-critical accretion model though, in that it is able to grow a $10^{10}\, \Msun$ BH at $z\sim6$ while not affecting the majority of the less massive BHs.  However, when BHs are spun up due to coherent flows with a large duty cycle \citep{Volonteri2013}, the final BH mass is even closer to our fiducial model. 

We are mindful that the simulation results depend strongly on the physical models and numerical methods. We note that the different BH accretion model in the Simba simulation assumes that the rate scales linearly with the gas inflow rate, which may significantly boost the BHAR and make it easier for small BH seeds to grow more quickly. However, the observations suggest that the $z \sim 6$  quasars they do not follow a linear $BHAR$--$SFR$  correlation \citep{Venemans2018}, and that they do not follow the same $\Mbh - \Mstar$  correlation as local galaxies \citep{Pensabene2020} as produed by the Simba simulation. Our simulations show broad agreement with observations of $z \sim 6$, which is encouraging, but more work is needed to fully unravel the BH accretion and star formation at high redshifts.

\section{Conclusions}
\label{sec:summary}

We have performed a suite of 15 cosmological zoom-in simulations to test models of the formation of early quasars. We chose the most massive halo from a $\sim\rm{Gpc}^3$ volume, which has a mass of $\sim 10^{13}\, \Msun$ at $z=6$ and a history of multiple major mergers. We test BH seeds of $10^1 - 10^6\, \Msun$, and different accretion and feedback models including both radiatively-efficient thin-disk and super-critical slim-disk accretion. Here is a summary of our findings: 

\begin{itemize}

\item Abundant gas supply and efficient angular momentum transport by gravitational torques is critical to black hole growth and star formation of the quasars, but the eventual black hole mass depends strongly on the seed mass and radiative efficiency. 

\item In our fiducial model with Eddington-limited thin-disk accretion, a $10^5\, \Msun$ BH seed planted in the host halo of $10^{10}\, \Msun$ at $z \sim 19$ can grow to $\sim 7\times 10^8\, \Msun$ at $z \sim 7.5$, resembling the SMBH of ULAS J1342+0928, currently the most distant quasar at $z=7.54$, and to $\sim 4 \times 10^9\, \Msun$ at $z \sim 6$, consistent with the majority of luminous quasars in that epoch. 

\item Light or intermediate seeds of $\lesssim 10^4\, \Msun$ fail to grow to $10^9\, \Msun$ by $z \sim 6$ despite super-critical accretion, while heavy seeds of $\gtrsim 10^5\, \Msun$ in $10^{10}\, \Msun$ halos can in both thin- and slim-disk  models, provided that the duty cycle of near-Eddington accretion with an Eddington ratio $\lambda_{\rm Edd} \gtrsim  0.6$ is maintained at $\gtrsim 40\%$. 

\item A hyper-Eddington model with $\lambda_{\rm Edd} >> 1$, rather than boosting BH growth, suffers from dramatically reduced accretion because of strong feedback to produce $\sim 10^9\, \Msun$ SMBHs by $z \sim 6$, while near-Eddington or moderately super-Eddington accretion produces more massive BHs than the former by two orders of magnitude.  

\item The super-critical model may lead to moderate super-Eddington accretion with $\lambda_{\rm Edd} \lesssim  3$ for low-spin BHs due to low radiative efficiency $\sim 10^{-4} \lesssim \epsilon_{\rm r} \lesssim 0.1$, which facilitates rapid BH growth. However, for high-spin BHs, the advantage is diminished sharply by increased radiative feedback. A $10^5\, \Msun$ BH seed can grow to $\sim 8\times 10^9\, \Msun$ by $z \sim 6$ under super-critical accretion with no spin, but only to $\sim 6\times 10^8\, \Msun$ if it spins maximally.  

\item Most of the BH mass is assembled through gas accretion of the main progenitor other than through the merger of the less massive BHs. There is good agreement on the BH mass among various Bondi accretion models, as the BH growth is largely dominated by the Eddington-accretion phase.

\item The quasar host galaxy experiences strong starbursts with star formation rates of $10^2 - 10^3\, \Msun/yr$ triggered by violent galaxy interactions. The host galaxy is massive with a stellar mass of $5.9 \times 10^{11}\, \Msun$, yet it is highly compact, with 2/3 of the stellar mass concentrating within 2.5 kpc from the galaxy center. 

\end{itemize}

Overall, our simulations show that the first $10^8 - 10^{9}\, \Msun$ SMBHs can grow from heavy seeds of $\gtrsim 10^4\, \Msun$ via rapid, self-regulated accretion booted by gas-rich mergers, even though we employ much stronger galactic feedback akin to the outflow model in the Illustris simulation than that in the \citet{Li2007} simulation. This suggests that the BH seeds play a critical role in the growth of SMBHs, and that more observations and theoretical works are needed to identify the formation mechanisms of potential heavy BH seeds and their subsequent evolution.

\section*{Acknowledgments}
We thank Phil Hopkins for a private version of \Gizmo code, and we thank Marta Volonteri, Rosa Valiante, Laura Pentericci and Luca Graziani for valuable discussions. QZ is supported by the McWilliams Fellowship from The McWilliams Center for Cosmology at Carnegie Mellon University. YL acknowledges support from NSF grants AST-1412719 and MRI-1626251.  YL and RS acknowledge support from the Amaldi Research Center funded by the MIUR program \lq \lq Dipartimento di Eccellenza\rq \rq \, (CUP:B81I18001170001), and from the INFN TEONGRAV specific initiative. HY acknowledges support from MEXT/JSPS KAKENHI Grant Number 17H04827, 20H04724, and NAOJ ALMA Scientific Research Grant Numbers 2019-11A. The numerical computations and data analysis in this paper have been carried out on the CyberLAMP cluster supported by MRI-1626251, operated and maintained by the Institute for CyberScience at the Pennsylvania State University. The Institute for Gravitation and the Cosmos is supported by the Eberly College of Science and the Office of the Senior Vice President for Research at the Pennsylvania State University. 

\section*{Data availability}
The data generated in this research will be shared on reasonable
request to the corresponding author.


\begin{thebibliography}{}
\makeatletter
\relax
\def\mn@urlcharsother{\let\do\@makeother \do\$\do\&\do\#\do\^\do\_\do\%\do\~}
\def\mn@doi{\begingroup\mn@urlcharsother \@ifnextchar [ {\mn@doi@}
  {\mn@doi@[]}}
\def\mn@doi@[#1]#2{\def\@tempa{#1}\ifx\@tempa\@empty \href
  {http://dx.doi.org/#2} {doi:#2}\else \href {http://dx.doi.org/#2} {#1}\fi
  \endgroup}
\def\mn@eprint#1#2{\mn@eprint@#1:#2::\@nil}
\def\mn@eprint@arXiv#1{\href {http://arxiv.org/abs/#1} {{\tt arXiv:#1}}}
\def\mn@eprint@dblp#1{\href {http://dblp.uni-trier.de/rec/bibtex/#1.xml}
  {dblp:#1}}
\def\mn@eprint@#1:#2:#3:#4\@nil{\def\@tempa {#1}\def\@tempb {#2}\def\@tempc
  {#3}\ifx \@tempc \@empty \let \@tempc \@tempb \let \@tempb \@tempa \fi \ifx
  \@tempb \@empty \def\@tempb {arXiv}\fi \@ifundefined
  {mn@eprint@\@tempb}{\@tempb:\@tempc}{\expandafter \expandafter \csname
  mn@eprint@\@tempb\endcsname \expandafter{\@tempc}}}

\bibitem[\protect\citeauthoryear{{Abramowicz} \& {Fragile}}{{Abramowicz} \&
  {Fragile}}{2013}]{Abramowicz2013}
{Abramowicz} M.~A.,  {Fragile} P.~C.,  2013, \mn@doi [Living Reviews in
  Relativity] {10.12942/lrr-2013-1}, \href
  {http://adsabs.harvard.edu/abs/2013LRR....16....1A} {16, 1}

\bibitem[\protect\citeauthoryear{{Abramowicz}, {Czerny}, {Lasota}  \&
  {Szuszkiewicz}}{{Abramowicz} et~al.}{1988}]{Abramowicz1988}
{Abramowicz} M.~A.,  {Czerny} B.,  {Lasota} J.~P.,   {Szuszkiewicz} E.,  1988,
  \mn@doi [\apj] {10.1086/166683}, \href
  {http://adsabs.harvard.edu/abs/1988ApJ...332..646A} {332, 646}

\bibitem[\protect\citeauthoryear{{Agarwal}, {Khochfar}, {Johnson}, {Neistein},
  {Dalla Vecchia}  \& {Livio}}{{Agarwal} et~al.}{2012}]{Agarwal2012}
{Agarwal} B.,  {Khochfar} S.,  {Johnson} J.~L.,  {Neistein} E.,  {Dalla
  Vecchia} C.,   {Livio} M.,  2012, \mn@doi [\mnras]
  {10.1111/j.1365-2966.2012.21651.x}, \href
  {http://adsabs.harvard.edu/abs/2012MNRAS.425.2854A} {425, 2854}

\bibitem[\protect\citeauthoryear{{Agarwal}, {Davis}, {Khochfar}, {Natarajan}
  \& {Dunlop}}{{Agarwal} et~al.}{2013}]{Agarwal2013}
{Agarwal} B.,  {Davis} A.~J.,  {Khochfar} S.,  {Natarajan} P.,   {Dunlop}
  J.~S.,  2013, \mn@doi [\mnras] {10.1093/mnras/stt696}, \href
  {https://ui.adsabs.harvard.edu/abs/2013MNRAS.432.3438A} {432, 3438}

\bibitem[\protect\citeauthoryear{{Agertz} et~al.,}{{Agertz}
  et~al.}{2007}]{Agertz2007}
{Agertz} O.,  et~al., 2007, \mn@doi [\mnras]
  {10.1111/j.1365-2966.2007.12183.x}, \href
  {https://ui.adsabs.harvard.edu/abs/2007MNRAS.380..963A} {380, 963}

\bibitem[\protect\citeauthoryear{{Alvarez}, {Wise}  \& {Abel}}{{Alvarez}
  et~al.}{2009}]{Alvarez2009}
{Alvarez} M.~A.,  {Wise} J.~H.,   {Abel} T.,  2009, \mn@doi [\apjl]
  {10.1088/0004-637X/701/2/L133}, \href
  {http://adsabs.harvard.edu/abs/2009ApJ...701L.133A} {701, L133}

\bibitem[\protect\citeauthoryear{{Angl{\'e}s-Alc{\'a}zar}, {Dav{\'e}},
  {Faucher-Gigu{\`e}re}, {{\"O}zel}  \& {Hopkins}}{{Angl{\'e}s-Alc{\'a}zar}
  et~al.}{2017}]{AnglesAlcazar2017}
{Angl{\'e}s-Alc{\'a}zar} D.,  {Dav{\'e}} R.,  {Faucher-Gigu{\`e}re} C.-A.,
  {{\"O}zel} F.,   {Hopkins} P.~F.,  2017, \mn@doi [\mnras]
  {10.1093/mnras/stw2565}, \href
  {http://adsabs.harvard.edu/abs/2017MNRAS.464.2840A} {464, 2840}

\bibitem[\protect\citeauthoryear{{Ba{\~n}ados} et~al.,}{{Ba{\~n}ados}
  et~al.}{2016}]{Banados2016}
{Ba{\~n}ados} E.,  et~al., 2016, \mn@doi [\apjs] {10.3847/0067-0049/227/1/11},
  \href {https://ui.adsabs.harvard.edu/abs/2016ApJS..227...11B} {227, 11}

\bibitem[\protect\citeauthoryear{{Ba{\~n}ados} et~al.,}{{Ba{\~n}ados}
  et~al.}{2018}]{Banados2018}
{Ba{\~n}ados} E.,  et~al., 2018, \mn@doi [\nat] {10.1038/nature25180}, \href
  {https://ui.adsabs.harvard.edu/abs/2018Natur.553..473B} {553, 473}

\bibitem[\protect\citeauthoryear{{Barnes} \& {Hernquist}}{{Barnes} \&
  {Hernquist}}{1996}]{Barnes1996}
{Barnes} J.~E.,  {Hernquist} L.,  1996, \mn@doi [\apj] {10.1086/177957}, \href
  {http://adsabs.harvard.edu/abs/1996ApJ...471..115B} {471, 115}

\bibitem[\protect\citeauthoryear{{Bauer} \& {Springel}}{{Bauer} \&
  {Springel}}{2012}]{Bauer2012}
{Bauer} A.,  {Springel} V.,  2012, \mn@doi [\mnras]
  {10.1111/j.1365-2966.2012.21058.x}, \href
  {https://ui.adsabs.harvard.edu/abs/2012MNRAS.423.2558B} {423, 2558}

\bibitem[\protect\citeauthoryear{{Becerra}, {Greif}, {Springel}  \&
  {Hernquist}}{{Becerra} et~al.}{2015}]{Becerra2015}
{Becerra} F.,  {Greif} T.~H.,  {Springel} V.,   {Hernquist} L.~E.,  2015,
  \mn@doi [\mnras] {10.1093/mnras/stu2284}, \href
  {http://adsabs.harvard.edu/abs/2015MNRAS.446.2380B} {446, 2380}

\bibitem[\protect\citeauthoryear{{Becerra}, {Marinacci}, {Bromm}  \&
  {Hernquist}}{{Becerra} et~al.}{2018}]{Becerra2018}
{Becerra} F.,  {Marinacci} F.,  {Bromm} V.,   {Hernquist} L.~E.,  2018, \mn@doi
  [\mnras] {10.1093/mnras/sty2210}, \href
  {https://ui.adsabs.harvard.edu/abs/2018MNRAS.480.5029B} {480, 5029}

\bibitem[\protect\citeauthoryear{{Begelman}, {Volonteri}  \& {Rees}}{{Begelman}
  et~al.}{2006}]{Begelman2006}
{Begelman} M.~C.,  {Volonteri} M.,   {Rees} M.~J.,  2006, \mn@doi [\mnras]
  {10.1111/j.1365-2966.2006.10467.x}, \href
  {http://adsabs.harvard.edu/abs/2006MNRAS.370..289B} {370, 289}

\bibitem[\protect\citeauthoryear{{Bekenstein}}{{Bekenstein}}{1973}]{Bekenstein1973}
{Bekenstein} J.~D.,  1973, \mn@doi [\apj] {10.1086/152255}, \href
  {http://adsabs.harvard.edu/abs/1973ApJ...183..657B} {183, 657}

\bibitem[\protect\citeauthoryear{{Blecha} \& {Loeb}}{{Blecha} \&
  {Loeb}}{2008}]{Blecha2008}
{Blecha} L.,  {Loeb} A.,  2008, \mn@doi [\mnras]
  {10.1111/j.1365-2966.2008.13790.x}, \href
  {http://adsabs.harvard.edu/abs/2008MNRAS.390.1311B} {390, 1311}

\bibitem[\protect\citeauthoryear{{Blecha}, {Cox}, {Loeb}  \&
  {Hernquist}}{{Blecha} et~al.}{2011}]{Blecha2011}
{Blecha} L.,  {Cox} T.~J.,  {Loeb} A.,   {Hernquist} L.,  2011, \mn@doi
  [\mnras] {10.1111/j.1365-2966.2010.18042.x}, \href
  {http://adsabs.harvard.edu/abs/2011MNRAS.412.2154B} {412, 2154}

\bibitem[\protect\citeauthoryear{{Bondi} \& {Hoyle}}{{Bondi} \&
  {Hoyle}}{1944}]{Bondi1944}
{Bondi} H.,  {Hoyle} F.,  1944, \mn@doi [\mnras] {10.1093/mnras/104.5.273},
  \href {http://adsabs.harvard.edu/abs/1944MNRAS.104..273B} {104, 273}

\bibitem[\protect\citeauthoryear{{Booth} \& {Schaye}}{{Booth} \&
  {Schaye}}{2009}]{Booth2009}
{Booth} C.~M.,  {Schaye} J.,  2009, \mn@doi [\mnras]
  {10.1111/j.1365-2966.2009.15043.x}, \href
  {http://adsabs.harvard.edu/abs/2009MNRAS.398...53B} {398, 53}

\bibitem[\protect\citeauthoryear{{Bourne}, {Zubovas}  \& {Nayakshin}}{{Bourne}
  et~al.}{2015}]{Bourne2015}
{Bourne} M.~A.,  {Zubovas} K.,   {Nayakshin} S.,  2015, \mn@doi [\mnras]
  {10.1093/mnras/stv1730}, \href
  {http://adsabs.harvard.edu/abs/2015MNRAS.453.1829B} {453, 1829}

\bibitem[\protect\citeauthoryear{{Bromm} \& {Loeb}}{{Bromm} \&
  {Loeb}}{2003}]{Bromm2003}
{Bromm} V.,  {Loeb} A.,  2003, \mn@doi [\apj] {10.1086/377529}, \href
  {http://adsabs.harvard.edu/abs/2003ApJ...596...34B} {596, 34}

\bibitem[\protect\citeauthoryear{{Calura}, {Gilli}, {Vignali}, {Pozzi},
  {Pipino}  \& {Matteucci}}{{Calura} et~al.}{2014}]{Calura2014}
{Calura} F.,  {Gilli} R.,  {Vignali} C.,  {Pozzi} F.,  {Pipino} A.,
  {Matteucci} F.,  2014, \mn@doi [\mnras] {10.1093/mnras/stt2329}, \href
  {http://adsabs.harvard.edu/abs/2014MNRAS.438.2765C} {438, 2765}

\bibitem[\protect\citeauthoryear{{Chon} \& {Hosokawa}}{{Chon} \&
  {Hosokawa}}{2019}]{Chon2019}
{Chon} S.,  {Hosokawa} T.,  2019, \mn@doi [\mnras] {10.1093/mnras/stz1824},
  \href {https://ui.adsabs.harvard.edu/abs/2019MNRAS.488.2658C} {488, 2658}

\bibitem[\protect\citeauthoryear{{Chon}, {Hosokawa}  \& {Yoshida}}{{Chon}
  et~al.}{2018}]{Chon2018}
{Chon} S.,  {Hosokawa} T.,   {Yoshida} N.,  2018, \mn@doi [\mnras]
  {10.1093/mnras/sty086}, \href
  {https://ui.adsabs.harvard.edu/abs/2018MNRAS.475.4104C} {475, 4104}

\bibitem[\protect\citeauthoryear{{Costa}, {Sijacki}, {Trenti}  \&
  {Haehnelt}}{{Costa} et~al.}{2014}]{Costa2014}
{Costa} T.,  {Sijacki} D.,  {Trenti} M.,   {Haehnelt} M.~G.,  2014, \mn@doi
  [\mnras] {10.1093/mnras/stu101}, \href
  {http://adsabs.harvard.edu/abs/2014MNRAS.439.2146C} {439, 2146}

\bibitem[\protect\citeauthoryear{{Costa}, {Sijacki}  \& {Haehnelt}}{{Costa}
  et~al.}{2015}]{Costa2015}
{Costa} T.,  {Sijacki} D.,   {Haehnelt} M.~G.,  2015, \mn@doi [\mnras]
  {10.1093/mnrasl/slu193}, \href
  {http://adsabs.harvard.edu/abs/2015MNRAS.448L..30C} {448, L30}

\bibitem[\protect\citeauthoryear{{Crain} et~al.,}{{Crain}
  et~al.}{2015}]{Crain2015}
{Crain} R.~A.,  et~al., 2015, \mn@doi [\mnras] {10.1093/mnras/stv725}, \href
  {https://ui.adsabs.harvard.edu/abs/2015MNRAS.450.1937C} {450, 1937}

\bibitem[\protect\citeauthoryear{{Curtis} \& {Sijacki}}{{Curtis} \&
  {Sijacki}}{2015}]{Curtis2015}
{Curtis} M.,  {Sijacki} D.,  2015, \mn@doi [\mnras] {10.1093/mnras/stv2246},
  \href {http://adsabs.harvard.edu/abs/2015MNRAS.454.3445C} {454, 3445}

\bibitem[\protect\citeauthoryear{{Curtis} \& {Sijacki}}{{Curtis} \&
  {Sijacki}}{2016a}]{Curtis2016}
{Curtis} M.,  {Sijacki} D.,  2016a, \mn@doi [\mnras] {10.1093/mnrasl/slv199},
  \href {http://adsabs.harvard.edu/abs/2016MNRAS.457L..34C} {457, L34}

\bibitem[\protect\citeauthoryear{{Curtis} \& {Sijacki}}{{Curtis} \&
  {Sijacki}}{2016b}]{Curtis2016b}
{Curtis} M.,  {Sijacki} D.,  2016b, \mn@doi [\mnras] {10.1093/mnras/stw1944},
  \href {http://adsabs.harvard.edu/abs/2016MNRAS.463...63C} {463, 63}

\bibitem[\protect\citeauthoryear{{Dav{\'e}}, {Angl{\'e}s-Alc{\'a}zar},
  {Narayanan}, {Li}, {Rafieferantsoa}  \& {Appleby}}{{Dav{\'e}}
  et~al.}{2019}]{Dave2019}
{Dav{\'e}} R.,  {Angl{\'e}s-Alc{\'a}zar} D.,  {Narayanan} D.,  {Li} Q.,
  {Rafieferantsoa} M.~H.,   {Appleby} S.,  2019, \mn@doi [\mnras]
  {10.1093/mnras/stz937}, \href
  {https://ui.adsabs.harvard.edu/abs/2019MNRAS.486.2827D} {486, 2827}

\bibitem[\protect\citeauthoryear{{Davies}, {Hennawi}  \& {Eilers}}{{Davies}
  et~al.}{2019}]{Davies2019}
{Davies} F.~B.,  {Hennawi} J.~F.,   {Eilers} A.-C.,  2019, \mn@doi [\apjl]
  {10.3847/2041-8213/ab42e3}, \href
  {https://ui.adsabs.harvard.edu/abs/2019ApJ...884L..19D} {884, L19}

\bibitem[\protect\citeauthoryear{{Davis}, {Efstathiou}, {Frenk}  \&
  {White}}{{Davis} et~al.}{1985}]{Davis1985}
{Davis} M.,  {Efstathiou} G.,  {Frenk} C.~S.,   {White} S.~D.~M.,  1985,
  \mn@doi [\apj] {10.1086/163168}, \href
  {https://ui.adsabs.harvard.edu/abs/1985ApJ...292..371D} {292, 371}

\bibitem[\protect\citeauthoryear{{DeGraf} \& {Sijacki}}{{DeGraf} \&
  {Sijacki}}{2020}]{DeGraf2020}
{DeGraf} C.,  {Sijacki} D.,  2020, \mn@doi [\mnras] {10.1093/mnras/stz3309},
  \href {https://ui.adsabs.harvard.edu/abs/2020MNRAS.491.4973D} {491, 4973}

\bibitem[\protect\citeauthoryear{{Decarli} et~al.,}{{Decarli}
  et~al.}{2018}]{Decarli2018}
{Decarli} R.,  et~al., 2018, \mn@doi [\apj] {10.3847/1538-4357/aaa5aa}, \href
  {https://ui.adsabs.harvard.edu/abs/2018ApJ...854...97D} {854, 97}

\bibitem[\protect\citeauthoryear{{Di Matteo}, {Springel}  \& {Hernquist}}{{Di
  Matteo} et~al.}{2005}]{DiMatteo2005}
{Di Matteo} T.,  {Springel} V.,   {Hernquist} L.,  2005, \mn@doi [\nat]
  {10.1038/nature03335}, \href
  {http://adsabs.harvard.edu/abs/2005Natur.433..604D} {433, 604}

\bibitem[\protect\citeauthoryear{{Di Matteo}, {Colberg}, {Springel},
  {Hernquist}  \& {Sijacki}}{{Di Matteo} et~al.}{2008}]{DiMatteo2008}
{Di Matteo} T.,  {Colberg} J.,  {Springel} V.,  {Hernquist} L.,   {Sijacki} D.,
   2008, \mn@doi [\apj] {10.1086/524921}, \href
  {http://adsabs.harvard.edu/abs/2008ApJ...676...33D} {676, 33}

\bibitem[\protect\citeauthoryear{{Di Matteo}, {Khandai}, {DeGraf}, {Feng},
  {Croft}, {Lopez}  \& {Springel}}{{Di Matteo} et~al.}{2012}]{DiMatteo2012}
{Di Matteo} T.,  {Khandai} N.,  {DeGraf} C.,  {Feng} Y.,  {Croft} R.~A.~C.,
  {Lopez} J.,   {Springel} V.,  2012, \mn@doi [\apjl]
  {10.1088/2041-8205/745/2/L29}, \href
  {http://adsabs.harvard.edu/abs/2012ApJ...745L..29D} {745, L29}

\bibitem[\protect\citeauthoryear{{Dolag}, {Komatsu}  \& {Sunyaev}}{{Dolag}
  et~al.}{2016}]{Dolag2016}
{Dolag} K.,  {Komatsu} E.,   {Sunyaev} R.,  2016, \mn@doi [\mnras]
  {10.1093/mnras/stw2035}, \href
  {https://ui.adsabs.harvard.edu/abs/2016MNRAS.463.1797D} {463, 1797}

\bibitem[\protect\citeauthoryear{{Dubois}, {Devriendt}, {Teyssier}  \&
  {Slyz}}{{Dubois} et~al.}{2011}]{Dubois2011}
{Dubois} Y.,  {Devriendt} J.,  {Teyssier} R.,   {Slyz} A.,  2011, \mn@doi
  [\mnras] {10.1111/j.1365-2966.2011.19381.x}, \href
  {http://adsabs.harvard.edu/abs/2011MNRAS.417.1853D} {417, 1853}

\bibitem[\protect\citeauthoryear{{Dubois}, {Pichon}, {Haehnelt}, {Kimm},
  {Slyz}, {Devriendt}  \& {Pogosyan}}{{Dubois} et~al.}{2012}]{Dubois2012}
{Dubois} Y.,  {Pichon} C.,  {Haehnelt} M.,  {Kimm} T.,  {Slyz} A.,  {Devriendt}
  J.,   {Pogosyan} D.,  2012, \mn@doi [\mnras]
  {10.1111/j.1365-2966.2012.21160.x}, \href
  {http://adsabs.harvard.edu/abs/2012MNRAS.423.3616D} {423, 3616}

\bibitem[\protect\citeauthoryear{{Dubois}, {Pichon}, {Devriendt}, {Silk},
  {Haehnelt}, {Kimm}  \& {Slyz}}{{Dubois} et~al.}{2013}]{Dubois2013}
{Dubois} Y.,  {Pichon} C.,  {Devriendt} J.,  {Silk} J.,  {Haehnelt} M.,  {Kimm}
  T.,   {Slyz} A.,  2013, \mn@doi [\mnras] {10.1093/mnras/sts224}, \href
  {http://adsabs.harvard.edu/abs/2013MNRAS.428.2885D} {428, 2885}

\bibitem[\protect\citeauthoryear{{Dubois} et~al.,}{{Dubois}
  et~al.}{2014}]{Dubois2014}
{Dubois} Y.,  et~al., 2014, \mn@doi [\mnras] {10.1093/mnras/stu1227}, \href
  {https://ui.adsabs.harvard.edu/abs/2014MNRAS.444.1453D} {444, 1453}

\bibitem[\protect\citeauthoryear{{Eggen}, {Lynden-Bell}  \& {Sandage}}{{Eggen}
  et~al.}{1962}]{ELS1962}
{Eggen} O.~J.,  {Lynden-Bell} D.,   {Sandage} A.~R.,  1962, \mn@doi [\apj]
  {10.1086/147433}, \href {http://adsabs.harvard.edu/abs/1962ApJ...136..748E}
  {136, 748}

\bibitem[\protect\citeauthoryear{{Fan} et~al.,}{{Fan} et~al.}{2001}]{Fan2001}
{Fan} X.,  et~al., 2001, \mn@doi [\aj] {10.1086/324111}, \href
  {https://ui.adsabs.harvard.edu/abs/2001AJ....122.2833F} {122, 2833}

\bibitem[\protect\citeauthoryear{{Faucher-Gigu{\`e}re}, {Kere{\v s}},
  {Dijkstra}, {Hernquist}  \& {Zaldarriaga}}{{Faucher-Gigu{\`e}re}
  et~al.}{2010}]{Faucher2010}
{Faucher-Gigu{\`e}re} C.-A.,  {Kere{\v s}} D.,  {Dijkstra} M.,  {Hernquist} L.,
    {Zaldarriaga} M.,  2010, \mn@doi [\apj] {10.1088/0004-637X/725/1/633},
  \href {http://adsabs.harvard.edu/abs/2010ApJ...725..633F} {725, 633}

\bibitem[\protect\citeauthoryear{{Feng}, {Di Matteo}, {Croft}  \&
  {Khandai}}{{Feng} et~al.}{2014}]{Feng2014}
{Feng} Y.,  {Di Matteo} T.,  {Croft} R.,   {Khandai} N.,  2014, \mn@doi
  [\mnras] {10.1093/mnras/stu432}, \href
  {http://adsabs.harvard.edu/abs/2014MNRAS.440.1865F} {440, 1865}

\bibitem[\protect\citeauthoryear{{Feng}, {Di-Matteo}, {Croft}, {Bird},
  {Battaglia}  \& {Wilkins}}{{Feng} et~al.}{2016}]{Feng2016}
{Feng} Y.,  {Di-Matteo} T.,  {Croft} R.~A.,  {Bird} S.,  {Battaglia} N.,
  {Wilkins} S.,  2016, \mn@doi [\mnras] {10.1093/mnras/stv2484}, \href
  {https://ui.adsabs.harvard.edu/abs/2016MNRAS.455.2778F} {455, 2778}

\bibitem[\protect\citeauthoryear{{Ferrara}, {Salvadori}, {Yue}  \&
  {Schleicher}}{{Ferrara} et~al.}{2014}]{Ferrara2014}
{Ferrara} A.,  {Salvadori} S.,  {Yue} B.,   {Schleicher} D.,  2014, \mn@doi
  [\mnras] {10.1093/mnras/stu1280}, \href
  {https://ui.adsabs.harvard.edu/abs/2014MNRAS.443.2410F} {443, 2410}

\bibitem[\protect\citeauthoryear{{Freitag}, {Rasio}  \& {Baumgardt}}{{Freitag}
  et~al.}{2006a}]{Freitag2006a}
{Freitag} M.,  {Rasio} F.~A.,   {Baumgardt} H.,  2006a, \mn@doi [\mnras]
  {10.1111/j.1365-2966.2006.10095.x}, \href
  {http://adsabs.harvard.edu/abs/2006MNRAS.368..121F} {368, 121}

\bibitem[\protect\citeauthoryear{{Freitag}, {G{\"u}rkan}  \& {Rasio}}{{Freitag}
  et~al.}{2006b}]{Freitag2006b}
{Freitag} M.,  {G{\"u}rkan} M.~A.,   {Rasio} F.~A.,  2006b, \mn@doi [\mnras]
  {10.1111/j.1365-2966.2006.10096.x}, \href
  {http://adsabs.harvard.edu/abs/2006MNRAS.368..141F} {368, 141}

\bibitem[\protect\citeauthoryear{{Fukushima}, {Hosokawa}, {Chiaki}, {Omukai},
  {Yoshida}  \& {Kuiper}}{{Fukushima} et~al.}{2020}]{Fukushima2020}
{Fukushima} H.,  {Hosokawa} T.,  {Chiaki} G.,  {Omukai} K.,  {Yoshida} N.,
  {Kuiper} R.,  2020, \mn@doi [\mnras] {10.1093/mnras/staa1994}, \href
  {https://ui.adsabs.harvard.edu/abs/2020MNRAS.497..829F} {497, 829}

\bibitem[\protect\citeauthoryear{{Gaspari}, {Ruszkowski}  \& {Oh}}{{Gaspari}
  et~al.}{2013}]{Gaspari2013}
{Gaspari} M.,  {Ruszkowski} M.,   {Oh} S.~P.,  2013, \mn@doi [\mnras]
  {10.1093/mnras/stt692}, \href
  {http://adsabs.harvard.edu/abs/2013MNRAS.432.3401G} {432, 3401}

\bibitem[\protect\citeauthoryear{{Gaspari}, {Brighenti}  \& {Temi}}{{Gaspari}
  et~al.}{2015}]{Gaspari2015}
{Gaspari} M.,  {Brighenti} F.,   {Temi} P.,  2015, \mn@doi [\aap]
  {10.1051/0004-6361/201526151}, \href
  {http://adsabs.harvard.edu/abs/2015A%26A...579A..62G} {579, A62}

\bibitem[\protect\citeauthoryear{{Genel} et~al.,}{{Genel}
  et~al.}{2014}]{Genel2014}
{Genel} S.,  et~al., 2014, \mn@doi [\mnras] {10.1093/mnras/stu1654}, \href
  {https://ui.adsabs.harvard.edu/abs/2014MNRAS.445..175G} {445, 175}

\bibitem[\protect\citeauthoryear{{Genzel} et~al.,}{{Genzel}
  et~al.}{2010}]{Genzel2010}
{Genzel} R.,  et~al., 2010, \mn@doi [\mnras]
  {10.1111/j.1365-2966.2010.16969.x}, \href
  {https://ui.adsabs.harvard.edu/abs/2010MNRAS.407.2091G} {407, 2091}

\bibitem[\protect\citeauthoryear{{Glazebrook} et~al.,}{{Glazebrook}
  et~al.}{2017}]{Glazebrook2017}
{Glazebrook} K.,  et~al., 2017, \mn@doi [\nat] {10.1038/nature21680}, \href
  {http://adsabs.harvard.edu/abs/2017Natur.544...71G} {544, 71}

\bibitem[\protect\citeauthoryear{{Glover}}{{Glover}}{2015a}]{Glover2015a}
{Glover} S. C.~O.,  2015a, \mn@doi [\mnras] {10.1093/mnras/stv1059}, \href
  {https://ui.adsabs.harvard.edu/abs/2015MNRAS.451.2082G} {451, 2082}

\bibitem[\protect\citeauthoryear{{Glover}}{{Glover}}{2015b}]{Glover2015b}
{Glover} S. C.~O.,  2015b, \mn@doi [\mnras] {10.1093/mnras/stv1781}, \href
  {https://ui.adsabs.harvard.edu/abs/2015MNRAS.453.2901G} {453, 2901}

\bibitem[\protect\citeauthoryear{{Greene}, {Strader}  \& {Ho}}{{Greene}
  et~al.}{2020}]{Greene2020}
{Greene} J.~E.,  {Strader} J.,   {Ho} L.~C.,  2020, \mn@doi [\araa]
  {10.1146/annurev-astro-032620-021835}, \href
  {https://ui.adsabs.harvard.edu/abs/2020ARA&A..58..257G} {58, 257}

\bibitem[\protect\citeauthoryear{{Haemmerl{\'e}}, {Mayer}, {Klessen},
  {Hosokawa}, {Madau}  \& {Bromm}}{{Haemmerl{\'e}}
  et~al.}{2020}]{Haemmerle2020}
{Haemmerl{\'e}} L.,  {Mayer} L.,  {Klessen} R.~S.,  {Hosokawa} T.,  {Madau} P.,
    {Bromm} V.,  2020, \mn@doi [\ssr] {10.1007/s11214-020-00673-y}, \href
  {https://ui.adsabs.harvard.edu/abs/2020SSRv..216...48H} {216, 48}

\bibitem[\protect\citeauthoryear{{Hahn} \& {Abel}}{{Hahn} \&
  {Abel}}{2011}]{Hahn2011}
{Hahn} O.,  {Abel} T.,  2011, \mn@doi [\mnras]
  {10.1111/j.1365-2966.2011.18820.x}, \href
  {http://adsabs.harvard.edu/abs/2011MNRAS.415.2101H} {415, 2101}

\bibitem[\protect\citeauthoryear{{Hirano} \& {Bromm}}{{Hirano} \&
  {Bromm}}{2017}]{Hirano2017}
{Hirano} S.,  {Bromm} V.,  2017, \mn@doi [\mnras] {10.1093/mnras/stx1220},
  \href {https://ui.adsabs.harvard.edu/abs/2017MNRAS.470..898H} {470, 898}

\bibitem[\protect\citeauthoryear{{Ho}}{{Ho}}{2007}]{Ho2007}
{Ho} L.~C.,  2007, \mn@doi [\apj] {10.1086/521917}, \href
  {http://adsabs.harvard.edu/abs/2007ApJ...669..821H} {669, 821}

\bibitem[\protect\citeauthoryear{{Hopkins}}{{Hopkins}}{2013}]{Hopkins2013}
{Hopkins} P.~F.,  2013, \mn@doi [\mnras] {10.1093/mnras/sts210}, \href
  {http://adsabs.harvard.edu/abs/2013MNRAS.428.2840H} {428, 2840}

\bibitem[\protect\citeauthoryear{{Hopkins}}{{Hopkins}}{2015}]{Hopkins2015}
{Hopkins} P.~F.,  2015, \mn@doi [\mnras] {10.1093/mnras/stv195}, \href
  {https://ui.adsabs.harvard.edu/abs/2015MNRAS.450...53H} {450, 53}

\bibitem[\protect\citeauthoryear{{Hopkins} \& {Quataert}}{{Hopkins} \&
  {Quataert}}{2010}]{Hopkins2010}
{Hopkins} P.~F.,  {Quataert} E.,  2010, \mn@doi [\mnras]
  {10.1111/j.1365-2966.2010.17064.x}, \href
  {http://adsabs.harvard.edu/abs/2010MNRAS.407.1529H} {407, 1529}

\bibitem[\protect\citeauthoryear{{Hopkins}, {Torrey}, {Faucher-Gigu{\`e}re},
  {Quataert}  \& {Murray}}{{Hopkins} et~al.}{2016}]{Hopkins2016}
{Hopkins} P.~F.,  {Torrey} P.,  {Faucher-Gigu{\`e}re} C.-A.,  {Quataert} E.,
  {Murray} N.,  2016, \mn@doi [\mnras] {10.1093/mnras/stw289}, \href
  {http://adsabs.harvard.edu/abs/2016MNRAS.458..816H} {458, 816}

\bibitem[\protect\citeauthoryear{{Huang}, {Ni}, {Feng}  \& {Di Matteo}}{{Huang}
  et~al.}{2020}]{Huang2019}
{Huang} K.-W.,  {Ni} Y.,  {Feng} Y.,   {Di Matteo} T.,  2020, \mn@doi [\mnras]
  {10.1093/mnras/staa1515}, \href
  {https://ui.adsabs.harvard.edu/abs/2020MNRAS.496....1H} {496, 1}

\bibitem[\protect\citeauthoryear{{Inayoshi} \& {Haiman}}{{Inayoshi} \&
  {Haiman}}{2016}]{Inayoshi2016}
{Inayoshi} K.,  {Haiman} Z.,  2016, \mn@doi [\apj]
  {10.3847/0004-637X/828/2/110}, \href
  {http://adsabs.harvard.edu/abs/2016ApJ...828..110I} {828, 110}

\bibitem[\protect\citeauthoryear{{Inayoshi}, {Visbal}  \&
  {Kashiyama}}{{Inayoshi} et~al.}{2015}]{Inayoshi2015}
{Inayoshi} K.,  {Visbal} E.,   {Kashiyama} K.,  2015, \mn@doi [\mnras]
  {10.1093/mnras/stv1654}, \href
  {http://adsabs.harvard.edu/abs/2015MNRAS.453.1692I} {453, 1692}

\bibitem[\protect\citeauthoryear{{Inayoshi}, {Visbal}  \& {Haiman}}{{Inayoshi}
  et~al.}{2020}]{Inayoshi2020}
{Inayoshi} K.,  {Visbal} E.,   {Haiman} Z.,  2020, \mn@doi [\araa]
  {10.1146/annurev-astro-120419-014455}, \href
  {https://ui.adsabs.harvard.edu/abs/2019arXiv191105791I} {58, 27}

\bibitem[\protect\citeauthoryear{{Jeon}, {Pawlik}, {Greif}, {Glover}, {Bromm},
  {Milosavljevi{\'c}}  \& {Klessen}}{{Jeon} et~al.}{2012}]{Jeon2012}
{Jeon} M.,  {Pawlik} A.~H.,  {Greif} T.~H.,  {Glover} S. C.~O.,  {Bromm} V.,
  {Milosavljevi{\'c}} M.,   {Klessen} R.~S.,  2012, \mn@doi [\apj]
  {10.1088/0004-637X/754/1/34}, \href
  {https://ui.adsabs.harvard.edu/abs/2012ApJ...754...34J} {754, 34}

\bibitem[\protect\citeauthoryear{{Jiang} et~al.,}{{Jiang}
  et~al.}{2016}]{Jiang2016}
{Jiang} L.,  et~al., 2016, \mn@doi [\apj] {10.3847/1538-4357/833/2/222}, \href
  {http://adsabs.harvard.edu/abs/2016ApJ...833..222J} {833, 222}

\bibitem[\protect\citeauthoryear{{Johnson}, {Khochfar}, {Greif}  \&
  {Durier}}{{Johnson} et~al.}{2011}]{Johnson2011}
{Johnson} J.~L.,  {Khochfar} S.,  {Greif} T.~H.,   {Durier} F.,  2011, \mn@doi
  [\mnras] {10.1111/j.1365-2966.2010.17491.x}, \href
  {http://adsabs.harvard.edu/abs/2011MNRAS.410..919J} {410, 919}

\bibitem[\protect\citeauthoryear{{Katz}, {Sijacki}  \& {Haehnelt}}{{Katz}
  et~al.}{2015}]{Katz2015}
{Katz} H.,  {Sijacki} D.,   {Haehnelt} M.~G.,  2015, \mn@doi [\mnras]
  {10.1093/mnras/stv1048}, \href
  {http://adsabs.harvard.edu/abs/2015MNRAS.451.2352K} {451, 2352}

\bibitem[\protect\citeauthoryear{{Kaviraj} et~al.,}{{Kaviraj}
  et~al.}{2017}]{Kaviraj2017}
{Kaviraj} S.,  et~al., 2017, \mn@doi [\mnras] {10.1093/mnras/stx126}, \href
  {http://adsabs.harvard.edu/abs/2017MNRAS.tmp..224K} {}

\bibitem[\protect\citeauthoryear{{Kelley}, {Blecha}  \& {Hernquist}}{{Kelley}
  et~al.}{2017}]{Kelley2017}
{Kelley} L.~Z.,  {Blecha} L.,   {Hernquist} L.,  2017, \mn@doi [\mnras]
  {10.1093/mnras/stw2452}, \href
  {https://ui.adsabs.harvard.edu/abs/2017MNRAS.464.3131K} {464, 3131}

\bibitem[\protect\citeauthoryear{{Khandai}, {Feng}, {DeGraf}, {Di Matteo}  \&
  {Croft}}{{Khandai} et~al.}{2012}]{Khandai2012}
{Khandai} N.,  {Feng} Y.,  {DeGraf} C.,  {Di Matteo} T.,   {Croft} R.~A.~C.,
  2012, \mn@doi [\mnras] {10.1111/j.1365-2966.2012.21047.x}, \href
  {http://adsabs.harvard.edu/abs/2012MNRAS.423.2397K} {423, 2397}

\bibitem[\protect\citeauthoryear{{Khandai}, {Di Matteo}, {Croft}, {Wilkins},
  {Feng}, {Tucker}, {DeGraf}  \& {Liu}}{{Khandai} et~al.}{2015}]{Khandai2015}
{Khandai} N.,  {Di Matteo} T.,  {Croft} R.,  {Wilkins} S.,  {Feng} Y.,
  {Tucker} E.,  {DeGraf} C.,   {Liu} M.-S.,  2015, \mn@doi [\mnras]
  {10.1093/mnras/stv627}, \href
  {http://adsabs.harvard.edu/abs/2015MNRAS.450.1349K} {450, 1349}

\bibitem[\protect\citeauthoryear{{Kim} et~al.,}{{Kim} et~al.}{2009}]{Kim2009}
{Kim} S.,  et~al., 2009, \mn@doi [\apj] {10.1088/0004-637X/695/2/809}, \href
  {https://ui.adsabs.harvard.edu/abs/2009ApJ...695..809K} {695, 809}

\bibitem[\protect\citeauthoryear{{Komatsu} et~al.,}{{Komatsu}
  et~al.}{2009}]{Komatsu2009}
{Komatsu} E.,  et~al., 2009, \mn@doi [\apjs] {10.1088/0067-0049/180/2/330},
  \href {https://ui.adsabs.harvard.edu/abs/2009ApJS..180..330K} {180, 330}

\bibitem[\protect\citeauthoryear{{Kormendy} \& {Ho}}{{Kormendy} \&
  {Ho}}{2013}]{Kormendy2013}
{Kormendy} J.,  {Ho} L.~C.,  2013, \mn@doi [\araa]
  {10.1146/annurev-astro-082708-101811}, \href
  {http://adsabs.harvard.edu/abs/2013ARA%26A..51..511K} {51, 511}

\bibitem[\protect\citeauthoryear{{Latif} \& {Ferrara}}{{Latif} \&
  {Ferrara}}{2016}]{Latif2016}
{Latif} M.~A.,  {Ferrara} A.,  2016, \mn@doi [\pasa] {10.1017/pasa.2016.41},
  \href {https://ui.adsabs.harvard.edu/abs/2016PASA...33...51L} {33, e051}

\bibitem[\protect\citeauthoryear{{Lauer}, {Tremaine}, {Richstone}  \&
  {Faber}}{{Lauer} et~al.}{2007}]{Lauer2007}
{Lauer} T.~R.,  {Tremaine} S.,  {Richstone} D.,   {Faber} S.~M.,  2007, \mn@doi
  [\apj] {10.1086/522083}, \href
  {http://adsabs.harvard.edu/abs/2007ApJ...670..249L} {670, 249}

\bibitem[\protect\citeauthoryear{{Leitherer} et~al.,}{{Leitherer}
  et~al.}{1999}]{Leitherer1999}
{Leitherer} C.,  et~al., 1999, \mn@doi [\apjs] {10.1086/313233}, \href
  {https://ui.adsabs.harvard.edu/abs/1999ApJS..123....3L} {123, 3}

\bibitem[\protect\citeauthoryear{{Leitherer} et~al.,}{{Leitherer}
  et~al.}{2011}]{Leitherer2011}
{Leitherer} C.,  et~al., 2011, {Starburst99: Synthesis Models for Galaxies with
  Active Star Formation} (\mn@eprint {ascl} {1104.003})

\bibitem[\protect\citeauthoryear{{Leroy}, {Walter}, {Brinks}, {Bigiel}, {de
  Blok}, {Madore}  \& {Thornley}}{{Leroy} et~al.}{2008}]{Leroy2008}
{Leroy} A.~K.,  {Walter} F.,  {Brinks} E.,  {Bigiel} F.,  {de Blok} W.~J.~G.,
  {Madore} B.,   {Thornley} M.~D.,  2008, \mn@doi [\aj]
  {10.1088/0004-6256/136/6/2782}, \href
  {http://adsabs.harvard.edu/abs/2008AJ....136.2782L} {136, 2782}

\bibitem[\protect\citeauthoryear{{Li} et~al.,}{{Li} et~al.}{2007}]{Li2007}
{Li} Y.,  et~al., 2007, \mn@doi [\apj] {10.1086/519297}, \href
  {https://ui.adsabs.harvard.edu/abs/2007ApJ...665..187L} {665, 187}

\bibitem[\protect\citeauthoryear{{Li} et~al.,}{{Li} et~al.}{2008}]{Li2008}
{Li} Y.,  et~al., 2008, \mn@doi [\apj] {10.1086/529364}, \href
  {https://ui.adsabs.harvard.edu/abs/2008ApJ...678...41L} {678, 41}

\bibitem[\protect\citeauthoryear{{Luo}, {Shlosman}, {Nagamine}  \&
  {Fang}}{{Luo} et~al.}{2020}]{Luo2020}
{Luo} Y.,  {Shlosman} I.,  {Nagamine} K.,   {Fang} T.,  2020, \mn@doi [\mnras]
  {10.1093/mnras/staa153}, \href
  {https://ui.adsabs.harvard.edu/abs/2020MNRAS.492.4917L} {492, 4917}

\bibitem[\protect\citeauthoryear{{Lupi}, {Haardt}, {Dotti}, {Fiacconi}, {Mayer}
   \& {Madau}}{{Lupi} et~al.}{2016}]{Lupi2016}
{Lupi} A.,  {Haardt} F.,  {Dotti} M.,  {Fiacconi} D.,  {Mayer} L.,   {Madau}
  P.,  2016, \mn@doi [\mnras] {10.1093/mnras/stv2877}, \href
  {http://adsabs.harvard.edu/abs/2016MNRAS.456.2993L} {456, 2993}

\bibitem[\protect\citeauthoryear{{Lupi}, {Volonteri}, {Decarli}, {Bovino},
  {Silk}  \& {Bergeron}}{{Lupi} et~al.}{2019}]{Lupi2019}
{Lupi} A.,  {Volonteri} M.,  {Decarli} R.,  {Bovino} S.,  {Silk} J.,
  {Bergeron} J.,  2019, \mn@doi [\mnras] {10.1093/mnras/stz1959}, \href
  {https://ui.adsabs.harvard.edu/abs/2019MNRAS.488.4004L} {488, 4004}

\bibitem[\protect\citeauthoryear{{Madau}, {Haardt}  \& {Dotti}}{{Madau}
  et~al.}{2014}]{Madau2014}
{Madau} P.,  {Haardt} F.,   {Dotti} M.,  2014, \mn@doi [\apjl]
  {10.1088/2041-8205/784/2/L38}, \href
  {http://adsabs.harvard.edu/abs/2014ApJ...784L..38M} {784, L38}

\bibitem[\protect\citeauthoryear{{Marchesini} et~al.,}{{Marchesini}
  et~al.}{2010}]{Marchesini2010}
{Marchesini} D.,  et~al., 2010, \mn@doi [\apj] {10.1088/0004-637X/725/1/1277},
  \href {http://ads.ari.uni-heidelberg.de/abs/2010ApJ...725.1277M} {725, 1277}

\bibitem[\protect\citeauthoryear{{Matsuoka} et~al.,}{{Matsuoka}
  et~al.}{2018}]{Matsuoka2018}
{Matsuoka} Y.,  et~al., 2018, \mn@doi [\apj] {10.3847/1538-4357/aaee7a}, \href
  {https://ui.adsabs.harvard.edu/abs/2018ApJ...869..150M} {869, 150}

\bibitem[\protect\citeauthoryear{{Matsuoka} et~al.,}{{Matsuoka}
  et~al.}{2019}]{Matsuoka2019}
{Matsuoka} Y.,  et~al., 2019, \mn@doi [\apjl] {10.3847/2041-8213/ab0216}, \href
  {https://ui.adsabs.harvard.edu/abs/2019ApJ...872L...2M} {872, L2}

\bibitem[\protect\citeauthoryear{{Mayer} \& {Bonoli}}{{Mayer} \&
  {Bonoli}}{2019}]{Mayer2019}
{Mayer} L.,  {Bonoli} S.,  2019, \mn@doi [Reports on Progress in Physics]
  {10.1088/1361-6633/aad6a5}, \href
  {https://ui.adsabs.harvard.edu/abs/2019RPPh...82a6901M} {82, 016901}

\bibitem[\protect\citeauthoryear{{Mayer}, {Kazantzidis}, {Escala}  \&
  {Callegari}}{{Mayer} et~al.}{2010}]{Mayer2010}
{Mayer} L.,  {Kazantzidis} S.,  {Escala} A.,   {Callegari} S.,  2010, \mn@doi
  [\nat] {10.1038/nature09294}, \href
  {https://ui.adsabs.harvard.edu/abs/2010Natur.466.1082M} {466, 1082}

\bibitem[\protect\citeauthoryear{{Mazzucchelli}, {Ba{\~n}ados}, {Decarli},
  {Farina}, {Venemans}, {Walter}  \& {Overzier}}{{Mazzucchelli}
  et~al.}{2017}]{Mazzucchelli2017}
{Mazzucchelli} C.,  {Ba{\~n}ados} E.,  {Decarli} R.,  {Farina} E.~P.,
  {Venemans} B.~P.,  {Walter} F.,   {Overzier} R.,  2017, \mn@doi [\apj]
  {10.3847/1538-4357/834/1/83}, \href
  {https://ui.adsabs.harvard.edu/abs/2017ApJ...834...83M} {834, 83}

\bibitem[\protect\citeauthoryear{{Mobasher} et~al.,}{{Mobasher}
  et~al.}{2005}]{Mobasher2005}
{Mobasher} B.,  et~al., 2005, \mn@doi [\apj] {10.1086/497626}, \href
  {http://ads.ari.uni-heidelberg.de/abs/2005ApJ...635..832M} {635, 832}

\bibitem[\protect\citeauthoryear{{Mortlock} et~al.,}{{Mortlock}
  et~al.}{2011}]{Mortlock2011}
{Mortlock} D.~J.,  et~al., 2011, \mn@doi [\nat] {10.1038/nature10159}, \href
  {https://ui.adsabs.harvard.edu/abs/2011Natur.474..616M} {474, 616}

\bibitem[\protect\citeauthoryear{{Muratov}, {Kere{\v s}},
  {Faucher-Gigu{\`e}re}, {Hopkins}, {Quataert}  \& {Murray}}{{Muratov}
  et~al.}{2015}]{Muratov2015}
{Muratov} A.~L.,  {Kere{\v s}} D.,  {Faucher-Gigu{\`e}re} C.-A.,  {Hopkins}
  P.~F.,  {Quataert} E.,   {Murray} N.,  2015, \mn@doi [\mnras]
  {10.1093/mnras/stv2126}, \href
  {http://adsabs.harvard.edu/abs/2015MNRAS.454.2691M} {454, 2691}

\bibitem[\protect\citeauthoryear{{Narayan}, {Sa{\`I}{\textsection}dowski}  \&
  {Soria}}{{Narayan} et~al.}{2017}]{Narayan2017}
{Narayan} R.,  {Sa{\`I}{\textsection}dowski} A.,   {Soria} R.,  2017, \mn@doi
  [\mnras] {10.1093/mnras/stx1027}, \href
  {https://ui.adsabs.harvard.edu/abs/2017MNRAS.469.2997N} {469, 2997}

\bibitem[\protect\citeauthoryear{{Negri} \& {Volonteri}}{{Negri} \&
  {Volonteri}}{2017}]{Negri2016}
{Negri} A.,  {Volonteri} M.,  2017, \mn@doi [\mnras] {10.1093/mnras/stx362},
  \href {https://ui.adsabs.harvard.edu/abs/2017MNRAS.467.3475N} {467, 3475}

\bibitem[\protect\citeauthoryear{{Nelson}, {Vogelsberger}, {Genel}, {Sijacki},
  {Kere{\v s}}, {Springel}  \& {Hernquist}}{{Nelson} et~al.}{2013}]{Nelson2013}
{Nelson} D.,  {Vogelsberger} M.,  {Genel} S.,  {Sijacki} D.,  {Kere{\v s}} D.,
  {Springel} V.,   {Hernquist} L.,  2013, \mn@doi [\mnras]
  {10.1093/mnras/sts595}, \href
  {http://adsabs.harvard.edu/abs/2013MNRAS.429.3353N} {429, 3353}

\bibitem[\protect\citeauthoryear{{Onoue} et~al.,}{{Onoue}
  et~al.}{2019}]{Onoue2019}
{Onoue} M.,  et~al., 2019, \mn@doi [\apj] {10.3847/1538-4357/ab29e9}, \href
  {https://ui.adsabs.harvard.edu/abs/2019ApJ...880...77O} {880, 77}

\bibitem[\protect\citeauthoryear{{Pacucci}, {Volonteri}  \&
  {Ferrara}}{{Pacucci} et~al.}{2015}]{Pacucci2015a}
{Pacucci} F.,  {Volonteri} M.,   {Ferrara} A.,  2015, \mn@doi [\mnras]
  {10.1093/mnras/stv1465}, \href
  {https://ui.adsabs.harvard.edu/abs/2015MNRAS.452.1922P} {452, 1922}

\bibitem[\protect\citeauthoryear{{Pacucci}, {Natarajan}, {Volonteri},
  {Cappelluti}  \& {Urry}}{{Pacucci} et~al.}{2017}]{Pacucci2017}
{Pacucci} F.,  {Natarajan} P.,  {Volonteri} M.,  {Cappelluti} N.,   {Urry}
  C.~M.,  2017, \mn@doi [\apjl] {10.3847/2041-8213/aa9aea}, \href
  {https://ui.adsabs.harvard.edu/abs/2017ApJ...850L..42P} {850, L42}

\bibitem[\protect\citeauthoryear{{Pelupessy}, {Di Matteo}  \&
  {Ciardi}}{{Pelupessy} et~al.}{2007}]{Pelupessy2007}
{Pelupessy} F.~I.,  {Di Matteo} T.,   {Ciardi} B.,  2007, \mn@doi [\apj]
  {10.1086/519235}, \href {http://adsabs.harvard.edu/abs/2007ApJ...665..107P}
  {665, 107}

\bibitem[\protect\citeauthoryear{{Pensabene}, {Carniani}, {Perna}, {Cresci},
  {Decarli}, {Maiolino}  \& {Marconi}}{{Pensabene}
  et~al.}{2020}]{Pensabene2020}
{Pensabene} A.,  {Carniani} S.,  {Perna} M.,  {Cresci} G.,  {Decarli} R.,
  {Maiolino} R.,   {Marconi} A.,  2020, \mn@doi [\aap]
  {10.1051/0004-6361/201936634}, \href
  {https://ui.adsabs.harvard.edu/abs/2020A&A...637A..84P} {637, A84}

\bibitem[\protect\citeauthoryear{{Pezzulli}, {Valiante}  \&
  {Schneider}}{{Pezzulli} et~al.}{2016}]{Pezzulli2016}
{Pezzulli} E.,  {Valiante} R.,   {Schneider} R.,  2016, \mn@doi [\mnras]
  {10.1093/mnras/stw505}, \href
  {http://adsabs.harvard.edu/abs/2016MNRAS.458.3047P} {458, 3047}

\bibitem[\protect\citeauthoryear{{Pezzulli}, {Valiante}, {Orofino},
  {Schneider}, {Gallerani}  \& {Sbarrato}}{{Pezzulli}
  et~al.}{2017}]{Pezzulli2017}
{Pezzulli} E.,  {Valiante} R.,  {Orofino} M.~C.,  {Schneider} R.,  {Gallerani}
  S.,   {Sbarrato} T.,  2017, \mn@doi [\mnras] {10.1093/mnras/stw3243}, \href
  {http://adsabs.harvard.edu/abs/2017MNRAS.466.2131P} {466, 2131}

\bibitem[\protect\citeauthoryear{{Pillepich} et~al.,}{{Pillepich}
  et~al.}{2018a}]{Pillepich2017}
{Pillepich} A.,  et~al., 2018a, \mn@doi [\mnras] {10.1093/mnras/stx2656}, \href
  {https://ui.adsabs.harvard.edu/abs/2018MNRAS.473.4077P} {473, 4077}

\bibitem[\protect\citeauthoryear{{Pillepich} et~al.,}{{Pillepich}
  et~al.}{2018b}]{Pillepich2018b}
{Pillepich} A.,  et~al., 2018b, \mn@doi [\mnras] {10.1093/mnras/stx3112}, \href
  {https://ui.adsabs.harvard.edu/abs/2018MNRAS.475..648P} {475, 648}

\bibitem[\protect\citeauthoryear{{Planck Collaboration} et~al.,}{{Planck
  Collaboration} et~al.}{2016}]{Planck2016}
{Planck Collaboration} et~al., 2016, \mn@doi [\aap]
  {10.1051/0004-6361/201525830}, \href
  {https://ui.adsabs.harvard.edu/abs/2016A&A...594A..13P} {594, A13}

\bibitem[\protect\citeauthoryear{{Portegies Zwart} \& {McMillan}}{{Portegies
  Zwart} \& {McMillan}}{2000}]{Zwart2000}
{Portegies Zwart} S.~F.,  {McMillan} S.~L.~W.,  2000, \mn@doi [\apjl]
  {10.1086/312422}, \href {http://adsabs.harvard.edu/abs/2000ApJ...528L..17P}
  {528, L17}

\bibitem[\protect\citeauthoryear{{Portegies Zwart} \& {McMillan}}{{Portegies
  Zwart} \& {McMillan}}{2002}]{Zwart2002}
{Portegies Zwart} S.~F.,  {McMillan} S.~L.~W.,  2002, \mn@doi [\apj]
  {10.1086/341798}, \href {http://adsabs.harvard.edu/abs/2002ApJ...576..899P}
  {576, 899}

\bibitem[\protect\citeauthoryear{{Reed} et~al.,}{{Reed}
  et~al.}{2017}]{Reed2017}
{Reed} S.~L.,  et~al., 2017, \mn@doi [\mnras] {10.1093/mnras/stx728}, \href
  {https://ui.adsabs.harvard.edu/abs/2017MNRAS.468.4702R} {468, 4702}

\bibitem[\protect\citeauthoryear{{Rees}}{{Rees}}{1978}]{Rees1978}
{Rees} M.~J.,  1978, \mn@doi [\physscr] {10.1088/0031-8949/17/3/010}, \href
  {http://adsabs.harvard.edu/abs/1978PhyS...17..193R} {17, 193}

\bibitem[\protect\citeauthoryear{{Regan}, {Visbal}, {Wise}, {Haiman},
  {Johansson}  \& {Bryan}}{{Regan} et~al.}{2017}]{Regan2017}
{Regan} J.~A.,  {Visbal} E.,  {Wise} J.~H.,  {Haiman} Z.,  {Johansson} P.~H.,
  {Bryan} G.~L.,  2017, \mn@doi [Nature Astronomy] {10.1038/s41550-017-0075},
  \href {https://ui.adsabs.harvard.edu/abs/2017NatAs...1E..75R} {1, 0075}

\bibitem[\protect\citeauthoryear{{Regan}, {Downes}, {Volonteri}, {Beckmann},
  {Lupi}, {Trebitsch}  \& {Dubois}}{{Regan} et~al.}{2019}]{Regan2019}
{Regan} J.~A.,  {Downes} T.~P.,  {Volonteri} M.,  {Beckmann} R.,  {Lupi} A.,
  {Trebitsch} M.,   {Dubois} Y.,  2019, \mn@doi [\mnras]
  {10.1093/mnras/stz1045}, \href
  {https://ui.adsabs.harvard.edu/abs/2019MNRAS.486.3892R} {486, 3892}

\bibitem[\protect\citeauthoryear{{Reines} \& {Volonteri}}{{Reines} \&
  {Volonteri}}{2015}]{Reines2015}
{Reines} A.~E.,  {Volonteri} M.,  2015, \mn@doi [\apj]
  {10.1088/0004-637X/813/2/82}, \href
  {https://ui.adsabs.harvard.edu/abs/2015ApJ...813...82R} {813, 82}

\bibitem[\protect\citeauthoryear{{Ricarte} \& {Natarajan}}{{Ricarte} \&
  {Natarajan}}{2018}]{Ricarte2018}
{Ricarte} A.,  {Natarajan} P.,  2018, \mn@doi [\mnras] {10.1093/mnras/sty2448},
  \href {https://ui.adsabs.harvard.edu/abs/2018MNRAS.481.3278R} {481, 3278}

\bibitem[\protect\citeauthoryear{{Ryu}, {Tanaka}, {Perna}  \& {Haiman}}{{Ryu}
  et~al.}{2016}]{Ryu2016}
{Ryu} T.,  {Tanaka} T.~L.,  {Perna} R.,   {Haiman} Z.,  2016, \mn@doi [\mnras]
  {10.1093/mnras/stw1241}, \href
  {https://ui.adsabs.harvard.edu/abs/2016MNRAS.460.4122R} {460, 4122}

\bibitem[\protect\citeauthoryear{{Sakurai}, {Yoshida}, {Fujii}  \&
  {Hirano}}{{Sakurai} et~al.}{2017}]{Sakurai2017}
{Sakurai} Y.,  {Yoshida} N.,  {Fujii} M.~S.,   {Hirano} S.,  2017, \mn@doi
  [\mnras] {10.1093/mnras/stx2044}, \href
  {https://ui.adsabs.harvard.edu/abs/2017MNRAS.472.1677S} {472, 1677}

\bibitem[\protect\citeauthoryear{{S{\c a}dowski}}{{S{\c
  a}dowski}}{2009}]{Scdowski2009}
{S{\c a}dowski} A.,  2009, \mn@doi [\apjs] {10.1088/0067-0049/183/2/171}, \href
  {http://adsabs.harvard.edu/abs/2009ApJS..183..171S} {183, 171}

\bibitem[\protect\citeauthoryear{{S{\c a}dowski}, {Lasota}, {Abramowicz}  \&
  {Narayan}}{{S{\c a}dowski} et~al.}{2016}]{Sadowski2016a}
{S{\c a}dowski} A.,  {Lasota} J.-P.,  {Abramowicz} M.~A.,   {Narayan} R.,
  2016, \mn@doi [\mnras] {10.1093/mnras/stv2854}, \href
  {http://adsabs.harvard.edu/abs/2016MNRAS.456.3915S} {456, 3915}

\bibitem[\protect\citeauthoryear{{Schaye} et~al.,}{{Schaye}
  et~al.}{2015}]{Schaye2015}
{Schaye} J.,  et~al., 2015, \mn@doi [\mnras] {10.1093/mnras/stu2058}, \href
  {https://ui.adsabs.harvard.edu/abs/2015MNRAS.446..521S} {446, 521}

\bibitem[\protect\citeauthoryear{{Shakura} \& {Sunyaev}}{{Shakura} \&
  {Sunyaev}}{1973}]{Shakura1973}
{Shakura} N.~I.,  {Sunyaev} R.~A.,  1973, \aap, \href
  {http://adsabs.harvard.edu/abs/1973A%26A....24..337S} {24, 337}

\bibitem[\protect\citeauthoryear{{Shen} et~al.,}{{Shen}
  et~al.}{2019}]{Shen2019}
{Shen} Y.,  et~al., 2019, \mn@doi [\apj] {10.3847/1538-4357/ab03d9}, \href
  {https://ui.adsabs.harvard.edu/abs/2019ApJ...873...35S} {873, 35}

\bibitem[\protect\citeauthoryear{{Sijacki}, {Springel}, {Di Matteo}  \&
  {Hernquist}}{{Sijacki} et~al.}{2007}]{Sijacki2007}
{Sijacki} D.,  {Springel} V.,  {Di Matteo} T.,   {Hernquist} L.,  2007, \mn@doi
  [\mnras] {10.1111/j.1365-2966.2007.12153.x}, \href
  {http://adsabs.harvard.edu/abs/2007MNRAS.380..877S} {380, 877}

\bibitem[\protect\citeauthoryear{{Sijacki}, {Springel}  \&
  {Haehnelt}}{{Sijacki} et~al.}{2009}]{Sijacki2009}
{Sijacki} D.,  {Springel} V.,   {Haehnelt} M.~G.,  2009, \mn@doi [\mnras]
  {10.1111/j.1365-2966.2009.15452.x}, \href
  {http://adsabs.harvard.edu/abs/2009MNRAS.400..100S} {400, 100}

\bibitem[\protect\citeauthoryear{{Sijacki}, {Vogelsberger}, {Genel},
  {Springel}, {Torrey}, {Snyder}, {Nelson}  \& {Hernquist}}{{Sijacki}
  et~al.}{2015}]{Sijacki2015}
{Sijacki} D.,  {Vogelsberger} M.,  {Genel} S.,  {Springel} V.,  {Torrey} P.,
  {Snyder} G.~F.,  {Nelson} D.,   {Hernquist} L.,  2015, \mn@doi [\mnras]
  {10.1093/mnras/stv1340}, \href
  {http://adsabs.harvard.edu/abs/2015MNRAS.452..575S} {452, 575}

\bibitem[\protect\citeauthoryear{{Simpson}, {Mortlock}, {Warren}, {Cantalupo},
  {Hewett}, {McLure}, {McMahon}  \& {Venemans}}{{Simpson}
  et~al.}{2014}]{Simpson2014}
{Simpson} C.,  {Mortlock} D.,  {Warren} S.,  {Cantalupo} S.,  {Hewett} P.,
  {McLure} R.,  {McMahon} R.,   {Venemans} B.,  2014, \mn@doi [\mnras]
  {10.1093/mnras/stu1116}, \href
  {http://adsabs.harvard.edu/abs/2014MNRAS.442.3454S} {442, 3454}

\bibitem[\protect\citeauthoryear{{S{\k{a}}dowski}}{{S{\k{a}}dowski}}{2009}]{Sadowski2009}
{S{\k{a}}dowski} A.,  2009, \mn@doi [\apjs] {10.1088/0067-0049/183/2/171},
  \href {https://ui.adsabs.harvard.edu/abs/2009ApJS..183..171S} {183, 171}

\bibitem[\protect\citeauthoryear{{Smidt}, {Whalen}, {Johnson}, {Surace}  \&
  {Li}}{{Smidt} et~al.}{2018}]{Smidt2018}
{Smidt} J.,  {Whalen} D.~J.,  {Johnson} J.~L.,  {Surace} M.,   {Li} H.,  2018,
  \mn@doi [\apj] {10.3847/1538-4357/aad7b8}, \href
  {https://ui.adsabs.harvard.edu/abs/2018ApJ...865..126S} {865, 126}

\bibitem[\protect\citeauthoryear{{Smith}, {Wise}, {O'Shea}, {Norman}  \&
  {Khochfar}}{{Smith} et~al.}{2015}]{Smith2015}
{Smith} B.~D.,  {Wise} J.~H.,  {O'Shea} B.~W.,  {Norman} M.~L.,   {Khochfar}
  S.,  2015, \mn@doi [\mnras] {10.1093/mnras/stv1509}, \href
  {https://ui.adsabs.harvard.edu/abs/2015MNRAS.452.2822S} {452, 2822}

\bibitem[\protect\citeauthoryear{{Smith} et~al.,}{{Smith}
  et~al.}{2017}]{Smith2017}
{Smith} B.~D.,  et~al., 2017, \mn@doi [\mnras] {10.1093/mnras/stw3291}, \href
  {http://adsabs.harvard.edu/abs/2017MNRAS.466.2217S} {466, 2217}

\bibitem[\protect\citeauthoryear{{Smith}, {Regan}, {Downes}, {Norman}, {O'Shea}
   \& {Wise}}{{Smith} et~al.}{2018}]{Smith2018}
{Smith} B.~D.,  {Regan} J.~A.,  {Downes} T.~P.,  {Norman} M.~L.,  {O'Shea}
  B.~W.,   {Wise} J.~H.,  2018, \mn@doi [\mnras] {10.1093/mnras/sty2103}, \href
  {https://ui.adsabs.harvard.edu/abs/2018MNRAS.480.3762S} {480, 3762}

\bibitem[\protect\citeauthoryear{{Soltan}}{{Soltan}}{1982}]{Soltan1982}
{Soltan} A.,  1982, \mn@doi [\mnras] {10.1093/mnras/200.1.115}, \href
  {https://ui.adsabs.harvard.edu/abs/1982MNRAS.200..115S} {200, 115}

\bibitem[\protect\citeauthoryear{{Spergel} et~al.,}{{Spergel}
  et~al.}{2003}]{Spergel2003}
{Spergel} D.~N.,  et~al., 2003, \mn@doi [\apjs] {10.1086/377226}, \href
  {https://ui.adsabs.harvard.edu/abs/2003ApJS..148..175S} {148, 175}

\bibitem[\protect\citeauthoryear{{Springel} \& {Hernquist}}{{Springel} \&
  {Hernquist}}{2003}]{Springel2003}
{Springel} V.,  {Hernquist} L.,  2003, \mn@doi [\mnras]
  {10.1046/j.1365-8711.2003.06206.x}, \href
  {https://ui.adsabs.harvard.edu/abs/2003MNRAS.339..289S} {339, 289}

\bibitem[\protect\citeauthoryear{{Springel}, {Di Matteo}  \&
  {Hernquist}}{{Springel} et~al.}{2005a}]{Springel2005model}
{Springel} V.,  {Di Matteo} T.,   {Hernquist} L.,  2005a, \mn@doi [\mnras]
  {10.1111/j.1365-2966.2005.09238.x}, \href
  {https://ui.adsabs.harvard.edu/abs/2005MNRAS.361..776S} {361, 776}

\bibitem[\protect\citeauthoryear{{Springel} et~al.,}{{Springel}
  et~al.}{2005b}]{Springel2005qso}
{Springel} V.,  et~al., 2005b, \mn@doi [\nat] {10.1038/nature03597}, \href
  {https://ui.adsabs.harvard.edu/abs/2005Natur.435..629S} {435, 629}

\bibitem[\protect\citeauthoryear{{Springel} et~al.,}{{Springel}
  et~al.}{2018}]{Springel2018}
{Springel} V.,  et~al., 2018, \mn@doi [\mnras] {10.1093/mnras/stx3304}, \href
  {https://ui.adsabs.harvard.edu/abs/2018MNRAS.475..676S} {475, 676}

\bibitem[\protect\citeauthoryear{{Stacy}, {Bromm}  \& {Lee}}{{Stacy}
  et~al.}{2016}]{Stacy2016}
{Stacy} A.,  {Bromm} V.,   {Lee} A.~T.,  2016, \mn@doi [\mnras]
  {10.1093/mnras/stw1728}, \href
  {https://ui.adsabs.harvard.edu/abs/2016MNRAS.462.1307S} {462, 1307}

\bibitem[\protect\citeauthoryear{{Stefanon} et~al.,}{{Stefanon}
  et~al.}{2015}]{Stefanon2015}
{Stefanon} M.,  et~al., 2015, \mn@doi [\apj] {10.1088/0004-637X/803/1/11},
  \href {http://ads.ari.uni-heidelberg.de/abs/2015ApJ...803...11S} {803, 11}

\bibitem[\protect\citeauthoryear{{Steinhardt}, {Capak}, {Masters}  \&
  {Speagle}}{{Steinhardt} et~al.}{2016}]{Steinhardt2016}
{Steinhardt} C.~L.,  {Capak} P.,  {Masters} D.,   {Speagle} J.~S.,  2016,
  \mn@doi [\apj] {10.3847/0004-637X/824/1/21}, \href
  {http://adsabs.harvard.edu/abs/2016ApJ...824...21S} {824, 21}

\bibitem[\protect\citeauthoryear{{Sugimura}, {Hosokawa}, {Yajima}  \&
  {Omukai}}{{Sugimura} et~al.}{2017}]{Sugimura2017}
{Sugimura} K.,  {Hosokawa} T.,  {Yajima} H.,   {Omukai} K.,  2017, \mn@doi
  [\mnras] {10.1093/mnras/stx769}, \href
  {https://ui.adsabs.harvard.edu/abs/2017MNRAS.469...62S} {469, 62}

\bibitem[\protect\citeauthoryear{{Sugimura}, {Matsumoto}, {Hosokawa}, {Hirano}
  \& {Omukai}}{{Sugimura} et~al.}{2020}]{Sugimura2020}
{Sugimura} K.,  {Matsumoto} T.,  {Hosokawa} T.,  {Hirano} S.,   {Omukai} K.,
  2020, \mn@doi [\apjl] {10.3847/2041-8213/ab7d37}, \href
  {https://ui.adsabs.harvard.edu/abs/2020ApJ...892L..14S} {892, L14}

\bibitem[\protect\citeauthoryear{{Tagawa}, {Haiman}  \& {Kocsis}}{{Tagawa}
  et~al.}{2020}]{Tagawa2019}
{Tagawa} H.,  {Haiman} Z.,   {Kocsis} B.,  2020, \mn@doi [\apj]
  {10.3847/1538-4357/ab7922}, \href
  {https://ui.adsabs.harvard.edu/abs/2020ApJ...892...36T} {892, 36}

\bibitem[\protect\citeauthoryear{{Takeo}, {Inayoshi}, {Ohsuga}, {Takahashi}  \&
  {Mineshige}}{{Takeo} et~al.}{2019}]{Takeo2019}
{Takeo} E.,  {Inayoshi} K.,  {Ohsuga} K.,  {Takahashi} H.~R.,   {Mineshige} S.,
   2019, \mn@doi [\mnras] {10.1093/mnras/stz1899}, \href
  {https://ui.adsabs.harvard.edu/abs/2019MNRAS.488.2689T} {488, 2689}

\bibitem[\protect\citeauthoryear{{Tanaka} \& {Haiman}}{{Tanaka} \&
  {Haiman}}{2009}]{Tanaka2009}
{Tanaka} T.,  {Haiman} Z.,  2009, \mn@doi [\apj]
  {10.1088/0004-637X/696/2/1798}, \href
  {https://ui.adsabs.harvard.edu/abs/2009ApJ...696.1798T} {696, 1798}

\bibitem[\protect\citeauthoryear{{Tanaka}, {Perna}  \& {Haiman}}{{Tanaka}
  et~al.}{2012}]{Tanaka2012}
{Tanaka} T.,  {Perna} R.,   {Haiman} Z.,  2012, \mn@doi [\mnras]
  {10.1111/j.1365-2966.2012.21539.x}, \href
  {https://ui.adsabs.harvard.edu/abs/2012MNRAS.425.2974T} {425, 2974}

\bibitem[\protect\citeauthoryear{{Tenneti}, {Wilkins}, {Di Matteo}, {Croft}  \&
  {Feng}}{{Tenneti} et~al.}{2019}]{Tenneti2019}
{Tenneti} A.,  {Wilkins} S.~M.,  {Di Matteo} T.,  {Croft} R. A.~C.,   {Feng}
  Y.,  2019, \mn@doi [\mnras] {10.1093/mnras/sty3161}, \href
  {https://ui.adsabs.harvard.edu/abs/2019MNRAS.483.1388T} {483, 1388}

\bibitem[\protect\citeauthoryear{{Thomas}, {Dav{\'e}}, {Angl{\'e}s-Alc{\'a}zar}
   \& {Jarvis}}{{Thomas} et~al.}{2019}]{Thomas2019}
{Thomas} N.,  {Dav{\'e}} R.,  {Angl{\'e}s-Alc{\'a}zar} D.,   {Jarvis} M.,
  2019, \mn@doi [\mnras] {10.1093/mnras/stz1703}, \href
  {https://ui.adsabs.harvard.edu/abs/2019MNRAS.487.5764T} {487, 5764}

\bibitem[\protect\citeauthoryear{{Tohline}, {Simonson}  \&
  {Caldwell}}{{Tohline} et~al.}{1982}]{Tohline1982}
{Tohline} J.~E.,  {Simonson} G.~F.,   {Caldwell} N.,  1982, \mn@doi [\apj]
  {10.1086/159537}, \href {http://adsabs.harvard.edu/abs/1982ApJ...252...92T}
  {252, 92}

\bibitem[\protect\citeauthoryear{{Townsend}}{{Townsend}}{2009}]{Townsend2009}
{Townsend} R.~H.~D.,  2009, \mn@doi [\apjs] {10.1088/0067-0049/181/2/391},
  \href {http://adsabs.harvard.edu/abs/2009ApJS..181..391T} {181, 391}

\bibitem[\protect\citeauthoryear{{Tremblay} et~al.,}{{Tremblay}
  et~al.}{2016}]{Tremblay2016}
{Tremblay} G.~R.,  et~al., 2016, \mn@doi [\nat] {10.1038/nature17969}, \href
  {http://adsabs.harvard.edu/abs/2016Natur.534..218T} {534, 218}

\bibitem[\protect\citeauthoryear{{Valiante}, {Schneider}, {Salvadori}  \&
  {Gallerani}}{{Valiante} et~al.}{2014}]{Valiante2014}
{Valiante} R.,  {Schneider} R.,  {Salvadori} S.,   {Gallerani} S.,  2014,
  \mn@doi [\mnras] {10.1093/mnras/stu1613}, \href
  {http://adsabs.harvard.edu/abs/2014MNRAS.444.2442V} {444, 2442}

\bibitem[\protect\citeauthoryear{{Valiante}, {Schneider}, {Volonteri}  \&
  {Omukai}}{{Valiante} et~al.}{2016}]{Valiante2016}
{Valiante} R.,  {Schneider} R.,  {Volonteri} M.,   {Omukai} K.,  2016, \mn@doi
  [\mnras] {10.1093/mnras/stw225}, \href
  {http://adsabs.harvard.edu/abs/2016MNRAS.457.3356V} {457, 3356}

\bibitem[\protect\citeauthoryear{{Valiante}, {Agarwal}, {Habouzit}  \&
  {Pezzulli}}{{Valiante} et~al.}{2017}]{Valiante2017}
{Valiante} R.,  {Agarwal} B.,  {Habouzit} M.,   {Pezzulli} E.,  2017, \mn@doi
  [\pasa] {10.1017/pasa.2017.25}, \href
  {https://ui.adsabs.harvard.edu/abs/2017PASA...34...31V} {34, e031}

\bibitem[\protect\citeauthoryear{{Venemans} et~al.,}{{Venemans}
  et~al.}{2015}]{Venemans2015}
{Venemans} B.~P.,  et~al., 2015, \mn@doi [\mnras] {10.1093/mnras/stv1774},
  \href {http://adsabs.harvard.edu/abs/2015MNRAS.453.2259V} {453, 2259}

\bibitem[\protect\citeauthoryear{{Venemans} et~al.,}{{Venemans}
  et~al.}{2018}]{Venemans2018}
{Venemans} B.~P.,  et~al., 2018, \mn@doi [\apj] {10.3847/1538-4357/aadf35},
  \href {https://ui.adsabs.harvard.edu/abs/2018ApJ...866..159V} {866, 159}

\bibitem[\protect\citeauthoryear{{Vito} et~al.,}{{Vito}
  et~al.}{2019}]{Vito2019}
{Vito} F.,  et~al., 2019, \mn@doi [\aap] {10.1051/0004-6361/201936217}, \href
  {https://ui.adsabs.harvard.edu/abs/2019A&A...630A.118V} {630, A118}

\bibitem[\protect\citeauthoryear{{Vogelsberger}, {Genel}, {Sijacki}, {Torrey},
  {Springel}  \& {Hernquist}}{{Vogelsberger} et~al.}{2013}]{Vogelsberger2013}
{Vogelsberger} M.,  {Genel} S.,  {Sijacki} D.,  {Torrey} P.,  {Springel} V.,
  {Hernquist} L.,  2013, \mn@doi [\mnras] {10.1093/mnras/stt1789}, \href
  {http://adsabs.harvard.edu/abs/2013MNRAS.436.3031V} {436, 3031}

\bibitem[\protect\citeauthoryear{{Vogelsberger} et~al.,}{{Vogelsberger}
  et~al.}{2014a}]{Vogelsberger2014b}
{Vogelsberger} M.,  et~al., 2014a, \mn@doi [\mnras] {10.1093/mnras/stu1536},
  \href {http://adsabs.harvard.edu/abs/2014MNRAS.444.1518V} {444, 1518}

\bibitem[\protect\citeauthoryear{{Vogelsberger} et~al.,}{{Vogelsberger}
  et~al.}{2014b}]{Vogelsberger2014}
{Vogelsberger} M.,  et~al., 2014b, \mn@doi [\mnras] {10.1093/mnras/stu1536},
  \href {https://ui.adsabs.harvard.edu/abs/2014MNRAS.444.1518V} {444, 1518}

\bibitem[\protect\citeauthoryear{{Vogelsberger} et~al.,}{{Vogelsberger}
  et~al.}{2014c}]{Vogelsberger2014a}
{Vogelsberger} M.,  et~al., 2014c, \mn@doi [\nat] {10.1038/nature13316}, \href
  {https://ui.adsabs.harvard.edu/abs/2014Natur.509..177V} {509, 177}

\bibitem[\protect\citeauthoryear{{Volonteri}}{{Volonteri}}{2010}]{Volonteri2010}
{Volonteri} M.,  2010, \mn@doi [\aapr] {10.1007/s00159-010-0029-x}, \href
  {https://ui.adsabs.harvard.edu/abs/2010A&ARv..18..279V} {18, 279}

\bibitem[\protect\citeauthoryear{{Volonteri} \& {Bellovary}}{{Volonteri} \&
  {Bellovary}}{2012}]{Volonteri2012a}
{Volonteri} M.,  {Bellovary} J.,  2012, \mn@doi [Reports on Progress in
  Physics] {10.1088/0034-4885/75/12/124901}, \href
  {https://ui.adsabs.harvard.edu/abs/2012RPPh...75l4901V} {75, 124901}

\bibitem[\protect\citeauthoryear{{Volonteri} \& {Rees}}{{Volonteri} \&
  {Rees}}{2006}]{Volonteri2006}
{Volonteri} M.,  {Rees} M.~J.,  2006, \mn@doi [\apj] {10.1086/507444}, \href
  {http://adsabs.harvard.edu/abs/2006ApJ...650..669V} {650, 669}

\bibitem[\protect\citeauthoryear{{Volonteri} \& {Stark}}{{Volonteri} \&
  {Stark}}{2011}]{Volonteri2011a}
{Volonteri} M.,  {Stark} D.~P.,  2011, \mn@doi [\mnras]
  {10.1111/j.1365-2966.2011.19391.x}, \href
  {https://ui.adsabs.harvard.edu/abs/2011MNRAS.417.2085V} {417, 2085}

\bibitem[\protect\citeauthoryear{{Volonteri}, {Sikora}, {Lasota}  \&
  {Merloni}}{{Volonteri} et~al.}{2013}]{Volonteri2013}
{Volonteri} M.,  {Sikora} M.,  {Lasota} J.-P.,   {Merloni} A.,  2013, \mn@doi
  [\apj] {10.1088/0004-637X/775/2/94}, \href
  {http://adsabs.harvard.edu/abs/2013ApJ...775...94V} {775, 94}

\bibitem[\protect\citeauthoryear{{Volonteri}, {Silk}  \& {Dubus}}{{Volonteri}
  et~al.}{2015}]{Volonteri2015}
{Volonteri} M.,  {Silk} J.,   {Dubus} G.,  2015, \mn@doi [\apj]
  {10.1088/0004-637X/804/2/148}, \href
  {http://adsabs.harvard.edu/abs/2015ApJ...804..148V} {804, 148}

\bibitem[\protect\citeauthoryear{{Wang} et~al.,}{{Wang}
  et~al.}{2010}]{Wang2010}
{Wang} R.,  et~al., 2010, \mn@doi [\apj] {10.1088/0004-637X/714/1/699}, \href
  {http://adsabs.harvard.edu/abs/2010ApJ...714..699W} {714, 699}

\bibitem[\protect\citeauthoryear{{Wang} et~al.,}{{Wang}
  et~al.}{2013}]{Wang2013}
{Wang} R.,  et~al., 2013, \mn@doi [\apj] {10.1088/0004-637X/773/1/44}, \href
  {http://adsabs.harvard.edu/abs/2013ApJ...773...44W} {773, 44}

\bibitem[\protect\citeauthoryear{{Wang} et~al.,}{{Wang}
  et~al.}{2019}]{Wang2019}
{Wang} F.,  et~al., 2019, \mn@doi [\apj] {10.3847/1538-4357/ab2be5}, \href
  {https://ui.adsabs.harvard.edu/abs/2019ApJ...884...30W} {884, 30}

\bibitem[\protect\citeauthoryear{{Weinberger} et~al.,}{{Weinberger}
  et~al.}{2017}]{Weinberger2017}
{Weinberger} R.,  et~al., 2017, \mn@doi [\mnras] {10.1093/mnras/stw2944}, \href
  {https://ui.adsabs.harvard.edu/abs/2017MNRAS.465.3291W} {465, 3291}

\bibitem[\protect\citeauthoryear{{Weinberger} et~al.,}{{Weinberger}
  et~al.}{2018}]{Weinberger2018}
{Weinberger} R.,  et~al., 2018, \mn@doi [\mnras] {10.1093/mnras/sty1733}, \href
  {https://ui.adsabs.harvard.edu/abs/2018MNRAS.479.4056W} {479, 4056}

\bibitem[\protect\citeauthoryear{{Wiklind}, {Dickinson}, {Ferguson},
  {Giavalisco}, {Mobasher}, {Grogin}  \& {Panagia}}{{Wiklind}
  et~al.}{2008}]{Wiklind2008}
{Wiklind} T.,  {Dickinson} M.,  {Ferguson} H.~C.,  {Giavalisco} M.,  {Mobasher}
  B.,  {Grogin} N.~A.,   {Panagia} N.,  2008, \mn@doi [\apj] {10.1086/524919},
  \href {http://ads.ari.uni-heidelberg.de/abs/2008ApJ...676..781W} {676, 781}

\bibitem[\protect\citeauthoryear{{Willott} et~al.,}{{Willott}
  et~al.}{2010}]{Willott2010}
{Willott} C.~J.,  et~al., 2010, \mn@doi [\aj] {10.1088/0004-6256/139/3/906},
  \href {https://ui.adsabs.harvard.edu/abs/2010AJ....139..906W} {139, 906}

\bibitem[\protect\citeauthoryear{{Wise}, {Regan}, {O'Shea}, {Norman}, {Downes}
  \& {Xu}}{{Wise} et~al.}{2019}]{Wise2019}
{Wise} J.~H.,  {Regan} J.~A.,  {O'Shea} B.~W.,  {Norman} M.~L.,  {Downes}
  T.~P.,   {Xu} H.,  2019, \mn@doi [\nat] {10.1038/s41586-019-0873-4}, \href
  {https://ui.adsabs.harvard.edu/abs/2019Natur.566...85W} {566, 85}

\bibitem[\protect\citeauthoryear{{Woods}, {Heger}, {Whalen}, {Haemmerl{\'e}}
  \& {Klessen}}{{Woods} et~al.}{2017}]{Woods2017}
{Woods} T.~E.,  {Heger} A.,  {Whalen} D.~J.,  {Haemmerl{\'e}} L.,   {Klessen}
  R.~S.,  2017, \mn@doi [\apjl] {10.3847/2041-8213/aa7412}, \href
  {https://ui.adsabs.harvard.edu/abs/2017ApJ...842L...6W} {842, L6}

\bibitem[\protect\citeauthoryear{{Woods} et~al.,}{{Woods}
  et~al.}{2019}]{Woods2019}
{Woods} T.~E.,  et~al., 2019, \mn@doi [\pasa] {10.1017/pasa.2019.14}, \href
  {https://ui.adsabs.harvard.edu/abs/2019PASA...36...27W} {36, e027}

\bibitem[\protect\citeauthoryear{{Wu} et~al.,}{{Wu} et~al.}{2015}]{Wu2015}
{Wu} X.-B.,  et~al., 2015, \mn@doi [\nat] {10.1038/nature14241}, \href
  {http://adsabs.harvard.edu/abs/2015Natur.518..512W} {518, 512}

\bibitem[\protect\citeauthoryear{{Wurster} \& {Thacker}}{{Wurster} \&
  {Thacker}}{2013}]{Wurster2013}
{Wurster} J.,  {Thacker} R.~J.,  2013, \mn@doi [\mnras] {10.1093/mnras/stt346},
  \href {http://adsabs.harvard.edu/abs/2013MNRAS.431.2513W} {431, 2513}

\bibitem[\protect\citeauthoryear{{Yajima} \& {Khochfar}}{{Yajima} \&
  {Khochfar}}{2016}]{Yajima2016}
{Yajima} H.,  {Khochfar} S.,  2016, \mn@doi [\mnras] {10.1093/mnras/stw058},
  \href {http://adsabs.harvard.edu/abs/2016MNRAS.457.2423Y} {457, 2423}

\bibitem[\protect\citeauthoryear{{Yang} et~al.,}{{Yang}
  et~al.}{2019}]{Yang2019}
{Yang} J.,  et~al., 2019, \mn@doi [\aj] {10.3847/1538-3881/ab1be1}, \href
  {https://ui.adsabs.harvard.edu/abs/2019AJ....157..236Y} {157, 236}

\bibitem[\protect\citeauthoryear{{Yang} et~al.,}{{Yang}
  et~al.}{2020}]{Yang2020}
{Yang} J.,  et~al., 2020, \mn@doi [\apjl] {10.3847/2041-8213/ab9c26}, \href
  {https://ui.adsabs.harvard.edu/abs/2020ApJ...897L..14Y} {897, L14}

\bibitem[\protect\citeauthoryear{{Yi} et~al.,}{{Yi} et~al.}{2014}]{Yi2014}
{Yi} W.-M.,  et~al., 2014, \mn@doi [\apjl] {10.1088/2041-8205/795/2/L29}, \href
  {http://adsabs.harvard.edu/abs/2014ApJ...795L..29Y} {795, L29}

\bibitem[\protect\citeauthoryear{{Zhu} \& {Li}}{{Zhu} \& {Li}}{2016}]{Zhu2016}
{Zhu} Q.,  {Li} Y.,  2016, \mn@doi [\apj] {10.3847/0004-637X/831/1/52}, \href
  {http://adsabs.harvard.edu/abs/2016ApJ...831...52Z} {831, 52}

\bibitem[\protect\citeauthoryear{{Zhu}, {Hernquist}  \& {Li}}{{Zhu}
  et~al.}{2015}]{Zhu2015}
{Zhu} Q.,  {Hernquist} L.,   {Li} Y.,  2015, \mn@doi [\apj]
  {10.1088/0004-637X/800/1/6}, \href
  {https://ui.adsabs.harvard.edu/abs/2015ApJ...800....6Z} {800, 6}

\bibitem[\protect\citeauthoryear{{Zhu}, {Smith}  \& {Hernquist}}{{Zhu}
  et~al.}{2017}]{Zhu2017}
{Zhu} Q.,  {Smith} B.,   {Hernquist} L.,  2017, \mn@doi [\mnras]
  {10.1093/mnras/stx1346}, \href
  {http://adsabs.harvard.edu/abs/2017MNRAS.470.1017Z} {470, 1017}

\bibitem[\protect\citeauthoryear{{Zubovas}, {Bourne}  \& {Nayakshin}}{{Zubovas}
  et~al.}{2016}]{Zubovas2016}
{Zubovas} K.,  {Bourne} M.~A.,   {Nayakshin} S.,  2016, \mn@doi [\mnras]
  {10.1093/mnras/stv2971}, \href
  {http://adsabs.harvard.edu/abs/2016MNRAS.457..496Z} {457, 496}

\bibitem[\protect\citeauthoryear{{van de Voort}, {Davis}, {Kere{\v s}},
  {Quataert}, {Faucher-Gigu{\`e}re}  \& {Hopkins}}{{van de Voort}
  et~al.}{2015}]{vandeVoort2015}
{van de Voort} F.,  {Davis} T.~A.,  {Kere{\v s}} D.,  {Quataert} E.,
  {Faucher-Gigu{\`e}re} C.-A.,   {Hopkins} P.~F.,  2015, \mn@doi [\mnras]
  {10.1093/mnras/stv1217}, \href
  {http://adsabs.harvard.edu/abs/2015MNRAS.451.3269V} {451, 3269}

\makeatother
\end{thebibliography}

\bsp	
\label{lastpage}
\end{document}